\documentclass[review,hidelinks,onefignum,onetabnum]{siamart251216}

\usepackage{lipsum}
\usepackage{amsfonts}
\usepackage{graphicx}
\usepackage{epstopdf}

\usepackage{graphicx}
\usepackage{algorithm}
\usepackage{algpseudocode}
\usepackage[%
    prefixes-as-symbols=false,
    ]{siunitx}
\usepackage{tikz}
\usetikzlibrary{shapes,arrows}
\usetikzlibrary{decorations.pathreplacing}
\usetikzlibrary {arrows.meta}
\usetikzlibrary{calc}
\usepackage{pgffor}
\usepackage{tikz-3dplot}
\usepackage{listings}
\usetikzlibrary{decorations.pathreplacing}
\usepackage[braket, qm]{qcircuit}
\usepackage{listings}
\usepackage{cleveref}
\usepackage{lineno}
\usepackage{xcolor}
\usepackage{xspace}
\usepackage{upgreek}
\usepackage{subcaption}
\usepackage{svg}
\usepackage{multirow}
\usepackage{latex-bibliography/calin}
\usepackage{ifthen}
\usepackage{amssymb}
\usepackage[colorinlistoftodos,prependcaption,textsize=tiny]{todonotes}
\usepackage{pifont}

\ifpdf
  \DeclareGraphicsExtensions{.eps,.pdf,.png,.jpg}
\else
  \DeclareGraphicsExtensions{.eps}
\fi

\newsiamremark{remark}{Remark}
\newsiamremark{hypothesis}{Hypothesis}
\crefname{hypothesis}{Hypothesis}{Hypotheses}
\newsiamthm{claim}{Claim}
\newsiamremark{fact}{Fact}
\crefname{fact}{Fact}{Facts}

\headers{Zone-Agnostic Boundary Conditions in QLBM}{C. A. Georgescu and M. M\"{o}ller}

\title{Efficient and Expressive Boundary Conditions in Quantum Lattice Boltzmann Methods\thanks{Version of May 31st, 2026.
\funding{We gratefully acknowledge support from the joint research program \emph{Quantum Computational Fluid Dynamics} by Fujitsu Limited and Delft University of Technology, co-funded by the Netherlands Enterprise Agency under project number PPS23-3-03596728.}}}

\author{C\u{a}lin A. Georgescu\thanks{Delft University of Technology
  (\email{c.a.georgescu@tudelft.nl}, \url{https://gcalin.github.io}).}
\and Matthias M\"{o}ller\thanks{Delft University of Technology 
  (\email{m.moller@tudelft.nl}).}}

\usepackage{amsopn}

\nolinenumbers

\ifpdf
\hypersetup{
  pdftitle={Efficient and Expressive Boundary Conditions in Quantum Lattice Boltzmann Methods},
  pdfauthor={C. A. Georgescu and M. M\"{o}ller}
}
\fi

\begin{document}

\maketitle

% REQUIRED
\begin{abstract}
Quantum Lattice Boltzmann Methods (QLBM) have emerged as a promising candidate
for quantum realizations of computational fluid dynamics solvers.
However, despite intensive research into the QLBM in recent years,
methods for imposing boundary conditions remain limited both in terms
of efficiency and expressivity.
In this work, we introduce a new method for imposing simple
boundary conditions on QLBM that overcomes several limitations of
current approaches.
Our method forgoes the partitioning of the solid domain into segments and instead
applies a single, coherent operation on the entire boundary.
We show that our method requires fewer resources both
asymptotically and practically for bounce-back
and specular reflection boundary conditions.
\end{abstract}

% REQUIRED

% REQUIRED
% \begin{MSCcodes}
% 81P68, 76M28, 68Q12
% \end{MSCcodes}

\section{Introduction\label{sec:bp-1-intro}}

The advent of computational technology has
fundamentally revolutionized how countless industries
apply mathematics.
One branch that stands out thanks to its immense versatility
is computational fluid dynamics (CFD).
Thanks to advances in compute and CFD technologies,
fields such as the aerospace and automotive
industries have experienced
tremendous productivity boosts.
For instance, costly physical experiments that would require months
and multiple teams of engineers to prepare
in expensive facilities have been replaced by simulations
that take hours to complete and can be set up by 
a handful of practitioners \cite{mani2023perspective}.

Yet, despite impressive advances, CFD is not without its limitations.
One of the main shortcomings of CFD stems from its sheer computational cost.
For turbulent systems of high Reynolds numbers (Re), fully resolving
the flowfield by direct numerical simulations is estimated
to require $\textrm{Re}^{2.91}$ gridpoints,
while wall-resolved large eddy simulations (LES)
require $\textrm{Re}^{2.72}$ sites \cite{yang2021grid}.
Wall-modelled LES, estimated to scale as $\mathrm{Re}^{1.14}$ \cite{yang2021grid},
of a full aircraft has only
recently become computationally affordable,
but remains a frontier capability rather than
a standard practice \cite{bose2018wall, mani2023perspective}.

The intractable cost of many industrially
valuable CFD simulations has motivated researchers
to consider alternative computing paradigms.
One such direction that has received significant attention in
recent decades is quantum computing.
The research hypothesis is that
through the phenomena of superposition and entanglement,
quantum computers could perform CFD computations at
scales that are significantly greater than are possible classically.
This idea has sparked several directions, such as linear solvers
\cite{gaitan2020finding, ingelmann2024two},
variational algorithms \cite{kyriienko2021solving},
and reservoir computing \cite{pfeffer2022hybrid}.

Of the available quantum CFD (QCFD) algorithms,
one algorithm has features that make it a particularly
promising candidate for a quantum implementation:
the lattice Boltzmann method (LBM).
Unlike linear solvers, the algorithmic framework of the
LBM makes it such that its constituent
routines can be modelled and designed
structurally and independently.
In contrast to variational approaches, the LBM
requires no parameter tuning and offers clearer
convergence and accuracy guarantees.

Historically, the lattice Boltzmann method
emerged in the late 1980s
\cite{mcnamara1988boltzmann, higuera1989boltzmann, higuera1989lattice}
from the boolean lattice gas models of the 1970s and 1980s
\cite{hardy1973time,frisch1986lattice,frisch1987lattice}.
Over the ensuing decades, LBM has gained popularity among
practitioners for its ability to recover
Navier-Stokes quantities,
its ease of implementation,
and its naturally parallelizable implementation \cite{kruger2017lattice}.
Currently, state-of-the art LBM solvers scale to 
billions of gridpoints on frontier heterogeneous computer clusters 
\cite{kummerlander2025large, suffa2025large}.
In the remainder of this section, we provide a utilitarian 
introduction to the LBM and briefly review the
quantum LBM (QLBM) research landscape before introducing
our contribution.

\subsection{The Lattice Boltzmann Method and Quantum Computing}
\label{subsec:bp-1-lbm}

We first provide a brief overview of the key components of 
the classical LBM before discussing quantum implementations.
For a thorough treatment of the LBM,
we refer the reader to the works of
Succi \cite{succi2010lattice} and Kr\"{u}ger et al. \cite{kruger2017lattice}.
The Boltzmann Equation (BE) governs the physical behavior of fluids
at the mesoscopic scale, between Newtonian molecular dynamics
and the continuum of the Navier-Stokes equations.
The key component of the LBM is the distribution function $f$,
which is defined over the physical phase space and time.
The BE models the evolution of particle \emph{populations}
as changes to $f$ over time.
Discretizing the BE across physical space, velocity space, and time,
and assuming the commonplace Bhatnagar-Gross-Krook (BGK) model
\cite{bhatnagar1954model} for collision,
one obtains the lattice Boltzmann equation (LBE)

\begin{equation}
    f_v(\overbrace{\mathbf{x} + \boldsymbol{\Delta}({v})\Delta t}^\text{Streaming}, \underbrace{t + \Delta t}_\text{Timestep}) = \overbrace{f_v(\mathbf{x}, t) - \frac{\Delta t}{\tau} (f_v(\mathbf{x}, t) - \underbrace{f_v^{\text{eq}}(\mathbf{x}, t)}_\text{Equilibrium})}^\text{Collision},
    \label{eq:bp-lbe}
\end{equation}

\noindent with $\mathbf{x}$ the physical position,
$\boldsymbol{\Delta}: \{0, \ldots, q-1\} \to \mathbb{R}^{d}$
the \emph{velocity increment} or \emph{displacement},
$\Delta t$ the size of the timestep,
relaxation time $\tau$,
and equilibrium function $f_v^{\text{eq}}$.
The subscript $v$ indicates an index of a finite set of velocity directions
that each population travels with.
The LBM algorithm consists of \emph{streaming} and \emph{collision}.

While streaming is linear and has been shown to be
efficiently implementable on quantum computers by several means
\cite{todorova2020quantum,schalkers2024efficient,budinski2023efficient},
collision has proven a daunting hurdle in the realization of a practical QLBM.
The $f_v^{\text{eq}}$ term in \Cref{eq:bp-lbe} used for recovering Navier-Stokes
quantities requires computations that are both nonlinear
and irreversible, and hence difficult to model on quantum computers.
In an effort to effectively model collision, researchers have pursued
a plethora of directions, such as
linear combinations of unitaries \cite{ljubomir2022quantum, budinski2021quantum, childs2012hamiltonian},
dynamic circuits \cite{wawrzyniak2025dynamic, tiwari2025algorithmic},
machine learning \cite{luacuatucs2025surrogate,itani2025qml,zamora2026qml},
linearization techniques \cite{sanavio2024lattice,sanavio2025carleman,itani2022analysis,zamora2026carlemann,jennings2025end},
and block encodings \cite{duong2026quantum}.
Despite recent advances, all approaches in the literature currently incur some
significant drawback, be in the form of a number of qubits that scales with the number of timesteps,
modelling errors, or low probabilities of success.
As such, practical collision modeling remains
largely an open challenge in QLBM literature.
Yet, the joint effort to solve the collision problem
has in turn led to other, equally important
components of the algorithm, receiving less research attention.
In this work, we focus on one such crucial aspect of the QLBM: boundary conditions.

\subsection{Boundary Conditions in the LBM}

Boundary conditions (BCs) in the classical
LBM form a broad and nuanced research topic,
with a diverse ``zoo'' of options.
In this work, we focus on boundary conditions that describe the interaction
of fluid particles with rigid bodies, \ie, solid objects
suspended in the fluid domain.
Of these, two BCs that we specifically target in this work
are \emph{bounce-back} (BB) and \emph{specular reflection} (SR).
We specifically consider the case where the solid boundary is
placed halfway between gridpoints, which is standard in LBM literature \cite{kruger2017lattice}.
Algorithmically, the BC step for BB and SR
follows the streaming routine and ensures
that particles do not travel into the 
solid domain from one timestep to the next.
In keeping with the mesoscopic lens of the LBM,
these BCs are formulated at the population
level rather than in terms of macroscopic quantities.
In what follows, we provide a working example of BB and SR,
and refer the interested reader to Sections 5 and 11 of 
Kr\"{u}ger et al. \cite{kruger2017lattice} for a more in-depth treatment.

\begin{figure}[htbp]
    \centering
    \begin{subfigure}[b]{0.25\textwidth}
        \centering
        \centering
\begin{tikzpicture}
    % Center node with D2Q8 velocity arrows
    \velocityarrow{(0, 0)}{++(1.1,1.1)};
    \velocityarrow{(0, 0)}{++(-1.1,1.1)};
    \velocityarrow{(0, 0)}{++(1.1,-1.1)};
    \velocityarrow{(0, 0)}{++(-1.1,-1.1)};

    \dtwoqfour{(0,0)}{0.3cm}{4}{c}{0.8}{white}
    \dtwoqfourlabel{(0,0)}{0.3cm}{4}{1}{0.8}{white}{{0}}

    % Neighbor circles (same style as center, no arrows)
    % Cardinal: E, N, W, S
    \surroundingcircle{(1.25,0)}{0.15cm}{n1};   \colorcircle{(1.25,0)}{0.2cm}{lightgray}{3pt};
    \node[font=\scriptsize] at (1.25, 0) {$1$};
    \surroundingcircle{(0,1.25)}{0.15cm}{n2};   \colorcircle{(0,1.25)}{0.1cm}{lightgray}{3pt};
    \node[font=\scriptsize] at (0, 1.25) {$2$};
    \surroundingcircle{(-1.25,0)}{0.15cm}{n3};  \colorcircle{(-1.25,0)}{0.1cm}{lightgray}{3pt};
    \node[font=\scriptsize] at (-1.25, 0) {$3$};
    \surroundingcircle{(0,-1.25)}{0.15cm}{n4};  \colorcircle{(0,-1.25)}{0.1cm}{lightgray}{3pt};
    \node[font=\scriptsize] at (0, -1.25) {$4$};
    % Diagonal: NE, NW, SW, SE
    \surroundingcircle{(1.25,1.25)}{0.15cm}{n5};   \colorcircle{(1.25,1.25)}{0.1cm}{lightgray}{3pt};
    \node[font=\scriptsize] at (1.25, 1.25) {$5$};
    \surroundingcircle{(-1.25,1.25)}{0.15cm}{n6};  \colorcircle{(-1.25,1.25)}{0.1cm}{lightgray}{3pt};
    \node[font=\scriptsize] at (-1.25, 1.25) {$6$};
    \surroundingcircle{(-1.25,-1.25)}{0.15cm}{n7}; \colorcircle{(-1.25,-1.25)}{0.1cm}{lightgray}{3pt};
    \node[font=\scriptsize] at (-1.25, -1.25) {$7$};
    \surroundingcircle{(1.25,-1.25)}{0.15cm}{n8};  \colorcircle{(1.25,-1.25)}{0.1cm}{lightgray}{3pt};
    \node[font=\scriptsize] at (1.25, -1.25) {$8$};

\end{tikzpicture}
        \caption{D2Q9 stencil.}
        \label{fig:bp-intro-d2q9-stencil}
    \end{subfigure}
    \hfill
    \begin{subfigure}[b]{0.7\textwidth}
        \centering
        \providecommand{\diagBpBcsGrid}{%
    \draw[
      help lines,
      line width=0.4pt,
      color=gray!30,
      dashed
    ] (-8, -8) grid[step={($(4, 4) - (0, 0)$)}] (8, 8);

    % Background lattice points (4x4).
    \foreach \x in {-6,-2,2,6} {
        \foreach \y in {-6,-2,2,6} {
            \dtwoqeight{(\x,\y)}{0.1cm}{2}{0}{1}{white}
        }
    }

    % 2x2 obstacle square in the center.
    \fill[gray!18] (-4,-4) rectangle (4,4);

    \foreach \x/\y in {
        -2/-2,2/-2,
        -2/2,2/2
    } {
        \dtwoqeight{(\x,\y)}{0.1cm}{2}{0}{1}{gray}
    }

    \draw[black, ultra thick, rounded corners=2pt]
        (-4,-4) -- (-4,4) -- (4,4) -- (4,-4) -- cycle;
}

\centering
\begin{tikzpicture}[scale=0.1575]
    % Post-collision under bounce-back (left).
    \begin{scope}[shift={(0,0)}, local bounding box=gridBB]
        \diagBpBcsGrid
        \dtwoqeightcolor{(-6,-6)}{0.1cm}{2}{0}{1}{red}{black}{black}{black}{black}{black}{black}{red}{black}
        \dtwoqeightcolor{(-6,-2)}{0.1cm}{2}{0}{1}{blue}{black}{black}{black}{black}{black}{black}{blue}{black}
        \dtwoqeightcolor{(-2,-6)}{0.1cm}{2}{0}{1}{violet}{black}{black}{black}{black}{black}{black}{violet}{black}
    \end{scope}

    % Pre-collision state (middle). Particles approach the obstacle from
    % the left and from below.
    \begin{scope}[shift={(20,0)}, local bounding box=gridPre]
        \diagBpBcsGrid
        \dtwoqeightcolor{(-6,-6)}{0.1cm}{2}{0}{1}{red}{black}{black}{black}{black}{red}{black}{black}{black}
        \dtwoqeightcolor{(-6,-2)}{0.1cm}{2}{0}{1}{blue}{black}{black}{black}{black}{blue}{black}{black}{black}
        \dtwoqeightcolor{(-2,-6)}{0.1cm}{2}{0}{1}{violet}{black}{black}{black}{black}{violet}{black}{black}{black}
    \end{scope}

    % Post-collision under specular reflection (right).
    \begin{scope}[shift={(40,0)}, local bounding box=gridSR]
        \diagBpBcsGrid
        \dtwoqeightcolor{(-6,-6)}{0.1cm}{2}{0}{1}{red}{black}{black}{black}{black}{black}{black}{red}{black}
        \dtwoqeightcolor{(-6,2)}{0.1cm}{2}{0}{1}{blue}{black}{black}{black}{black}{black}{blue}{black}{black}
        \dtwoqeightcolor{(2,-6)}{0.1cm}{2}{0}{1}{violet}{black}{black}{black}{black}{black}{black}{black}{violet}
    \end{scope}

    % Dashed vertical separators between the three grids.
    \draw[densely dashed, gray!55, semithick] (10, -8) -- (10, 8);
    \draw[densely dashed, gray!55, semithick] (30, -8) -- (30, 8);

    \tikzset{connArrow/.style={-{Stealth[length=5pt, width=3.5pt]},shorten <=4pt, shorten >=4pt, semithick, black!80}}

    % BB arrow: from pre (middle) to post-BB (left), north-center to north-center.
    \draw[connArrow]
        (gridPre.north) to[bend left=-25]
        node[midway, above, font=\footnotesize] {BB}
        (gridBB.north);

    % SR arrow: from pre (middle) to post-SR (right), north-center to north-center.
    \draw[connArrow]
        (gridPre.north) to[bend right=-25]
        node[midway, above, font=\footnotesize] {SR}
        (gridSR.north);

\end{tikzpicture}
        \caption{Bounce-back (BB) and specular reflection (SR) example.}
        \label{fig:bp-intro-demo-bcs}
    \end{subfigure}
    \caption{Overview of halfway bounce-back and specular reflection boundary conditions.}
    \label{fig:bp-intro-demo}
\end{figure}
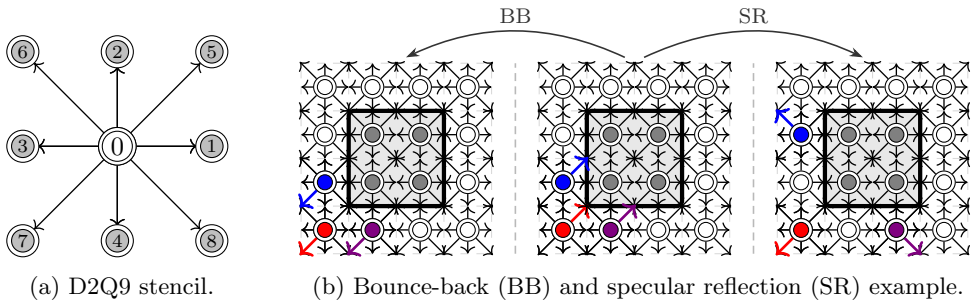

One can intuitively understand the BB and SR BCs
as descriptions of how solid domains alter the trajectory
of fluid particles.
Throughout this paper, we address BCs in terms of typical \dq{d}{q} discretizations
that describe $d$-dimensional systems with $q$ discrete velocity channels.
\Cref{fig:bp-intro-d2q9-stencil} shows the \dq{2}{9}
discretization that we use as a running example.
Velocity channels are labelled $0$ to $8$,
with $0$ the \emph{rest} population, and the remaining
$8$ labels connecting the gridpoint to its nearest
cardinal and diagonal neighbors.
Streaming simply moves particles from one gridpoint to one of its neighbors
on one of these channels.
\Cref{fig:bp-intro-demo} provides a visual depiction
of the BB and SR routines alter the streaming step in the presence of a solid domain.
The solid domain, which we henceforth label $\Omega$,
is composed of a square that envelops the four central gridpoints.
The middle grid shows the state of the system in the beginning of the timestep, with
three color-coded populations traveling towards $\Omega$.
The scope of the BCs is such that at the end of the timestep,
no population is physically located in $\Omega$,
and reflection rules apply to particles that interact with solid ``walls''.
The bounce-back rule simply inverts the direction of the population,
as shown in the left grid.
The specular reflection rule is more complex, as it requires
that, for a given velocity vector $\boldsymbol{v}$, only the directional
component of $\boldsymbol{v}$ that is normal to the contact surface
between the population and $\Omega$ is inverted.
This means that populations with traveling with the same velocity
get reflected differently depending which \emph{segment} (or \emph{zone})
of $\Omega$ they impinge on.

\subsection{Quantum algorithms for boundary conditions}

Implementing BCs for complex geometries efficiently is a crucial step
for numerous industrial applications of the LBM.
Examples include
bloodflow inside vasculature \cite{mazzeo2008hemelb},
unsteady flows around automotive models \cite{fares2006unsteady},
and wind in urban landscapes \cite{xu2025towards}.
Yet, despite their importance in classical simulations, BCs
in the QLBM remain relatively understudied.

Todorova and Steijl \cite{todorova2020quantum} are, to the best
of our knowledge, the first to describe boundary conditions in a QLBM setting.
They describe an algorithm for specular reflection performed
on axis-aligned solid objects (\ie, rectangles and cuboids)
in a complex velocity discretization.
While practical for simple use cases, this work assumes that the
number of gridpoints subject to BC treatment is constant with respect
to the size of the grid, and do not account for more complex geometries \cite{todorova2020quantum}.
Schalkers and M\"{o}ller \cite{schalkers2024efficient} describe
an algorithm for SR under the same discretization as \cite{todorova2020quantum},
and show that decomposing an axis-aligned region $\Omega$ into
$N_s$ segments and treating each segment sequentially
yields an algorithm that requires $\mathcal{O}(N_s n_g^2)$ gates,
with $n_g$ the number of grid qubits.
Crucially, this algorithm relaxes the previous assumption
that the number of gridpoints is constant, and thus allows
for arbitrarily long axis-aligned segments to be treated
at no additional cost in complexity.
The authors later adapted this approach to the bounce-back rule \cite{schalkers2024momentum},
and diagonal segments \cite{georgescu2025fully},
while Sayonee et al. \cite{ray2026quantum} developed an equivalent approach
for a different quantum register encoding, specific for the \dq{2}{5} and \dq{3}{7} discretizations.
Further works that focus on different encodings include 
the recent developments of Ueno et al. \cite{ueno2026demonstration}
for the linearized LBM, and the macroscopic
treatment of boundary conditions of Zamora et al. \cite{bastida2026quantumhardware},
both of which target different encodings than the widespread amplitude
encoding we assume throughout this work.

Despite recent advances, three fundamental limitations remain,
that compromise the practical applicability of such
boundary condition imposition techniques.
First is efficiency.
Though the \emph{segment-wise} (SW) decomposition of objects
allows for quantum primitives to efficiently address large axis-aligned
segments of $\Omega$, its sequential treatment of the $N_s$ segments
makes it such that the scaling is dominated by $N_s$.
For shapes with large number of segments (\ie, irregular boundaries),
$N_s$ can be up to $\mathcal{O}(2^{n_g})$, thus incurring an up-to-exponential
overhead that compromises the efficiency of the entire algorithm.
The second drawback of the SW approach is its expressivity.
For any shape, even when a closed-form description of it is known,
a classical preprocessing step is needed to partition $\Omega$ into
its constituent segments.
In doing so, one must consider how segments overlap, 
and design algorithms that cover all edge cases.
This comes at significant cost, both in the quantum circuits, and in classical preprocessing.
For instance, a simple cube object requires
up to $80$ distinct segments to implement \cite{georgescu2025qlbm},
which requires tedious programming and verification tasks.
Clearly, more complex shapes only exacerbate these issues.
We argue that if QLBM is to reach practical utility, 
more scalable and expressive boundary condition methods are necessary.

In this work, we introduce a novel boundary condition imposition algorithm
called the \emph{zone-agnostic} (ZA) method.
Unlike the SW approach, the ZA algorithm describes a single atomic operation
that, provided an oracular description of $\Omega$, applies
either BB or SR over the entire region.
By circumventing the SW decomposition, the ZA BC imposition
avoids the fundamental disadvantages we previously outlined.
We describe the ZA method for typical \dq{d}{q} discretizations,
and provide gate-level implementations for every described step.
We analyze the complexity of our algorithms in terms of one- and two-qubit
gates, and show how the ZA method outperforms the SW baseline both asymptotically and practically.
All algorithms are implemented in the open-source \qlbm~\cite{georgescu2025qlbm} library.

The remainder of the paper is structured as follows.
\Cref{sec:bp-2-method-za} describes the ZA method for
bounce-back (\Cref{subsec:bp-2-bb}) specular reflection (\Cref{subsec:bp-2-sr}), and 
analyzes feasible implementations of oracles (\Cref{subsec:bp-2-oracle}).
\Cref{sec:bp-3-results} empirically analyzes the correctness and cost
of the ZA method against the SW baseline.
Finally, \Cref{subsec:bp-4-conclusion} analyzes shortcomings,
discusses future directions, and concludes the paper.
The rest of this section briefly introduces nomenclature and variable names.

\subsection{Nomenclature and variables}

We assume an amplitude-based encoding

\begin{equation}
    \sum_{\boldsymbol{x}, v}\alpha_{\boldsymbol{x},v}\ket{\boldsymbol{x}}_{\mathrm{G}}\ket{v}_{\mathrm{V}}\ket{a_o}_{\mathrm{O}},
\end{equation}

\noindent with $\boldsymbol{x}$ the physical position on the grid,
$v$ the velocity index,
$\alpha_{\boldsymbol{x},j} \propto f(\boldsymbol{x})_v$
and $a_o$ an ancilla qubit initialized as $\ket{0}$
that indicates whether $\boldsymbol{x} \in \Omega$.
We assume the grid is composed of $N_g$ gridpoints,
compressed into $n_g = \lceil \log_2 N_g \rceil$ qubits.
Similarly, we assume a \dq{d}{q} discretization with
$n_q = \lceil \log_2 q \rceil$ qubits encoding the velocity register.
The amplitude of the basis state encodes its share of the global
probability distribution.
For a unitary operator $U$, we use the notation
$\mathrm{C}^{p}U$ to indicate the application of $U$ with $p$
control qubits in state $\ket{1}^{\otimes p}$.
% We use $v$ to indicate a velocity index and
% $\boldsymbol{\Delta}(v)$ the spatial
% increment associated with $v$ under, such as
% $\boldsymbol{\Delta}({v=5})=[1,1]^\mathrm{T}$ in \Cref{fig:bp-intro-d2q9-stencil}.

\section{Zone-Agnostic Imposition of Boundary Conditions}
\label{sec:bp-2-method-za}

This section introduces our novel Zone-Agnostic (ZA) method of 
imposing boundary conditions in the QLBM.
Fundamentally, the approach we use follows the same steps described by
Todorova and Steijl \cite{todorova2020quantum} and
Schalkers and M\"{o}ller \cite{schalkers2024efficient},
which is also commonplace in classical implementations.
The broad steps of the method are as follows:
(1) populations first stream into the solid domain,
(2) basis states encoding such populations are identified by an ancilla qubit,
(3) the direction of such populations is altered according to the BC,
(4) populations are streamed outside the solid domain,
and (5) the ancilla qubit is reset.
Where our approach differs from previous work is in the realization of the second and fifth steps.

To circumvent the segment-wise (SW) approach that
Schalkers and M\"{o}ller \cite{schalkers2024efficient}
introduced and its up to exponential overhead,
our ZA boundary conditions rely on two key ideas: an \emph{oracular}
description of the solid domain $\Omega$, and a \emph{trajectory-based inference}
technique for resetting the ancilla qubit(s).
Throughout the remainder of this section, we assume there exists a unitary
operator $\mathrm{U}_{\Omega}$, such that

\begin{equation}
    \mathrm{U}_{\Omega}\ket{\boldsymbol{x}}_{\mathrm{G}}\ket{v}_{\mathrm{V}}\ket{a_o}_{\mathrm{O}} = \ket{\boldsymbol{x}}_{\mathrm{G}}\ket{v}_{\mathrm{V}}\ket{a_o \oplus (\boldsymbol{x} \in \Omega)}_{\mathrm{O}},
\end{equation}

\noindent that is, $\mathrm{U}_{\Omega}$ toggles the object
ancilla qubit iff the grid position $\boldsymbol{x}$
is inside $\Omega$.
We refer to $\mathrm{U}_{\Omega}$ as the \emph{oracle}.
Equipped with such an oracle,
we can apply the appropriate reflection operator to the
selected populations, and reset the ancilla qubit without requiring to verify which
segment of the object it came into contact with.
In what follows, we first describe the
implementation of bounce-back and specular reflection boundary conditions
in \Cref{subsec:bp-2-bb} and \Cref{subsec:bp-2-sr},
before discussing implementations of $\mathrm{U}_{\Omega}$ in \Cref{subsec:bp-2-arithmetic}.

\subsection{Bounce-Back boundary conditions}
\label{subsec:bp-2-bb}

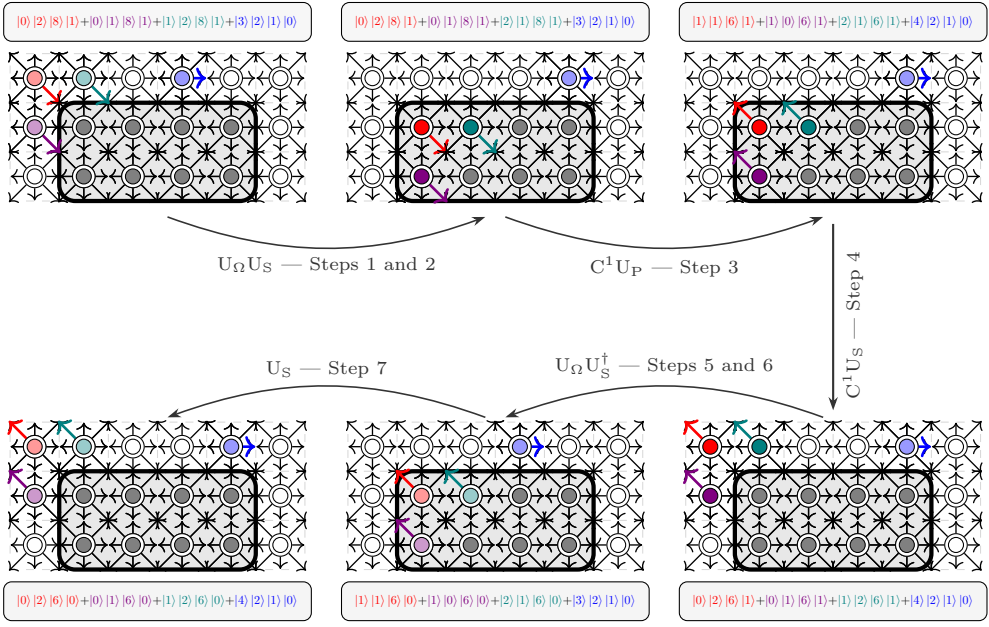
\begin{figure}
    \newcommand{\diagBpBcsGrid}{%
    \draw[
      help lines,
      line width=0.4pt,
      color=gray!30,
      dashed
    ] (-12, 0) grid[step={($(4, 4) - (0, 0)$)}] (12, 12);

    % Shade the geometry
    \fill[gray!18] (-8, 0) rectangle (8,8);

    \foreach [count=\ix from 0] \x in {-10,-6,-2,2,6,10} {
        \foreach [count=\iy from 0] \y in {2,6,10} {
            \dtwoqeight{(\x,\y)}{0.1cm}{2}{0}{1}{white}
        }
    }

    \foreach [count=\ix from 0] \x in {-6,-2,2,6} {
        \foreach [count=\iy from 0] \y in {2,6} {
            \dtwoqeight{(\x,\y)}{0.1cm}{2}{0}{1}{gray}
        }
    }

    % Highlight the obstacle
    \draw[black, ultra thick, rounded corners=6pt] (-8, 0) rectangle (8, 8);
}

% Identical per-grid title text and per-arrow caption text
\newcommand{\diagBpBcsArrowCaptionI}{$\mathrm{U}_{\Omega}\mathrm{U}_{\mathrm{S}}$ --- Steps \ref{alg:bp-2-bb:step1} and \ref{alg:bp-2-bb:step2}}
\newcommand{\diagBpBcsArrowCaptionII}{$\mathrm{C^1U}_{\mathrm{P}}$ --- Step \ref{alg:bp-2-bb:step3}}
\newcommand{\diagBpBcsArrowCaptionIII}{$\mathrm{C^1U}_{\mathrm{S}}$ --- Step \ref{alg:bp-2-bb:step4}}
\newcommand{\diagBpBcsArrowCaptionIV}{$\mathrm{U}_{\Omega}\mathrm{U}^\dagger_{\mathrm{S}}$ --- Steps \ref{alg:bp-2-bb:step5} and \ref{alg:bp-2-bb:step6}}
\newcommand{\diagBpBcsArrowCaptionV}{$\mathrm{U}_\mathrm{S}$ --- Step \ref{alg:bp-2-bb:step7}}

\centering
\begin{tikzpicture}[scale=0.1625]
    % \tikzset{diagBpBcsArrow/.style={->, thick, shorten <=6pt, shorten >=6pt}}
    % --- Row 1 (left to right) ---
    % Grid 1
    \begin{scope}[shift={(0,0)}, local bounding box=grid1]
        \diagBpBcsGrid
        \dtwoqeightcolor{(-10,10)}{0.1cm}{2}{0}{1}{red!40}{black}{black}{black}{black}{black}{black}{black}{red}
        \dtwoqeightcolor{(-10,6)}{0.1cm}{2}{0}{1}{violet!40}{black}{black}{black}{black}{black}{black}{black}{violet}
        \dtwoqeightcolor{(-6,10)}{0.1cm}{2}{0}{1}{teal!40}{black}{black}{black}{black}{black}{black}{black}{teal}
        \dtwoqeightcolor{(2,10)}{0.1cm}{2}{0}{1}{blue!40}{blue}{black}{black}{black}{black}{black}{black}{black}
    \end{scope}
    \node[anchor=south, yshift=4pt,
          fill=gray!6, draw=black, rounded corners=3pt, inner sep=5pt] at (grid1.north)
    {\scalebox{0.46}{$\color{red}{\ket{0}\ket{2}\ket{8}\ket{1}} \color{black}{+} \color{violet}{\ket{0}\ket{1}\ket{8}\ket{1}} \color{black}{+} \color{teal}{\ket{1}\ket{2}\ket{8}\ket{1}} \color{black}{+} \color{blue}{\ket{3}\ket{2}\ket{1}\ket{0}}$}};
    % % Grid 2
    \begin{scope}[shift={(27.5,0)}, local bounding box=grid2]
        \diagBpBcsGrid
        \dtwoqeightcolor{(-6,6)}{0.1cm}{2}{0}{1}{red}{black}{black}{black}{black}{black}{black}{black}{red}
        \dtwoqeightcolor{(-6,2)}{0.1cm}{2}{0}{1}{violet}{black}{black}{black}{black}{black}{black}{black}{violet}
        \dtwoqeightcolor{(-2,6)}{0.1cm}{2}{0}{1}{teal}{black}{black}{black}{black}{black}{black}{black}{teal}
        \dtwoqeightcolor{(6,10)}{0.1cm}{2}{0}{1}{blue!40}{blue}{black}{black}{black}{black}{black}{black}{black}
    \end{scope}
    \node[anchor=south, yshift=4pt,
          fill=gray!6, draw=black, rounded corners=3pt, inner sep=5pt] at (grid2.north)
    {\scalebox{0.46}{$\color{red}{\ket{0}\ket{2}\ket{8}\ket{1}} \color{black}{+} \color{violet}{\ket{0}\ket{1}\ket{8}\ket{1}} \color{black}{+} \color{teal}{\ket{2}\ket{1}\ket{8}\ket{1}} \color{black}{+} \color{blue}{\ket{3}\ket{2}\ket{1}\ket{0}}$}};

    % Grid 3
    \begin{scope}[shift={(55,0)}, local bounding box=grid3]
        \diagBpBcsGrid
        \dtwoqeightcolor{(-6,6)}{0.1cm}{2}{0}{1}{red}{black}{black}{black}{black}{black}{red}{black}{black}
        \dtwoqeightcolor{(-6,2)}{0.1cm}{2}{0}{1}{violet}{black}{black}{black}{black}{black}{violet}{black}{black}
        \dtwoqeightcolor{(-2,6)}{0.1cm}{2}{0}{1}{teal}{black}{black}{black}{black}{black}{teal}{black}{black}
        \dtwoqeightcolor{(6,10)}{0.1cm}{2}{0}{1}{blue!40}{blue}{black}{black}{black}{black}{black}{black}{black}
    \end{scope}
    \node[anchor=south, yshift=4pt,
          fill=gray!8, draw=black, rounded corners=3pt, inner sep=5pt] at (grid3.north)
    {\scalebox{0.46}{$\color{red}{\ket{1}\ket{1}\ket{6}\ket{1}} \color{black}{+} \color{violet}{\ket{1}\ket{0}\ket{6}\ket{1}} \color{black}{+} \color{teal}{\ket{2}\ket{1}\ket{6}\ket{1}} \color{black}{+} \color{blue}{\ket{4}\ket{2}\ket{1}\ket{0}}$}};

    % Grid 4
    \begin{scope}[shift={(55,-30)}, local bounding box=grid4]
        \diagBpBcsGrid
        \dtwoqeightcolor{(-10,10)}{0.1cm}{2}{0}{1}{red}{black}{black}{black}{black}{black}{red}{black}{black}
        \dtwoqeightcolor{(-10,6)}{0.1cm}{2}{0}{1}{violet}{black}{black}{black}{black}{black}{violet}{black}{black}
        \dtwoqeightcolor{(-6,10)}{0.1cm}{2}{0}{1}{teal}{black}{black}{black}{black}{black}{teal}{black}{black}
        \dtwoqeightcolor{(6,10)}{0.1cm}{2}{0}{1}{blue!40}{blue}{black}{black}{black}{black}{black}{black}{black}
    \end{scope}
    \node[anchor=north, yshift=-4pt,
          fill=gray!8, draw=black, rounded corners=3pt, inner sep=5pt] at (grid4.south)
    {\scalebox{0.46}{$\color{red}{\ket{0}\ket{2}\ket{6}\ket{1}} \color{black}{+} \color{violet}{\ket{0}\ket{1}\ket{6}\ket{1}} \color{black}{+} \color{teal}{\ket{1}\ket{2}\ket{6}\ket{1}} \color{black}{+} \color{blue}{\ket{4}\ket{2}\ket{1}\ket{0}}$}};

    % Grid 5
    \begin{scope}[shift={(27.5,-30)}, local bounding box=grid5]
        \diagBpBcsGrid
        \dtwoqeightcolor{(-6,6)}{0.1cm}{2}{0}{1}{red!40}{black}{black}{black}{black}{black}{red}{black}{black}
        \dtwoqeightcolor{(-6,2)}{0.1cm}{2}{0}{1}{violet!40}{black}{black}{black}{black}{black}{violet}{black}{black}
        \dtwoqeightcolor{(-2,6)}{0.1cm}{2}{0}{1}{teal!40}{black}{black}{black}{black}{black}{teal}{black}{black}
        \dtwoqeightcolor{(2,10)}{0.1cm}{2}{0}{1}{blue!40}{blue}{black}{black}{black}{black}{black}{black}{black}
    \end{scope}
    \node[anchor=north, yshift=-4pt,
          fill=gray!8, draw=black, rounded corners=3pt, inner sep=5pt] at (grid5.south)
    {\scalebox{0.46}{$\color{red}{\ket{1}\ket{1}\ket{6}\ket{0}} \color{black}{+} \color{violet}{\ket{1}\ket{0}\ket{6}\ket{0}} \color{black}{+} \color{teal}{\ket{2}\ket{1}\ket{6}\ket{0}} \color{black}{+} \color{blue}{\ket{3}\ket{2}\ket{1}\ket{0}}$}};

    % Grid 6
    \begin{scope}[shift={(0,-30)}, local bounding box=grid6]
        \diagBpBcsGrid
        \dtwoqeightcolor{(-10,10)}{0.1cm}{2}{0}{1}{red!40}{black}{black}{black}{black}{black}{red}{black}{black}
        \dtwoqeightcolor{(-10,6)}{0.1cm}{2}{0}{1}{violet!40}{black}{black}{black}{black}{black}{violet}{black}{black}
        \dtwoqeightcolor{(-6,10)}{0.1cm}{2}{0}{1}{teal!40}{black}{black}{black}{black}{black}{teal}{black}{black}
        \dtwoqeightcolor{(6,10)}{0.1cm}{2}{0}{1}{blue!40}{blue}{black}{black}{black}{black}{black}{black}{black}
    \end{scope}
    \node[anchor=north, yshift=-4pt,
          fill=gray!8, draw=black, rounded corners=3pt, inner sep=5pt] at (grid6.south)
    {\scalebox{0.46}{$\color{red}{\ket{0}\ket{2}\ket{6}\ket{0}} \color{black}{+} \color{violet}{\ket{0}\ket{1}\ket{6}\ket{0}} \color{black}{+} \color{teal}{\ket{1}\ket{2}\ket{6}\ket{0}} \color{black}{+} \color{blue}{\ket{4}\ket{2}\ket{1}\ket{0}}$}};

    % --- Connecting arrows ---
    \tikzset{connArrow/.style={-{Stealth[length=5pt, width=3.5pt]},shorten <=4pt, shorten >=4pt, semithick, black!80}}

    % Top row: arrows connecting at the bottom of grids
    \draw[connArrow] (0, -1) to[bend right=20]
        node[midway, below, font=\scriptsize] {\diagBpBcsArrowCaptionI} (27.5, -1);
    \draw[connArrow] (27.5, -1) to[bend right=20]
        node[midway, below, font=\scriptsize] {\diagBpBcsArrowCaptionII} (55, -1);

    % Transition: Grid 3 → Grid 4 (straight down)
    \draw[connArrow, thick]
        (55, -1) -- node[midway, below=3pt, font=\scriptsize, rotate=90] {\diagBpBcsArrowCaptionIII} (grid4.north);

    % Bottom row: arrows connecting at the top of grids (right to left)
    \draw[connArrow] (grid4.north) to[bend right=20]
        node[midway, above, font=\scriptsize] {\diagBpBcsArrowCaptionIV} (grid5.north);
    \draw[connArrow] (grid5.north) to[bend right=20]
        node[midway, above, font=\scriptsize] {\diagBpBcsArrowCaptionV} (grid6.north);

\end{tikzpicture}
    \caption{Example of zone-agnostic imposition of bounce-back boundary conditions on a $6\times 3$ \dq{2}{9}
    lattice with a rectangle obstacle spanning $[1,4]\times[0,1]$.\label{fig:bp-2-za-bb-example}}
\end{figure}

The zone-agnostic implementation of bounce-back boundary conditions
is formalized in the 7 steps of \Cref{alg:bp-2-bb}, and visually depicted on a
concrete grid with 6 key stages in \Cref{fig:bp-2-za-bb-example}.
Transitions in \Cref{fig:bp-2-za-bb-example} follow the steps
of the algorithm, with color-coded basis states indicating specific populations.
Each basis state represents a $\ket{x}\ket{y}\ket{v}\ket{a_o}$ component,
with $x$ and $y$ the positions on the grid, $v$ the velocity, 
and $a_o$ the state of the ancilla qubit.
Gridpoints are highlighted in bold colors if $a_o=1$ and in
shaded hues otherwise.

\begin{figure}
    \input{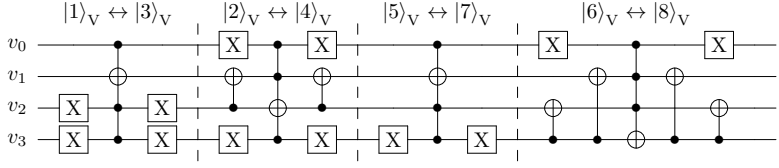}
    \caption{Quantum circuit implementing the \dq{2}{9} bounce-back reflection permutation.\label{fig:bp-2-za-bb-reflection}}
\end{figure}

To realize bounce-back boundary conditions, it suffices for the $a_o$ qubit
to encode \emph{if} a population has streamed into the object in the
previous propagation step -- precisely the information obtained by
applying $\mathrm{U}_{\Omega}$ following streaming
(Steps \ref{alg:bp-2-bb:step1} and \ref{alg:bp-2-bb:step2},
and Stage 2 of \Cref{fig:bp-2-za-bb-example}).
With this information available in the basis state, our ZA
algorithm applies the BB reflection operator
only on the subspace determined by $\ket{a_o}=\ket{1}$.
For the \dq{2}{9} stencil depicted in \Cref{fig:bp-intro-d2q9-stencil},
this operation consists of applying $4$ basis state transpositions:
$(1\leftrightarrow 3), (2\leftrightarrow 4), (5\leftrightarrow 7)$, and $(6\leftrightarrow 8)$.
\Cref{fig:bp-2-za-bb-reflection} depicts the quantum circuit performing this routine.
Once the velocities of the selected populations have been inverted,
we apply a controlled streaming step, which returns populations to the
gridpoint they streamed from (Stage 4 and Step \ref{alg:bp-2-bb:step4}).

At this point, all populations are in the correct location physically, but the
ancilla qubit is dirty.
To reset it, we first unstream \emph{all} populations before
applying the oracle again (Steps \ref{alg:bp-2-bb:step5} and \ref{alg:bp-2-bb:step6}, Stage 5).
These two operations return the crossing particles to $\Omega$, and hence the application
of $\mathrm{U}_\Omega$ sets $\ket{a_o}$ back to $\ket{0}$.
Finally, the application of a streaming operator returns to the same position and velocity
as in Stage 4, with a clean state in $a_o$.

\begin{algorithm}
\caption{Zone-Agnostic Bounceback Boundary Condition Imposition}
\label{alg:bp-2-bb}
\begin{algorithmic}[1]
\Require Oracle $\mathrm{U}_\Omega$, Streaming Operator $\mathrm{U}_{\mathrm{S}}$, Velocity Permutation Operator $\mathrm{U}_{\mathrm{P}}$ 
\State Stream all populations \label{alg:bp-2-bb:step1} \Comment{$\mathrm{U}_{\mathrm{S}}\ket{\boldsymbol{x_0}}\ket{v}\ket{0} = \ket{\boldsymbol{x_0}+\boldsymbol{\Delta}(v)}\ket{v}\ket{0}$}
\State Apply $\mathrm{U}_\Omega$ targeting the ancilla $a_o$ \label{alg:bp-2-bb:step2} \Comment{$\mathrm{U}_{\Omega}\ket{\boldsymbol{x}_1}\ket{v}\ket{0} = \ket{\boldsymbol{x}_1}\ket{v}\ket{\boldsymbol{x}\in\Omega}$}
\State Invert the velocity if $\boldsymbol{x} \in \Omega$\label{alg:bp-2-bb:step3}  \Comment{$\mathrm{C^1U}_{\mathrm{P}}\ket{\boldsymbol{x}_1}\ket{v}\ket{a_o} = \ket{\boldsymbol{x}_1}\ket{a_o\bar{v} + (1-a_o)v}\ket{a_o}$}
\State Stream on $\ket{a_o}=\ket{1}$ \label{alg:bp-2-bb:step4} \Comment{$\mathrm{C^{1}U}_{\mathrm{S}}\ket{\boldsymbol{x}_1}\ket{v'}\ket{a_o} = \ket{\boldsymbol{x}_1+b\boldsymbol{\Delta}({v'})}\ket{v'}\ket{a_o}$}
\State Inverse stream all populations \label{alg:bp-2-bb:step5} \Comment{$\mathrm{U}^\dagger_{\mathrm{S}}\ket{\boldsymbol{x}_2}\ket{v}\ket{a_o} = \ket{\boldsymbol{x}_2-\boldsymbol{\Delta}(v')}\ket{v'}\ket{a_o}$}
\State Apply $\mathrm{U}_\Omega$ targeting the ancilla $a_o$ \label{alg:bp-2-bb:step6} \Comment{$\mathrm{U}_{\Omega}\ket{\boldsymbol{x}_3}\ket{v'}\ket{a_o} = \ket{\boldsymbol{x}_3}\ket{v'}\ket{a_o\oplus(\boldsymbol{x}\in\Omega)}$}
\State Stream all populations \label{alg:bp-2-bb:step7} \Comment{$\mathrm{U}_{\mathrm{S}}\ket{\boldsymbol{x}_3}\ket{v'}\ket{0} = \ket{\boldsymbol{x}_3-\boldsymbol{\Delta}(v')=\boldsymbol{x}_4}\ket{v'}\ket{0}$}
\end{algorithmic}
\end{algorithm}

\paragraph{Assumptions and correctness}
One key assumption that the ZA-BB algorithm requires is
that at the beginning of the simulation, no population is inside $\Omega$.
Without this assumption, the controlled semantics of streaming and reflection
break down into undefined behavior.
There are no restrictions on the gridpoints within $\Omega$,
but one must be careful when composing objects from multiple subregions.
If $\Omega = \cup_k \Omega_k$, one must ensure that
$\forall k\neq j, \Omega_k \cap \Omega_j = \emptyset$,
otherwise applications of the composed oracle would lead to ill-formed states of the ancilla qubit.
We further assume that the velocity discretization only has
components valued $0$ and $\pm 1$, which includes
our \dq{2}{9} case, as well as \dq{3}{15},
\dq{3}{19}, and \dq{3}{27} \cite{kruger2017lattice}.
The correctness of the algorithms can be verified by independently analyzing the case where
particles hit the object, and the case when they do not.
\Cref{tab:bp-2-correctness-bb} provides the step-wise description
of the evolution of basis states falling into both of these categories.
We note that by definition, 

\begin{equation}
    \mathrm{U_P}\ket{v} = \ket{\bar{v}} \text{ s.t. } \boldsymbol{\Delta}(v) = -\boldsymbol{\Delta}(\bar{v}).
\end{equation}

\noindent Intuitively, the correctness of the non-crossing particles follows from the fact
that they are only affected by Steps \ref{alg:bp-2-bb:step1},
\ref{alg:bp-2-bb:step5}, and \ref{alg:bp-2-bb:step7},
which form a stream-unstream-stream sequence that reduces to a regular streaming step.
By the linearity of quantum computing, the cases outlined in \Cref{tab:bp-2-correctness-bb}
generalize to flow fields with arbitrarily many populations.

\begin{table}[h]
\centering
\setlength{\tabcolsep}{3pt}
\begin{tabular}{@{}cc|c|cc|cc@{}}
\hline
Step & Stage & Operator & $\ket{\psi_{\bar{\Omega} \to \Omega}}$ & Remark & $\ket{\psi_{\bar{\Omega} \to \bar{\Omega}}}$ & Remark \\
\hline
 & 1 & -- & $\ket{\boldsymbol{x}_0}\ket{v}\ket{0}$ &  -- & $\ket{\boldsymbol{x}_0}\ket{v}\ket{0}$ & -- \\
\ref{alg:bp-2-bb:step1} & -- & $\mathrm{U_S}$ & $\ket{\boldsymbol{x}_1}\ket{v}\ket{0}$ & $\boldsymbol{x}_1 = \boldsymbol{x}_0 + \boldsymbol{\Delta}(v)$ & $\ket{\boldsymbol{x}_1}\ket{v}\ket{0}$ & $\boldsymbol{x}_1 = \boldsymbol{x}_0 + \boldsymbol{\Delta}(v)$ \\
\ref{alg:bp-2-bb:step2} & 2 & $\mathrm{U_\Omega}$ & $\ket{\boldsymbol{x}_1}\ket{v}\ket{1}$ &  $\boldsymbol{x}_1 \in \Omega$ & $\ket{\boldsymbol{x}_1}\ket{v}\ket{0}$ & $\boldsymbol{x}_1 \notin \Omega$ \\
\ref{alg:bp-2-bb:step3} & 3 & $\mathrm{C^1U_P}$ & $\ket{\boldsymbol{x}_1}\ket{\bar{v}}\ket{1}$ &  $\ket{a_o}=\ket{1}$ & $\ket{\boldsymbol{x}_1}\ket{v}\ket{0}$ & $\ket{a_o}=\ket{0}$ \\
\ref{alg:bp-2-bb:step4} & 4 & $\mathrm{C^1U_S}$ & $\ket{\boldsymbol{x}_0}\ket{\bar{v}}\ket{1}$ &  $\boldsymbol{x}_0=\boldsymbol{x}_1+\boldsymbol{\Delta}({\bar{v}})$ & $\ket{\boldsymbol{x}_1}\ket{v}\ket{0}$ & $\ket{a_o}=\ket{0}$ \\
\ref{alg:bp-2-bb:step5} & -- & $\mathrm{U^\dagger_S}$ & $\ket{\boldsymbol{x}_1}\ket{\bar{v}}\ket{1}$ &  $\boldsymbol{x}_1=\boldsymbol{x}_0-\boldsymbol{\Delta}({\bar{v}})$ & $\ket{\boldsymbol{x}_0}\ket{v}\ket{0}$ & $\boldsymbol{x}_0=\boldsymbol{x}_1-\boldsymbol{\Delta}({v})$ \\
\ref{alg:bp-2-bb:step6} & 5 & $\mathrm{U_\Omega}$ & $\ket{\boldsymbol{x}_1}\ket{\bar{v}}\ket{0}$ &  $\boldsymbol{x}_1 \in \Omega$ & $\ket{\boldsymbol{x}_0}\ket{v}\ket{0}$ & $\boldsymbol{x}_0 \notin \Omega$ \\
\ref{alg:bp-2-bb:step7} & 6 & $\mathrm{U_S}$ & $\ket{\boldsymbol{x}_0}\ket{\bar{v}}\ket{0}$ & $\boldsymbol{x}_0=\boldsymbol{x}_1+\boldsymbol{\Delta}({\bar{v}})$ & $\ket{\boldsymbol{x}_1}\ket{v}\ket{0}$ & $\boldsymbol{x}_1=\boldsymbol{x}_0+\boldsymbol{\Delta}(v)$ \\

\hline
\end{tabular}
\caption{Case- and step-wise correctness check of \Cref{alg:bp-2-bb}. \label{tab:bp-2-correctness-bb}}
\end{table}

\paragraph{Complexity analysis}
The complexity of the ZA-BB algorithm hinges on the complexity
of the streaming ($\mathrm{U_S}$), reflection ($\mathrm{U_P}$), 
and oracle ($\mathrm{U_\Omega}$) operators.
The complexity of streaming is known to be quadratic
in $n_g$ in several realizations
\cite{todorova2020quantum, budinski2023efficient,schalkers2024efficient}.
In our implementation, we adapt the circuit developed by
Schalkers and M\"{o}ller  \cite{schalkers2024efficient}
by utilizing 2 Quantum Fourier Transform ($\qft$) blocks
and controlled phase gates that increase/decrease the value
of the basis states of each discrete velocity.
This is equivalent to $q$ sequential \emph{Draper adders} \cite{draper2000addition}
with $n_q$ additional controls on the phase gates.
Therefore, the complexity of the streaming operator used
in Steps \ref{alg:bp-2-bb:step1}, \ref{alg:bp-2-bb:step5}, and \ref{alg:bp-2-bb:step7},
as well as \ref{alg:bp-2-bb:step4}, which only requires one additional control on the 
phase gates, is $\mathcal{O}(qn_g^2)$.
For typical, symmetric \dq{d}{q} discretizations,
$\mathrm{U_P}$ requires $\lfloor q/2 \rfloor$ basis state transpositions, which
are known to be implementable with $\mathcal{O}(n_q)$ gates \cite{herbert2024almost}.
It follows that $\mathrm{U_P}$ is $\mathcal{O}(qn_q)$.
Finally, the algorithm requires exactly two, and therefore $\mathcal{O}(1)$,
applications of $\mathrm{U_\Omega}$,
the complexity of which we address in \Cref{subsec:bp-2-oracle}.

\subsection{Specular Reflection boundary conditions}
\label{subsec:bp-2-sr}

Implementing specular reflection requires additional information,
that is not necessary in the bounce-back case.
While for the BB BCs, it suffices to know \emph{if}
a population has crossed the fluid-solid boundary
in the previous time step, SR additionally
requires knowledge of \emph{why} the crossing happened.
Physically, this information governs
which component(s) of the velocity vector get reflected,
while from a quantum algorithm perspective, it is necessary
to ensure reversibility.
The challenge in realizing ZA-SR boundary conditions stems
from efficiently inferring this piece of information
without accessing the segment-wise breakdown of $\Omega$.
In this section, we first formalize the problem of identifying
why boundary crossing occurred,
before using this subroutine to construct the ZA-SR algorithm and analyzing its properties.

\subsubsection{Crossing factors identification}

The key to identifying why boundary crossing occurs lays in decomposing the velocity
vector into its $d$ scalar components.
Let $\mathcal{P}(\{1, \ldots, d\})$ be the powerset of indices
of the velocity vector.
We henceforth refer to members of $\mathcal{P}$ as \emph{factors}.
The problem of identifying why crossing happens reduces to
finding which subset(s) of factors, projected onto the velocity vector,
causes the streaming from $\bar{\Omega}$ into $\Omega$.
If we restrict our analysis to instances when crossing did, with certainty, happen,
we can remove the empty set from both spaces.
Thus, there are a total of $2^{2^d-1}-1$ such possible combinations.

To ensure uniqueness and reversibility, we prune this space by the following constraint:
the solution to the problem must be an \emph{irreducible} causal explanation.
Intuitively, this amounts to searching for the \emph{minimal} subset of factors that
explains the boundary crossing.
To formalize our problem, we can trim the set of solutions by searching over
the \emph{antichains} partially ordered set (poset)
$\left(\mathcal{P}(\{1, \ldots, d\}) \backslash \emptyset, \subseteq\right)$ instead.
An antichain is defined as a subset of a poset, such that any two distinct elements
are incomparable.
To help provide intuition as to why this formulation is sound,
let us analyze our $2$-dimensional running example. 

\paragraph{Example: 2-dimensional crossing factors}
In a \dq{2}{9} setting, all populations are attached to a velocity vector with 
$x$ and $y$ components.
The powerset of these components is

\begin{equation}
    \mathcal{P}(\{x, y\}) = \{ \emptyset, \{x\}, \{y\}, \{x, y\}\},
\end{equation}

\noindent from which we remove $\emptyset$ as we assume that crossing has in fact occurred.
Then, the space of all possible crossing factors is

\begin{equation}
    \begin{aligned}
        \mathcal{P}(\mathcal{P}(\{x, y\}) \backslash \emptyset) \backslash \emptyset & =
    \overbrace{\{\{\{x\}\}, \{\{y\}\}, \{\{x, y\}\}, \{\{x\},\{y\}\}\}}^{\text{Antichains of } \left(\mathcal{P}(\{1, \ldots, d\}) \backslash \emptyset, \subseteq\right)} \\
     & \cup \underbrace{\{ \{\{x\},\{x, y\}\}, \{\{y\},\{x, y\}\}, \{\{x\},\{y\}, \{x,y\}\}\}}_\text{Redundant explanations}.
    \end{aligned}
\end{equation}

Factors such as $\{\{x\},\{x, y\}\}$ include redundant information:
if $x$ itself is enough to cause the crossing, then $\{x, y\}$ provides no additional
information for cardinal-facing velocities, even if it itself is a valid explanation.
This aligns exactly with the definition of an antichain, as under our poset
construction $\{x\} \subseteq \{x, y\}$, and similarly for all other redundant explanation.
If for a set of factors $\mathcal{F}$,
$\forall s_1 \neq s_2 \in \mathcal{F},
(s_1 \subsetneq s_2) \land (s_2 \subsetneq s_1)$,
then $\mathcal{F}$ is an irreducible explanation for the boundary crossing,
and our task is to identify the set with the smallest cardinality.
Furthermore, one can reason about the difference between $\{\{x\},\{y\}\}$ and $\{\{x, y\}\}$
as ``the $x$ and the $y$ components both cause the boundary crossing independently and together'' and
``the $x$ and the $y$ components cause crossing together, but not independently''.

We formalize the procedure of identifying the crossing factors in \Cref{alg:bp-2-flag-crossing}.
We require two applications of this algorithm to realize specular reflection:
one to identify which factors caused the boundary
crossing in the previous step and another to flag
which factors would cause the crossing if the particle were to stream
with its current velocity.
We differentiate between the two cases by the parameter $\sigma$.

The algorithm proceeds by looping over all velocity states (Step \ref{alg:bp-2-flag-crossing:step1})
all causal explanations for the crossing (Step \ref{alg:bp-2-flag-crossing:step2}),
and all factors of the explanation (Step \ref{alg:bp-2-flag-crossing:step3}).
To verify whether each factor caused a crossing, 
the components of the factor are selected (and possibly inverted by $\sigma$) from the velocity
vector and the population is streamed
(Steps \ref{alg:bp-2-flag-crossing:step4} and \ref{alg:bp-2-flag-crossing:step5}).
Following the streaming step, crossing is verified by
applying the oracle (Step \ref{alg:bp-2-flag-crossing:step6}).
Once the information is encoded onto the ancilla qubits,
the streaming is undone to return the particle to its original location
and the factor's minimal property is verified
(Steps \ref{alg:bp-2-flag-crossing:step7} and \ref{alg:bp-2-flag-crossing:step8}).
Finally, the algorithm records this information onto a register of ancilla qubits.

\begin{algorithm}
\caption{\textsc{FlagCrossingFactors}$(\sigma)$ --- identify minimal antichains of dimensional components whose displacement $\sigma\,\pi_{T}(\boldsymbol{\Delta}(v))$ causes a boundary crossing.}
\label{alg:bp-2-flag-crossing}
\begin{algorithmic}[1]
\Require Sign $\sigma \in \{+1,-1\}$; Oracle $\mathrm{U}_\Omega$; Streaming Operator $\mathrm{U}_{\mathrm{S}}$; spatial dimension $d$; number of velocities $q$
\For{each velocity direction $v \in \{0, \ldots, q-1\}$, controlled on $\ket{a_o} = \ket{1}$} \label{alg:bp-2-flag-crossing:step1}
    \For{each antichain $\mathcal{A} \in \left(\mathcal{P}(\{1, \ldots, d\}) \setminus \emptyset,\, \subseteq \right)$} \label{alg:bp-2-flag-crossing:step2}
        \For{each minimal element $T_j \in \mathcal{A}$} \label{alg:bp-2-flag-crossing:step3}
            \State Project: $\pi_{T_j}(\boldsymbol{\Delta}(v)) = P_{T_j}\,\boldsymbol{\Delta}(v)$, with $P_{T_j} = \mathrm{diag}(\mathbf{1}_{i \in T_j})$ \label{alg:bp-2-flag-crossing:step4}
            \State Stream by $\sigma\,\pi_{T_j}(\boldsymbol{\Delta}(v))$ \label{alg:bp-2-flag-crossing:step5}
            \State Apply $\mathrm{U}_\Omega$ to test whether $\sigma\,\pi_{T_j}(\boldsymbol{\Delta}(v))$ causes boundary crossing \label{alg:bp-2-flag-crossing:step6}
            \State Stream by $-\sigma\,\pi_{T_j}(\boldsymbol{\Delta}(v))$ \Comment{Undo Step~\ref{alg:bp-2-flag-crossing:step5}} \label{alg:bp-2-flag-crossing:step7}
            \State Confirm that $\forall\, T' \subsetneq T_j,\ \sigma\,\pi_{T'}(\boldsymbol{\Delta}(v))$ does not cause crossing \label{alg:bp-2-flag-crossing:step8}
        \EndFor
        \If{all $T_j \in \mathcal{A}$ are sufficient \textbf{and} minimal} \label{alg:bp-2-flag-crossing:step9}
            \State Flag contributing components $\bigcup_{T_j \in \mathcal{A}} T_j$ onto ancillae \label{alg:bp-2-flag-crossing:step10}
        \EndIf
    \EndFor
\EndFor
\end{algorithmic}
\end{algorithm}

\begin{figure}
    \input{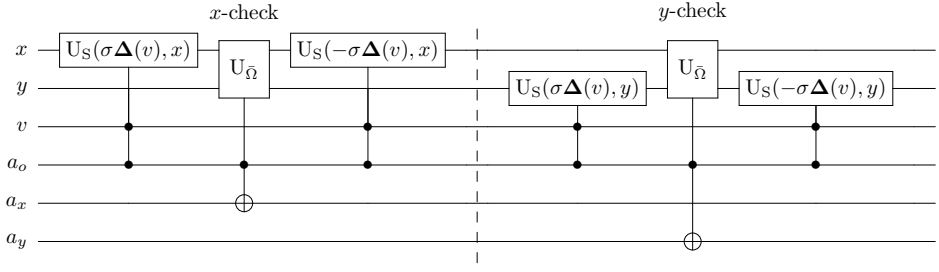}
    \caption{Quantum circuit implementing the inner loop of \Cref{alg:bp-2-flag-crossing} in $2$ dimensions.\label{fig:bp-2-za-sr-check}}
\end{figure}

To translate the enumerative definition of \Cref{alg:bp-2-flag-crossing}
into a quantum circuit, we make use of ancilla register as follows.
For our \dq{2}{9} implementation, we define a
register $\mathrm{SR}$ of two additional qubits labeled $a_x$ and $a_y$.
The purpose of these qubits
is to identify whether which component combinations caused the crossing.
The quantum circuit implementation of this procedure
for a single velocity index $v$ is given in \Cref{fig:bp-2-za-sr-check}.
The quantum algorithm first performs the projected streaming of the $x$ and $y$ components
sequentially, verifying whether each of these components cause crossing.
Since the particles are always inside the object at this stage, crossing is identified
by $\mathrm{U}_{\bar{\Omega}}$,
which can be implemented by $\mathrm{U}_{\Omega}$ and a single additional $\mathrm{X}$ gate.
At the end of the circuit, the state $\ket{a_xa_y}$ encodes precisely
which of the $4$ explanations caused ($\sigma=-1$) or would cause ($\sigma=1$) the crossing,
as laid out in \Cref{tab:bp-2-example-sr-check}.
Since the operation is controlled on the state of $\ket{a_o}=\ket{1}$,
the routine only ever runs for particles that have streamed into $\Omega$
at the time of execution.
This implementation requires exactly two applications of the oracle
and four blocks that are asymptotically equivalent to streaming.
In what follows, we use the crossing factor identification as a
subroutine to realize the full ZA specular reflection operator.

\begin{table}[h]
\centering
\begin{tabular}{@{}c|cccc@{}}
\hline
Stage & $\{\{x\}\}$ & $\{\{y\}\}$ & $\{\{x\},\{y\}\}$ & $\{\{x, y\}\}$ \\
\hline
Initial & $\ket{00}$ & $\ket{00}$ & $\ket{00}$ & $\ket{00}$\\
After $x$-check & $\ket{10}$ & $\ket{00}$ & $\ket{00}$ & $\ket{10}$\\
After $y$-check & $\ket{10}$ & $\ket{01}$ & $\ket{00}$ & $\ket{11}$\\
\hline
\end{tabular}
\caption{Step-wise correctness check of \Cref{alg:bp-2-flag-crossing} for $2d$ causal explanations. \label{tab:bp-2-example-sr-check}}
\end{table}

\subsubsection{End-to-end specular reflection}
\begin{figure}
    \input{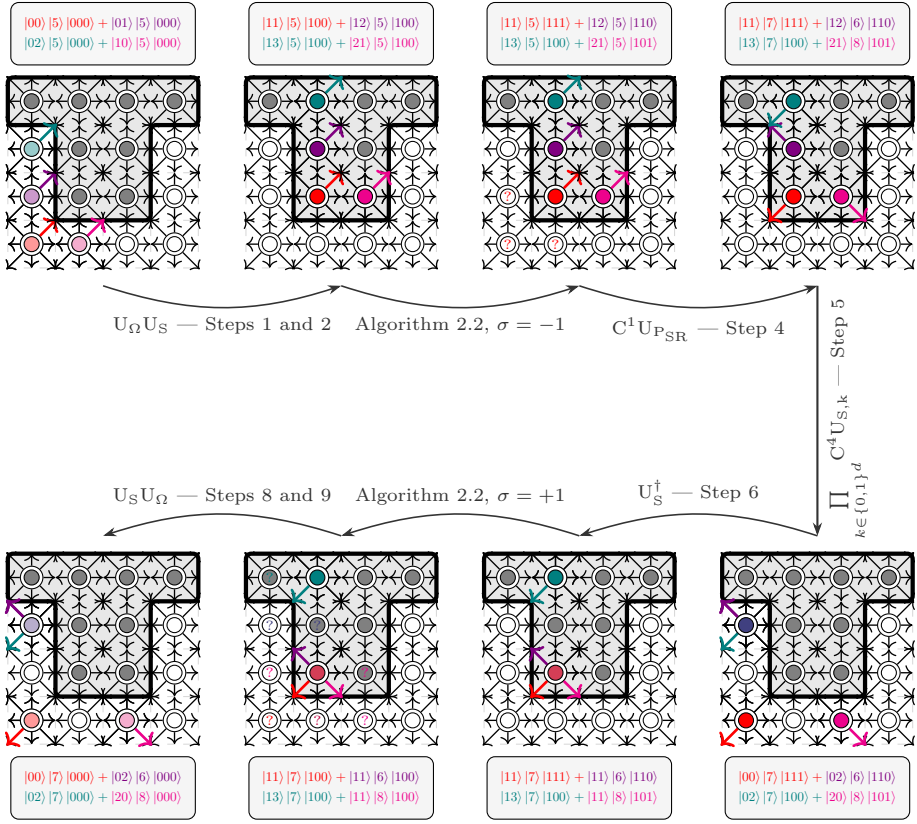}
    \caption{Example of zone-agnostic imposition of specular reflection
    boundary conditions on a $6\times 3$ \dq{2}{9} lattice with a
    rectangle obstacle spanning $[1,4]\times[0,1]$.\label{fig:bp-2-za-sr-example}}
\end{figure}

Equipped with \Cref{alg:bp-2-flag-crossing}, we can now construct
the full specular reflection routine.
We give the definition of the procedure in \Cref{alg:bp-2-sr}
and accompany it with a visual depiction in \Cref{fig:bp-2-za-sr-example}.
The example follows four populations traveling with velocity $\boldsymbol{\Delta}(v)=(1,1)$,
each of which is assigned a different causal explanation
for boundary crossing.
The algorithm begins identically to its bounce-back counterpart:
populations stream and those which have crossed into $\Omega$ 
are flagged through the application of the oracle
(Steps \ref{alg:bp-2-sr:step1} and \ref{alg:bp-2-sr:step2} and Stage 2).
Then, \Cref{alg:bp-2-flag-crossing} identifies which explanation
is responsible for the crossing (Step \ref{alg:bp-2-sr:step3}, Stage 3).
The possible origin of the particle highlighted in red, for instance, are marked
with question marks in this stage.
Following the application, a distinct reflection operator is applied based on the state
of the $a_x$ and $a_y$ qubits.

\begin{algorithm}
\caption{Zone-Agnostic Specular Reflection Boundary Condition Imposition}
\label{alg:bp-2-sr}
\begin{algorithmic}[1]
\Require Oracle $\mathrm{U}_\Omega$, Streaming Operator $\mathrm{U}_{\mathrm{S}}$
\State Stream all populations \label{alg:bp-2-sr:step1}
\State Apply $\mathrm{U}_\Omega$ targeting the ancilla $a_o$ \label{alg:bp-2-sr:step2}
\State \Call{FlagCrossingFactors}{$-1$}\label{alg:bp-2-sr:step3} \Comment{\Cref{alg:bp-2-flag-crossing}: identify components}
\State Reflect components that caused boundary crossing \label{alg:bp-2-sr:step4}
\State Stream inverted dimensional components \label{alg:bp-2-sr:step5}
\State Inverse stream all populations \label{alg:bp-2-sr:step6}
\State \Call{FlagCrossingFactors}{$+1$}\label{alg:bp-2-sr:step7} \Comment{\Cref{alg:bp-2-flag-crossing}: uncompute the flagging}
\State Apply $\mathrm{U}_\Omega$ targeting the ancilla $a_o$ \label{alg:bp-2-sr:step8}
\State Stream all populations \label{alg:bp-2-sr:step9}
\end{algorithmic}
\end{algorithm}

Intuitively, the algorithm uses the information provided
by the causal explanation imprinted onto the ancilla qubits
to apply the appropriate sequence of the reflection operator.
Though the reflection can again be reduced to a permutation, as outlined in
\Cref{fig:bp-2-za-bb-reflection}, in specular reflection we differentiate
between $x$- and $y$- component permutations.
The former consist of the transpositions
$\{(1\leftrightarrow 3), (5\leftrightarrow 6), (7\leftrightarrow 8)\}$ (\ie, reflection about the $x$-axis),
while the latter involve
$\{(2\leftrightarrow 4), (5\leftrightarrow 8), (6\leftrightarrow 7)\}$ (\ie, reflection about the $y$-axis).
Clearly, the two reflections can be composed to obtain exactly
physical space rotation of the bounce-back rule.
This rotation is precisely the one
required by the $\{\{x, y\}\}$ and $\{\{x\}, \{y\}\}$ explanations.
Following this procedure in Step \ref{alg:bp-2-sr:step4} and Stage 4,
the algorithm performs controlled
stream and uncontrolled inverse stream operations
(Steps and Stages \ref{alg:bp-2-sr:step5} and \ref{alg:bp-2-sr:step6}).

Resetting the ancilla qubits, however, requires additional care.
Unlike the bounce-back case, a simple application of the oracle does not suffice
to reset the $a_x$ and $a_y$ ancilla, and neither does an inverse application of
\Cref{alg:bp-2-sr:step3}.
Since the direction of some, but necessarily all velocity components has changed,
a different check is required to verify \emph{which} components have been affected.
This is exactly the difference that using $\sigma=+1$ in \Cref{alg:bp-2-flag-crossing}
allows for: when checking the positive direction, the qubit toggles
encode which factors \emph{would} cause crossing in the current velocity.
For the inverted velocity factors, this is easy to verify: the positive direction
in this phase is equivalent to the negative
direction that was flagged as a cause in \Cref{alg:bp-2-sr:step4}.
For the components that have not been inverted, it is important to notice that inverse
stream from Step \ref{alg:bp-2-sr:step6} implies that the check performed
by \Cref{alg:bp-2-flag-crossing} will not result in false positives:
if a component is inverse streamed from a position in $\Omega$,
the check performed on that component is equivalent to undoing the inverse stream,
and will therefore also land within $\Omega$, thus not causing a boundary crossing.
As before, the gridpoints queried during this check are highlighted with
color-coded question marks in Stage 7 of \Cref{fig:bp-2-za-sr-example}.
Finally, the algorithm concludes by applying the oracle to reset the $a_o$
qubit, and the populations are streamed to their final positions.

\paragraph{Assumptions and correctness}
The viability of the ZA-SR algorithm depends on three key factors:
the shape of the region $\Omega$,
the number of distinct regions being addressed,
and the velocity discretization.
Unlike the BB case, the SR stream-unstream loop makes it such that
particles travel to several other intermediate points
other than their initial gridpoint and the position within $\Omega$.
In cases with multiple obstacles, or with highly irregular obstacles,
one must ensure that no population is misflagged in the application
of \Cref{alg:bp-2-flag-crossing}.
We prove this property for \emph{unit-speed} $2$-dimensional
stencils in the Appendix of this work.
We leave a general assessment of the correctness
and limitations of the method
for more complex discretizations to future work.
% Due to complex velocity sets that can emerge in practice,
% for instance, \dq{3}{125} \cite{chikatamarla2009lattices},
% as well as the high complexity of practically useful obstacles,
% we postpone the general proof of correctness to future work.

\paragraph{Complexity analysis}
The difference between the ZA and SW realization
of specular reflection can be understood
as a difference in how information is queried.
The SW method sequentially identifies each segment
of the object, and applies a region-specific reflection operator.
By contrast, the ZA method identifies every possible 
cause for boundary crossing, and uses a cause-specific reflection circuit.
While the two reflections solve the same problem, the number of times
they are applied differs significantly.
Clearly, whether the ZA implementation offers any advantage
hinges on the scaling of the causal explanation implementation.

To analyze the complexity of \Cref{alg:bp-2-flag-crossing}, we can refer
to its established counterpart in combinatorics.
In particular, it is known that the number of antichains
on the subsets of a finite set with cardinality $n$ is
the $n^\text{th}$ Dedekind number, $\mathcal{D}(n)$ \cite{stanley2012,dedekind1897}.
Kleitman \cite{kleitman1969} shows that the scaling of this number
is doubly exponential in $n$, and Hansel \cite{hansel1966} has shown that
it is upper bounded by $D(n) \leq 3^{\binom{n}{\lfloor n/2 \rfloor}}$.
In the current body of literature, the largest known Dedekind number
is $\mathcal{D}(9)$, computed independently
by J{\"a}kel \cite{jakel2023} and Hirtum et al. \cite{vanhirtum2024}.

Establishing the implications of the scaling of ZA-SR
algorithm requires a certain degree of subjectivity.
The algorithm avoids an up-to-exponential
number of sequential boundary treatments by placing the burden
on a doubly exponential routine in the number of dimensions.
The advantage of the ZA implementation is positively
influenced by complex, low-dimensional shapes,
and drastically penalized by increasing the number of dimensions.
It is, however equally important to consider that typical
problems for which such boundary conditions would be applied are 2-
and 3-dimensional.
Even though $\mathcal{D}(9)$ contains 42 digits \cite{jakel2023,vanhirtum2024},
for practical problems $\mathcal{D}(3) = 20$, and therefore there are
$18$ possible antichain mappings of explanations for each discrete velocity.
Each antichain has at most $3$ components in $3d$.
Therefore, \Cref{alg:bp-2-flag-crossing} requires
at most $18 \cdot 3 \cdot 2 = 108$ modified streaming blocks and $54$
oracle applications.

We thus argue that the number of applications of the oracle
required to determine the cause of boundary crossing for practical applications
is $\mathcal{O}(q)$, with a multiplicative constant of at most $108$
when accounting for the forward and the backward passes of \Cref{alg:bp-2-flag-crossing} in 
\Cref{alg:bp-2-sr}.
Similarly, there are at most $216$ $\mathcal{O}(q\cdot n_g^2)$ additional
one- and two-qubit gates that perform modified streaming blocks.
The cost of the controlled velocity permutation does also not exceed
the cost of \Cref{alg:bp-2-flag-crossing}, as in the worst case,
as many permutations are applied as there are antichains.
We finally note that as the circuit depicted in \Cref{fig:bp-2-za-sr-check} demonstrates,
the number of oracle applications can be
reduced by exploiting the relation between factors, which we leave to future work.
By comparison, the best known $3d$ SW
breakdown of a cuboid object for SR uses $80$ edge cases,
each of which requires a streaming step and an
operation comparable to that of an oracle \cite{georgescu2025qlbm}.

\subsection{Oracle design and complexity}
\label{subsec:bp-2-oracle}

Our complexity analyses thus far have focused on counting the number 
of times our algorithms apply the oracle $\mathrm{U}_\Omega$,
which dominates the overall runtime of the reflection step.
Clearly, the ZA method is only efficient so long as the oracle itself
is efficiently implementable as a quantum circuit.
In this section, we first analyze the complexity of oracles encoding
shapes described in previous work, before discussing how oracles can be derived
from quantum arithmetic.

\subsubsection{Axis-aligned objects}

Both Todorova and Steijl \cite{todorova2020quantum}
and Schalkers and M\"{o}ller \cite{schalkers2024efficient}
describe implementations of specular reflection operators
against axis-aligned objects (\ie, rectangles and cuboids).
In particular, the SW breakdown of
Schalkers and M\"{o}ller \cite{schalkers2024efficient}
is implemented by a circuit with $2(d-1)$ ancilla qubits that works as follows.
Since each segment is axis-aligned and spans $d-1$ dimensions,
(a line segment in $2d$ and a bounded plane in $3d$), they can be
described by a lower bound and an upper bound in each dimensions.
The $2(d-1)$ qubits encode the information of whether the basis state
describing the physical position of a population
abides by the particular bound it is assigned to.
If all ancilla qubits are in state $\ket{1}$,
then the gridpoint is within the region of the segment.

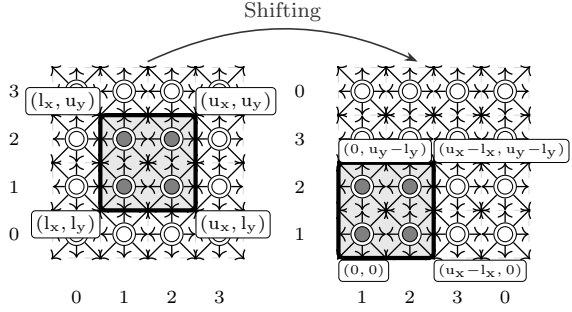
\begin{figure}[htbp]
    \centering
    \scalebox{1}{\centering
\begin{tikzpicture}[scale=0.1575]
    \tikzset{cornerLabel/.style={font=\scriptsize, inner sep=1.5pt, fill=white, draw=black, rounded corners=1.5pt, line width=0.3pt}}

    % --- Original specification (left) --------------------------------
    \begin{scope}[shift={(0,0)}, local bounding box=gridOrig]
        % 4x4 D2Q9 lattice with step=4: points at -6,-2,2,6.
        \draw[
          help lines,
          line width=0.4pt,
          color=gray!30,
          dashed
        ] (-8, -8) grid[step={($(4, 4) - (0, 0)$)}] (8, 8);

        % Background lattice points (D2Q9 stencils, white fill).
        \foreach \x in {-6,-2,2,6} {
            \foreach \y in {-6,-2,2,6} {
                \dtwoqeight{(\x,\y)}{0.1cm}{2}{0}{1}{white}
            }
        }

        % 2x2 square geometry centered on the lattice.
        \fill[gray!18] (-4,-4) rectangle (4,4);

        % Interior lattice points shaded gray.
        \foreach \x in {-2,2} {
            \foreach \y in {-2,2} {
                \dtwoqeight{(\x,\y)}{0.1cm}{2}{0}{1}{gray}
            }
        }

        % Object outline.
        \draw[black, ultra thick, rounded corners=2pt] (-4,-4) rectangle (4,4);

                % Axis labels: standard 0--3 indexing.
        \foreach \i/\lab in {0/0,1/1,2/2,3/3} {
            \node[anchor=north, font=\scriptsize] at ({-6+\i*4},-10) {$\lab$};
            \node[anchor=east,  font=\scriptsize] at (-10,{-6+\i*4}) {$\lab$};
        }

        % Corner annotations.
        \node[cornerLabel, anchor=north east] at (-4,-4) {$(\mathrm{l_x},\mathrm{l_y})$};
        \node[cornerLabel, anchor=south east] at (-4, 4) {$(\mathrm{l_x},\mathrm{u_y})$};
        \node[cornerLabel, anchor=north west] at ( 4,-4) {$(\mathrm{u_x},\mathrm{l_y})$};
        \node[cornerLabel, anchor=south west] at ( 4, 4) {$(\mathrm{u_x},\mathrm{u_y})$};

    \end{scope}

    % --- Shifted specification (right) --------------------------------
    \begin{scope}[shift={(24,0)}, local bounding box=gridShift]
        \draw[
          help lines,
          line width=0.4pt,
          color=gray!30,
          dashed
        ] (-8, -8) grid[step={($(4, 4) - (0, 0)$)}] (8, 8);

        % Background lattice points (D2Q9 stencils, white fill).
        \foreach \x in {-6,-2,2,6} {
            \foreach \y in {-6,-2,2,6} {
                \dtwoqeight{(\x,\y)}{0.1cm}{2}{0}{1}{white}
            }
        }

        % 2x2 square geometry shifted so its lower bounds sit at the lattice origin.
        \fill[gray!18] (-8,-8) rectangle (0,0);

        % Interior lattice points shaded gray.
        \foreach \x in {-6,-2} {
            \foreach \y in {-6,-2} {
                \dtwoqeight{(\x,\y)}{0.1cm}{2}{0}{1}{gray}
            }
        }

        % Object outline.
        \draw[black, ultra thick, rounded corners=2pt] (-8,-8) rectangle (0,0);

        % Corner annotations.
        \node[cornerLabel, anchor=south west] at (-8,-10) {\tiny{$(0,0)$}};
        \node[cornerLabel, anchor=north west] at (-8, 2) {\tiny{$(0,\mathrm{u_y}{-}\mathrm{l_y})$}};
        \node[cornerLabel, anchor=south west] at ( 0,-10) {\tiny{$(\mathrm{u_x}{-}\mathrm{l_x},0)$}};
        \node[cornerLabel, anchor=north west] at ( 0, 2) {\tiny{$(\mathrm{u_x}{-}\mathrm{l_x},\mathrm{u_y}{-}\mathrm{l_y})$}};

        % Axis labels: cyclically shifted to indicate the boundary shift.
        \foreach \i/\lab in {0/1,1/2,2/3,3/0} {
            \node[anchor=north, font=\scriptsize] at ({-6+\i*4},-10) {$\lab$};
            \node[anchor=east,  font=\scriptsize] at (-10,{-6+\i*4}) {$\lab$};
        }
    \end{scope}

    \tikzset{connArrow/.style={-{Stealth[length=5pt, width=3.5pt]}, shorten <=4pt, shorten >=4pt, semithick, black!80}}
    \draw[connArrow]
        (gridOrig.north) to[bend left=25]
        node[midway, above, font=\footnotesize] {Shifting}
        (gridShift.north);
\end{tikzpicture}}
    \caption{Grid shifting by adjusting lower bounds ($\mathrm{lb}$) and upper bounds ($\mathrm{ub}$).}
    \label{fig:bp-2-shifted-square}
\end{figure}

We can extend the approach
of Schalkers and M\"{o}ller \cite{schalkers2024efficient}
to encode axis-aligned objects as a whole, rather than their constituent segments,
at no additional qubit or gate cost.
We can construct such an oracle by manipulating the grid in $3$ steps:
shifting, upper-bounding and unshifting.
The core idea is built on the observation that if an object is specified by $d$
lower and upper bounds, we can shift the entire grid
by subtracting the lower bound modulo the size of the grid, such that the
object spans a rectangle or cuboid domain between the origin and 
the point specified by the upper bound in all dimensions.
\Cref{fig:bp-2-shifted-square} visually portrays a $2$-dimensional example
of shifting the original specification of a square
into its origin-anchored counterpart.

Using the shifting has the advantage that the lower bound
no longer needs to be checked, since the solid domain is guaranteed to start
at the origin.
As such, only the upper bound information needs to be imprinted onto 
one ancilla qubit per dimension, leading to $d$ ancilla qubits instead of
the $2(d-1)$ required for the segment-wise approach.
One can implement the grid shift by a simple subtraction step by a classically provided constant,
and the upper bound comparison operation
of Schalkers and M\"{o}ller \cite{schalkers2024efficient} can be
utilized without any modification.
The cost of both the subtraction and comparison steps
is known to be quadratic in the number of grid qubits \cite{schalkers2024efficient}.
The axis-aligned oracle operation is thus $\mathcal{O}(dn_g^2)$.

\subsubsection{Arithmetically-shaped volumes}
\label{subsec:bp-2-arithmetic}

The axis-aligned obstacles we previously addressed can be thought of
as some of the simplest objects of a much larger class:
\emph{arithmetically} specifiable shapes.
This class could generally (though not formally) be regarded
as all shapes that can be specified by defining a set
whose elements satisfy closed-form
logical predicates over arithmetic and expressions with
parameters known at circuit assembly time.
The expression for axis-aligned objects is 

\begin{equation}
    {\Large\land}_{0\leq j\leq d-1}\quad \boldsymbol{l}_j \leq \boldsymbol{x}_j \leq \boldsymbol{u}_j,
\end{equation}

\noindent with $\boldsymbol{l}$ and $\boldsymbol{u}$ the lower and upper
bounds in all dimensions, $\boldsymbol{x}$ the position on the grid, and $\land$ the
logical ``and'' operator..
Similarly, the sphere equation can be used to model circle-, sphere-, and cylinder-shaped
geometries within our oracular framework.

The question of \emph{which} arithmetic expressions can be efficiently
implemented as quantum circuits warrants some consideration.
Next to addition \cite{draper2000addition} and multiplication \cite{ruiz2017quantum},
H\"{a}ner et al. \cite{haner2018optimizing} show that the set of efficient
quantum arithmetic operations includes
the square root, Gaussians, hyperbolic tangent,
polynomials, exponential, as well as several trigonometric functions.
We note that some of these functions may require additional (clean) qubit registers:
for instance a polynomial of degree $p$
with variables encoded in $n$ qubits,
and coefficients requiring $b$ qubits to encode
requires $\mathcal{O}(p^2n + pb)$ ancillary qubits to evaluate
an exact integer expression \cite{haner2018optimizing}.

Crucially, however, both the number of qubits
and the number of 1- and 2-qubit gates required
to realize the arithmetic operations described by H\"{a}ner et al. \cite{haner2018optimizing}
scale polynomially in the number of grid qubits.
This is in contrast to the up-to-exponential scaling that the segmentwise
approach of Schalkers and M\"{o}ller \cite{schalkers2024efficient}
incurs for irregular shapes that require thin segments of the grid to be
treated sequentially.

\paragraph{Example: $x>y^2$}

\begin{figure}[htbp]
    \centering
    \resizebox{\linewidth}{!}{\centering
\begin{tikzpicture}[scale=0.15]
    % Distinct pastel fills, one per staircase segment.
    \definecolor{segA}{RGB}{255,210,210}
    \definecolor{segB}{RGB}{255,228,196}
    \definecolor{segC}{RGB}{255,245,205}
    \definecolor{segD}{RGB}{215,245,215}
    \definecolor{segE}{RGB}{200,235,235}
    \definecolor{segF}{RGB}{200,228,250}
    \definecolor{segG}{RGB}{225,210,245}
    \definecolor{segH}{RGB}{255,220,235}
    \definecolor{segI}{RGB}{240,230,200}
    \definecolor{segJ}{RGB}{220,240,225}
    \definecolor{segK}{RGB}{255,225,210}
    \definecolor{segL}{RGB}{235,225,250}

    % Segment background fills: maximal runs of columns sharing a common
    % staircase height for the region y < sqrt(x) on a 128 x 16 lattice.
    % Column 0 contributes no shaded points and is omitted.
    \fill[segA] (  0.5,-0.5) rectangle (  1.5,  0.5);   % col 1,        h=1
    \fill[segB] (  1.5,-0.5) rectangle (  4.5,  1.5);   % cols 2-4,     h=2
    \fill[segC] (  4.5,-0.5) rectangle (  9.5,  2.5);   % cols 5-9,     h=3
    \fill[segD] (  9.5,-0.5) rectangle ( 16.5,  3.5);   % cols 10-16,   h=4
    \fill[segE] ( 16.5,-0.5) rectangle ( 25.5,  4.5);   % cols 17-25,   h=5
    \fill[segF] ( 25.5,-0.5) rectangle ( 36.5,  5.5);   % cols 26-36,   h=6
    \fill[segG] ( 36.5,-0.5) rectangle ( 49.5,  6.5);   % cols 37-49,   h=7
    \fill[segH] ( 49.5,-0.5) rectangle ( 64.5,  7.5);   % cols 50-64,   h=8
    \fill[segI] ( 64.5,-0.5) rectangle ( 81.5,  8.5);   % cols 65-81,   h=9
    \fill[segJ] ( 81.5,-0.5) rectangle (100.5,  9.5);   % cols 82-100,  h=10
    \fill[segK] (100.5,-0.5) rectangle (121.5, 10.5);   % cols 101-121, h=11
    \fill[segL] (121.5,-0.5) rectangle (127.5, 11.5);   % cols 122-127, h=12

    % Dashed lattice grid.
    \draw[help lines, line width=0.20pt, color=gray!50, dotted]
        (-0.5,-0.5) grid[step=0.5] (127.5,15.5);

    % Vertical dashed separators between segments. Drawn after the help
    % grid so they sit on top of it, with a thicker, darker dash pattern
    % to remain visually distinct from the lattice gridlines.
    \foreach \xb in {0.5,1.5,4.5,9.5,16.5,25.5,36.5,49.5,64.5,81.5,100.5,121.5} {
        \draw[densely dashed, line width=0.6pt, gray!80]
            (\xb,-0.5) -- (\xb,15.5);
    }

    % Outer bounding box.
    \draw[black, thick] (-0.5,-0.5) rectangle (127.5,15.5);

    % Lattice gridpoints under the curve, drawn as grey dots.
    % We loop only over shaded points (per-column count = ceil(sqrt(x))).
    \foreach \x in {1,...,127} {
        \foreach \y in {0,...,15} {
            \fill[gray!75] (\x,\y) circle (0.18);
        }
    }

    % Continuous curve y = sqrt(x), drawn until it leaves the right edge
    % of the grid. Within the figure the curve never reaches the top.
    \draw[red, very thick, smooth, samples=200, domain=0:127.5]
        plot ({\x},{sqrt(\x)});

    % Staircase boundary of the shaded region.
    \draw[black, very thick]
        ( -0.5,-0.5) -- (  0.5,-0.5) -- (  0.5, 0.5) -- (  1.5, 0.5)
        -- (  1.5, 1.5) -- (  4.5, 1.5) -- (  4.5, 2.5) -- (  9.5, 2.5)
        -- (  9.5, 3.5) -- ( 16.5, 3.5) -- ( 16.5, 4.5) -- ( 25.5, 4.5)
        -- ( 25.5, 5.5) -- ( 36.5, 5.5) -- ( 36.5, 6.5) -- ( 49.5, 6.5)
        -- ( 49.5, 7.5) -- ( 64.5, 7.5) -- ( 64.5, 8.5) -- ( 81.5, 8.5)
        -- ( 81.5, 9.5) -- (100.5, 9.5) -- (100.5,10.5) -- (121.5,10.5)
        -- (121.5,11.5) -- (127.5,11.5);

    % Axis ticks: every 16 along x (very wide), every 4 along y.
    \foreach \i in {0,16,32,48,64,80,96,112,127} {
        \node[anchor=north, font=\scriptsize] at (\i,-1.4) {\i};
    }
    \foreach \i in {0,4,8,12,15} {
        \node[anchor=east,  font=\scriptsize] at (-1.4,\i) {\i};
    }

    % Curve label, anchored above the right end of the curve.
    \node[red, anchor=south] at (120,12.0) {$y=x^2$};
\end{tikzpicture}}
    \caption{Staircase discretization of the region $\Omega = \{ (x, y)~|~x > y^2 \}$ on a $128 \times 16$
    lattice. The area under the red curve determines the solid domain $\Omega$, with
    colors and dashed lines outlining a possible segment-wise breakdown.
    \label{fig:bp-3-ymon-staircase}}
\end{figure}
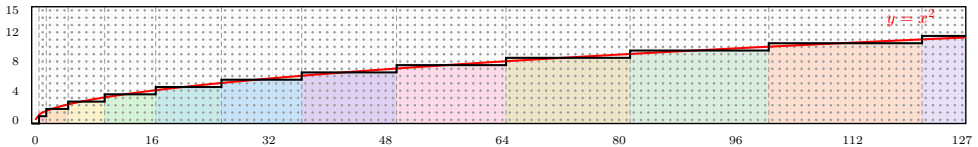

To help provide intuition the class of shapes that are arithmetically specifiable,
we consider a straightforward example.
Let us consider a $2$-dimensional lattice of $128\times 16$ gridpoints,
and let the region

\begin{equation}
    \Omega = \{ (x, y) \in \{0, \ldots, 127 \} \times \{0, \ldots, 15\}~|~ x>y^2 \}
\end{equation}

\noindent outline the solid domain.
\Cref{fig:bp-3-ymon-staircase} depicts the staircase approximation that emerges
from applying the predicate $x>y^2$ over all gridpoints of the grid.
In this case, $\Omega$ is the region underneath the red curve, which is subject to boundary conditions.
Applying the segment-wise paradigm to this region yields the $12$ distinct segments
highlighted in different colors and delimited with dashed lines in \Cref{fig:bp-3-ymon-staircase}.
Clearly, both the number of segments, as well as the number
of edge cases (\ie, at the intersection of segments) grows with the size of the grid with
the irregularity of the shape.
By contrast, one can realize the ZA counterpart with either two
or five applications of the oracle, depending
on whether BB or SR BCs are prescribed.
Though the number of oracle applications does
not scale with the size of the grid, the cost of each oracle does.
Our chosen operation requires one
multiplication and one comparison operation,
both of which are polylogarithmic in the size of the grid
\cite{ruiz2017quantum,schalkers2024efficient}. 
The following section
verifies the correctness and quantifies the cost of the ZA method
on both axis-aligned segments and simple, monomial-shaped geometries
such as the example considered in \Cref{fig:bp-3-ymon-staircase}.

\section{Results}
\label{sec:bp-3-results}

We assess the ZA algorithm in terms of two key measures: correctness and cost.
In \Cref{subsec:bp-3-1-equivalence}, we compare the results
of the ZA algorithm against the SW baseline and
show that the 2 implementations produce equivalent quantum states
on lattices with axis-aligned solid objects.
In \Cref{subsec:bp-3-2-cost}, we study the relative cost of the ZA
implementation against the SW counterpart, on both axis-aligned objects
and more complex shapes.
Both implementations are available in the \qlbm~library \cite{georgescu2025qlbm},
and we include all data and experiments in a replication package \cite{georgescu2026boundariesreplication}.

\subsection{Correctness}
\label{subsec:bp-3-1-equivalence}

To verify the correctness of our method, we compare the quantum
states produced by the ZA implementation against the baseline
SW realization described by
Schalkers and M\"{o}ller \cite{schalkers2024efficient}.
Our tests include only streaming and boundary conditions.
We deliberately choose to exclude the collision step
from the simulation, as it is a completely separate algorithmic step
which generally remains an open research question.
Its inclusion would only add unrelated noise to our simulations.
We also note that although correctness tests are empirical rather than exhaustive,
they cover all physically realizable cases in which the two methods are
fairly comparable in $2d$.
Further, exhaustive unit tests over all physical realizations
accompany our open-source implementation.

We perform the comparison over \dq{2}{9} grids with
$16\times 16$ and $32\times 32$ gridpoints.
For each grid we sample $k$ rectangular obstacles with $k \in \{1, 2, 4, 6, 8\}$,
which we assign either BB, SR, or mixed (half-BB and half-SR)
boundary conditions.\footnote{For the mixed case, we skip $k=1$.}
For each grid size, number of obstacles, and BC setting, we randomly sample five
sets of non-overlapping obstacle positions and execute the simulation for $50$ timesteps.
The initial conditions are set by means of an equal magnitude superposition
over the nine velocity directions, such that all possible cases occur in the simulation.
The populations are spread over the first two columns of the grid
(\ie, there are $32\cdot 9$ and $64\cdot 9$ distinct basis states, respectively).
In total, we simulate $7140$ timesteps.
Let the states for a given timestep, and configuration be $\ket{\uppsi_\textrm{SW}}$
and $\ket{\uppsi_\textrm{ZA}}$, respectively.
To assess whether the two quantum states produce physically equivalent
states, we compute $|\langle \uppsi_\textrm{SW} | \uppsi_\textrm{ZA} \rangle |^2$.

\begin{figure}[htbp]
  \centering
  \includegraphics[scale=0.5]{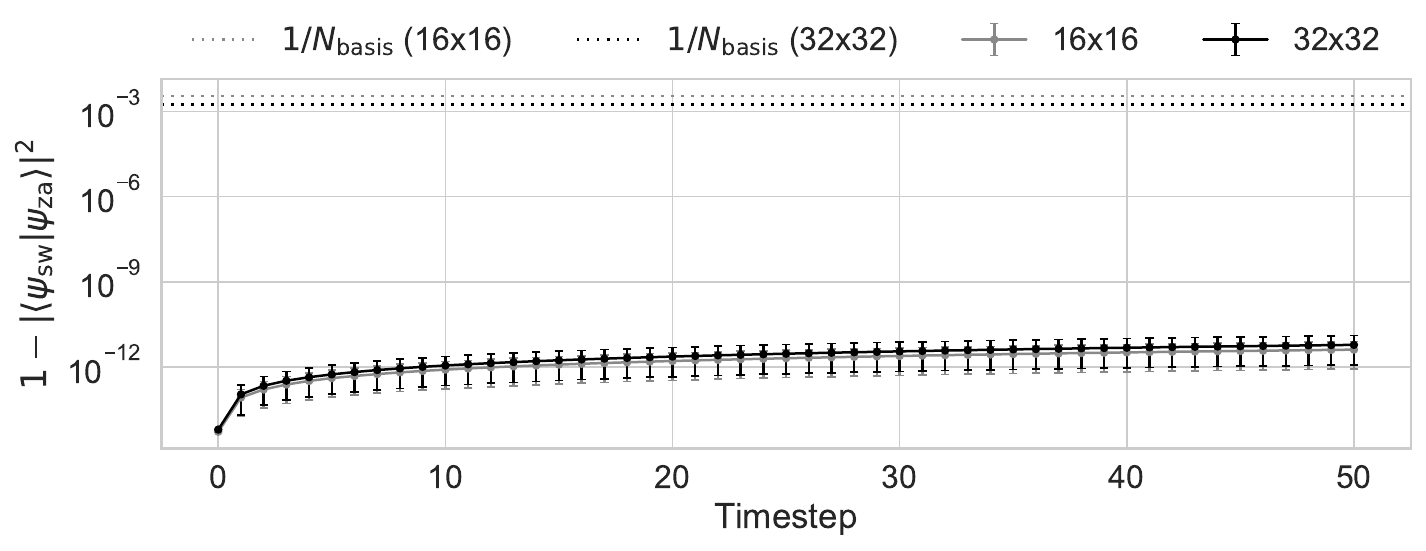}
  \caption{$95\%$ confidence interval for fidelity of
  quantum states obtained by the SW and ZA boundary conditions
  on $16\times 16$ and $32 \times 32$ grids
  for $50$ timestep of $5$ configurations with between $1$ and $8$ rectangle obstacles.}
  \label{fig:bp-3-equiv-vs-step}
\end{figure}

\Cref{fig:bp-3-equiv-vs-step} displays the 
$1-|\langle \uppsi_\textrm{SW} | \uppsi_\textrm{ZA} \rangle |^2$
value over all configurations and all timesteps.
The dashed lines indicate the minimum error
that would occur, should the quantum states differ by at least one basis state.
This is equivalent to checking whether at least one population
would be reflected differently between the two methods.
This threshold is the probability given by $1/N_\mathrm{basis}$, which is
$\approx 3.5\cdot 10^{-3}$ and $\approx 1.7\cdot 10^{-3}$ for
our $16\times 16$ and $32\times 32$ grids, respectively.

The largest deviation observed in our simulations
is $\approx 1.27\cdot 10^{-11}$, which occurred in one of the instances
associated with the most complex setting, with eight
obstacles prescribed with SR BCs on the $32\times 32$ grid.
This result implies that the highest deviation between the two methods
is almost eight orders of magnitude below our established threshold.
Though the deviation is low, one must also analyze
why the results shown in \Cref{fig:bp-3-equiv-vs-step}
indicate that the state fidelity decreases
as with the number of timesteps.
Importantly, this discrepancy could stem either from
an methodological error that repeatedly incorrectly
redistributes amplitudes among basis states, or from
numerical errors introduced by the quantum simulation software.

\begin{figure}[htbp]
  \centering
  \hfill
  \begin{subfigure}[b]{0.45\textwidth}
    \centering
    \includegraphics[scale=0.3]{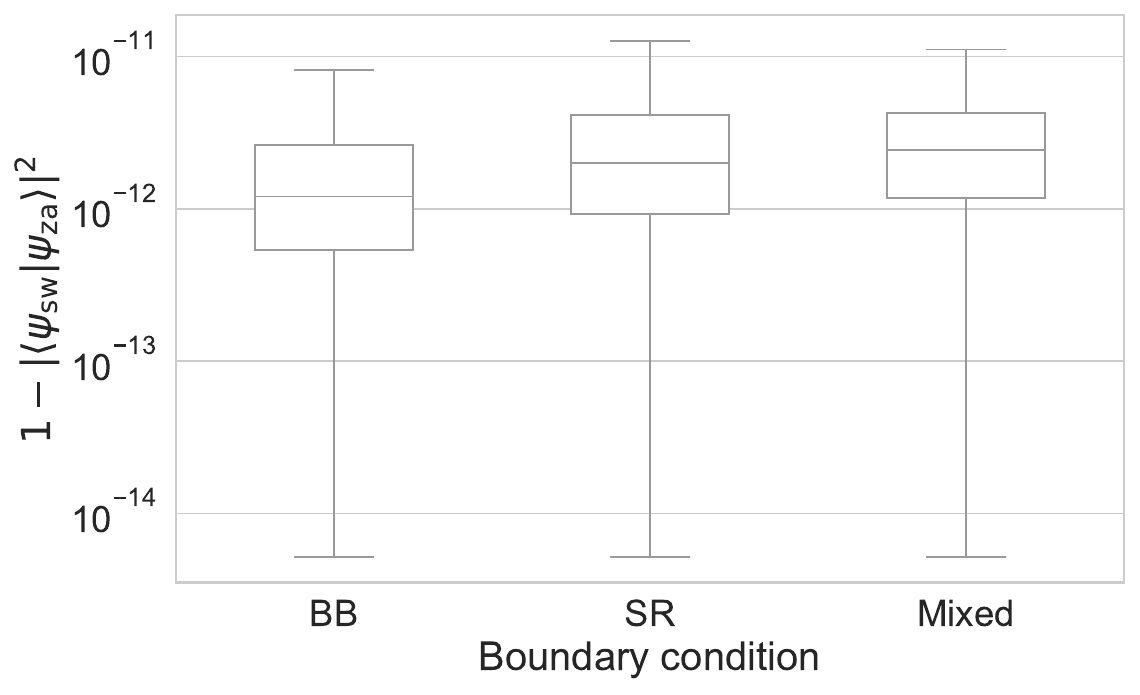}% placeholder
    \caption{Analysis by boundary condition type.}
    \label{fig:bp-3-equiv-vs-bc}
  \end{subfigure}
  \hfill
  \begin{subfigure}[b]{0.45\textwidth}
    \centering
    \includegraphics[scale=0.3]{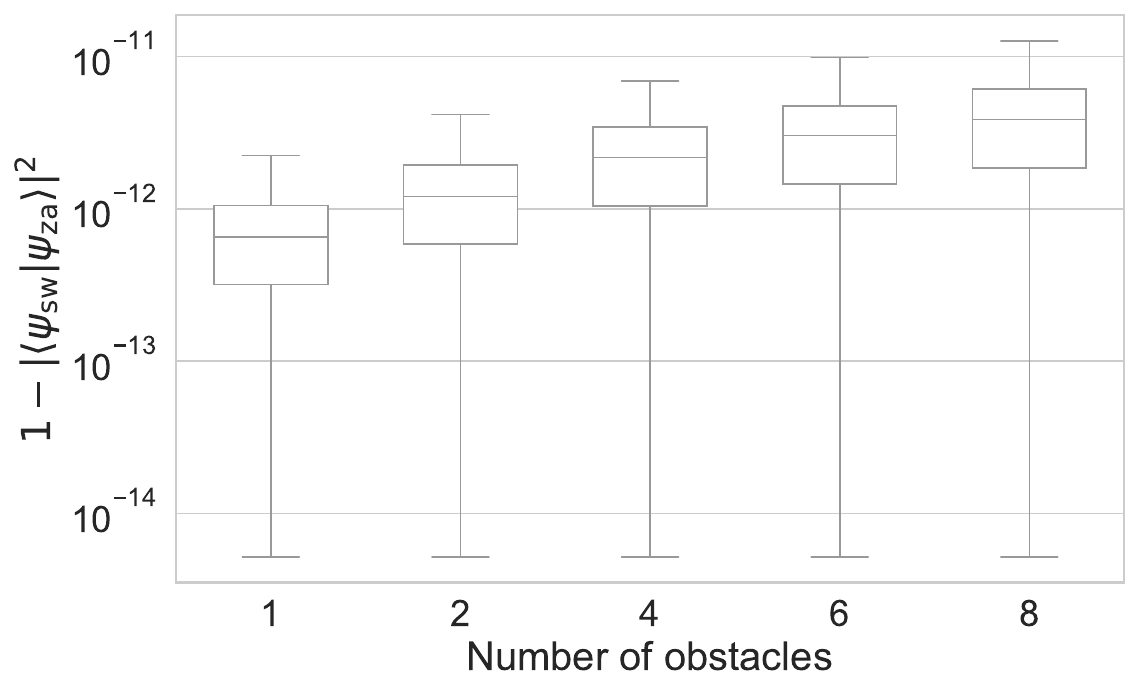}% placeholder
    \caption{Analysis by the number of objects.}
    \label{fig:bp-3-equiv-vs-obs}
  \end{subfigure}
  \hfill
  \caption{Fidelity of quantum states obtained by SW and ZA BCs as a function of
  BC type and number of rectangle objects on $16\times 16$ and $32 \times 32$ grids over $50$ timesteps.}
  \label{fig:bp-3-equiv-vs-bc-and-obs}
\end{figure}

To identify which of the two reasons is more likely to cause
the fidelity decrease, we further analyze the results by considering
boundary condition types and the number of objects,
as depicted in \Cref{fig:bp-3-equiv-vs-bc-and-obs}.
\Cref{fig:bp-3-equiv-vs-bc} indicates
that the fidelity is higher for BB-prescribed systems
and lower for the SR and mixed counterparts, while
\Cref{fig:bp-3-equiv-vs-obs} shows
that increasing the number of obstacles in the
system further increases the error.
All three analyses indicate that the
fidelity decreases as circuit complexity increases.
That is, more obstacles, more complex BCs, and more timesteps all contribute
to lower state fidelity.
This is consistent with the numerical error hypothesis, as increasing
circuit complexity requires more FLOPs, which contributes to the discrepancy.
Finally, the ZA circuits only consist of Hadamard, $\mathrm{CX}$, and phase gates,
none of which introduce the small amplitude deviations
we observe in the results.
Thus, we conclude that our results strongly indicate that the ZA
and SW boundary conditions produce physically equivalent results.
In what follows, we quantify the cost benefit
of the ZA approach over the SW counterpart.

\subsection{Cost}
\label{subsec:bp-3-2-cost}

To compare the cost of the ZA approach to the SW baseline,
we consider two cases.
First, we apply both methods to rectangular objects,
which are among the simplest $2$-dimensional shapes.
The reason for this is to gain insight whether there is a minimum
degree of complexity (\ie, a minimum number of segments) that a
shape ought to have for the SW approach to be practically cheaper.
The second case we consider is the $x>y^2$ predicate
described in \Cref{subsec:bp-2-arithmetic}.
The purpose of this comparison is to assess the cost difference between the 
two methods for shapes described by slightly more complex arithmetic predicates.

In both instances, we consider circuit depth as the cost measure, due
to it being a limiting factor for executing algorithms on quantum hardware
in the foreseeable future. We consider both bounce-back and specular reflection
BCs, and we transpile the circuits to the two gate sets:
Qiskit's default \textsc{AerSimulator} gate set,
and the gate set of the IBM Heron processor family.
The former provides an indication on what the difference is
with respect to today's simulation-based paradigms,
while the latter caters to real hardware gate sets.
All transpilations are performed through Qiskit's 
default transpiler, with optimization level 0.
The effect of circuit optimization techniques is outside the scope of this work.
In our analysis, we assume the qubits are all-to-all connected,
as nuanced hardware details are outside the scope of this work.

\subsection{Rectangular shapes}

To assess the difference between the ZA and SW methods on rectangular objects,
we consider square \dq{2}{9} grids with between $3$ and $14$ grid qubits per dimension,
\ie, with between $64$ and $2^{28} \approx 2.6 \cdot 10^8$ gridpoints.
We sample 10 randomly positioned objects and report mean values for depth.
We note that our analysis shows that variation is negligible and
that other measures of cost such as the number of gates, or $\mathrm{CX}$
gates scales very similarly.
Additional data is available in the replication package \cite{georgescu2026boundariesreplication}.

\begin{figure}[htbp]
  \centering
  \begin{subfigure}[b]{\textwidth}
    \centering
    \includegraphics[scale=0.5]{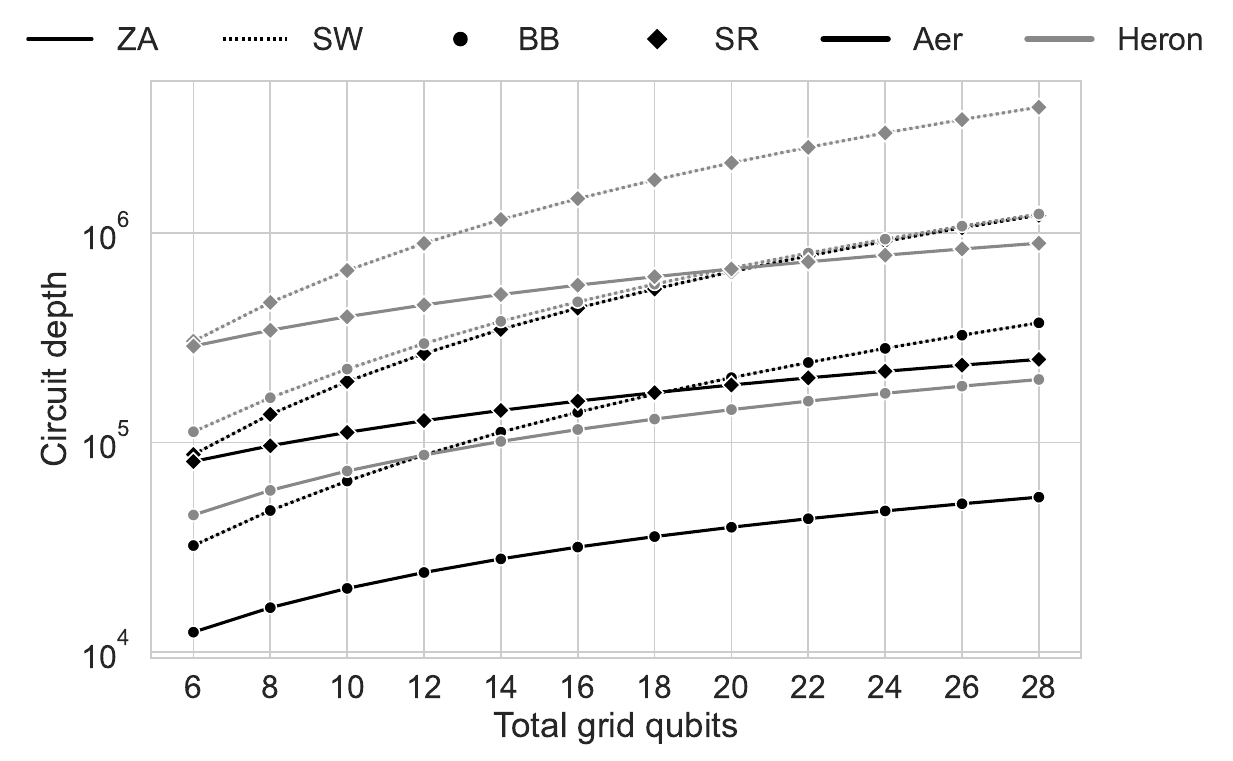}% placeholder
    \caption{Transpiled circuit depth.}
    \label{fig:bp-3-cost-cuboid-depth}
  \end{subfigure}
  \\
  \begin{subfigure}[b]{\textwidth}
    \centering
    \includegraphics[scale=0.5]{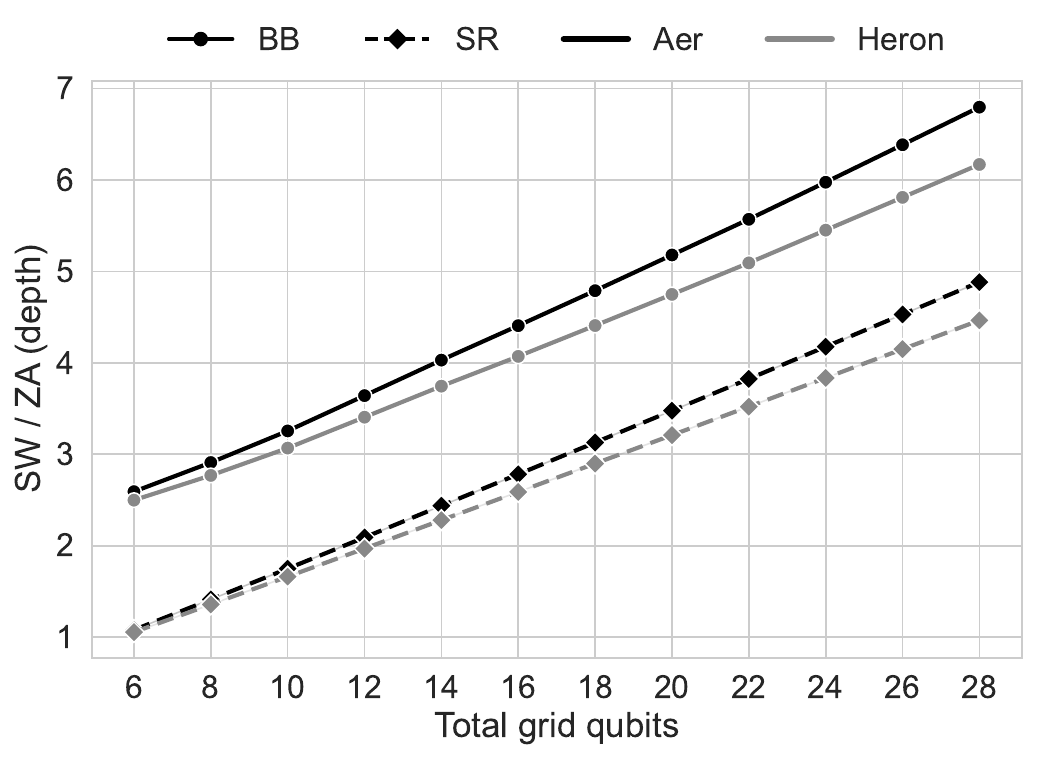}% placeholder
    \caption{Relative depth of SW compared to ZA.}
    \label{fig:bp-3-cost-cuboid-ratio}
  \end{subfigure}
\caption{Comparison of the circuit depth of the SW and ZA methods for rectangular shapes
  with bounce-back (BB) and specular reflection (SR) boundary conditions
on grid sizes between $8^2$ and $16384^2$, transpiled to the gate sets of
the Qiskit \textsc{AerSimulator} and IBM Heron processor.
All-to-all connectivity is assumed.}
  \label{fig:bp-3-cost-cuboid}
\end{figure}

\Cref{fig:bp-3-cost-cuboid} displays the results.
Notably, \Cref{fig:bp-3-cost-cuboid-depth} shows that the ZA realization (solid lines)
of the BCs outperforms its SW counterpart (dashed lines) in all tested instances.
One can explain this difference by comparing the number of $\mathcal{O}(n_g^2)$ operations
that each method utilizes.
This shows that, even for simple objects with a small number of distinct segments,
the ZA formulation generates significantly shallower circuits.
\Cref{fig:bp-3-cost-cuboid-ratio} emphasizes this difference by plotting
the ratio of the depth of the circuits belonging to the two methods.
The advantage is greater for BB than it is for SR,
which is explained by the additional steps 
the ZA-SR algorithm uses to reset the ancilla qubits.
Furthermore, the more expressive gate sets of the \textsc{AerSimulator}
yields a greater advantage as it allows for a more effective
decomposition of the common controlled phase gate.
Nonetheless, the relative advantage of the ZA algorithm increases
with the grid size for all tested combinations, yielding
circuits that are between $1/4$ and $1/7$ the depth of
the SW method for the largest grid size.

\subsection{Arithmetically specified shapes}

We consider the cost of applying SW and ZA BCs to the
shape $\Omega = \{ (x, y)~|~x > y^2 \}$ as depicted
in \Cref{fig:bp-3-ymon-staircase}.
For the SW approach, we partition $\Omega$ into rectangles
of the same height on the $y$ axis.
For the ZA approach, we employ $2n_{g_y}$ additional ancilla qubits
to allow for the computation of $y^2$ without overflow.
We compare the results on square \dq{2}{9} grids with between $3$ and $13$
grid qubits per dimension,
\ie, with between $64$ and $2^{26}$ total gridpoints.
The shape $\Omega$ associated with largest discretization consists of $90$
rectangular segments, and generally there are $\mathcal{O}(\sqrt{N_g})$.
We do not repeat the experiment, as there are no free parameters in this specification.

\begin{figure}[htbp]
  \centering
  \begin{subfigure}[b]{\textwidth}
    \centering
    \includegraphics[scale=0.5]{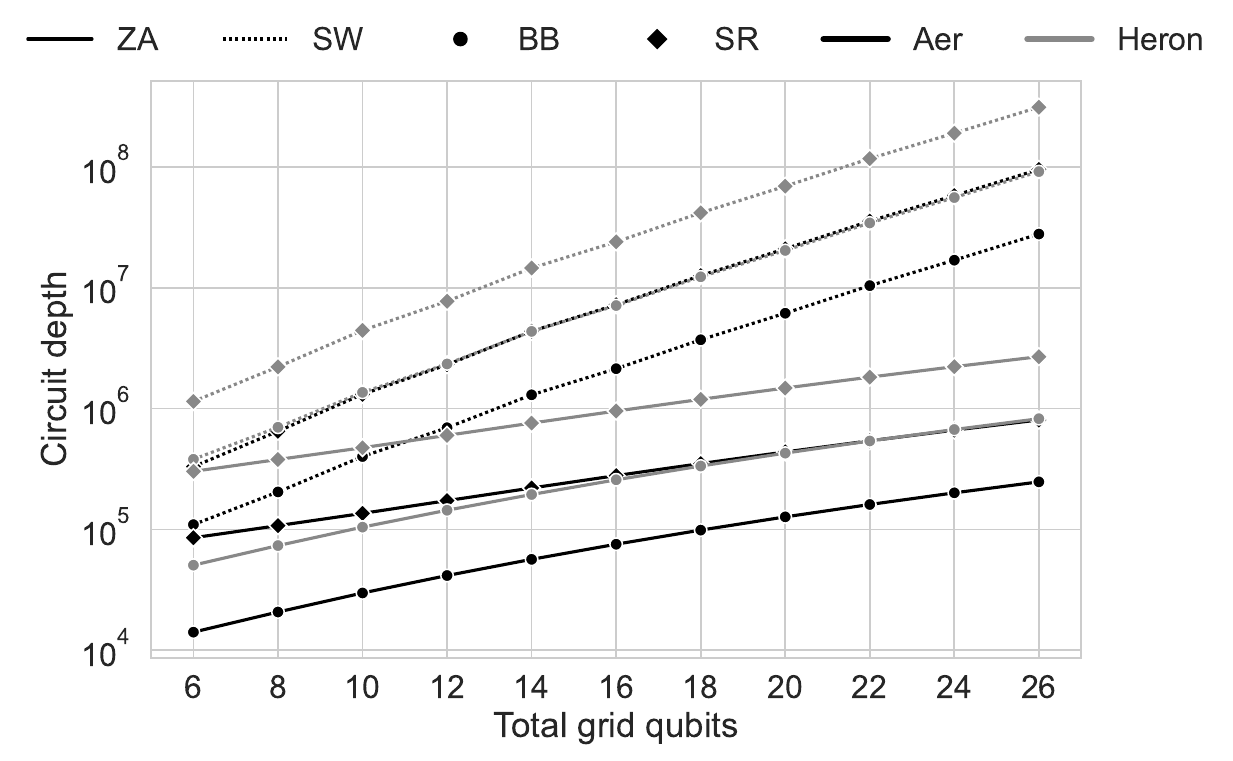}% placeholder
    \caption{Transpiled circuit depth.}
    \label{fig:bp-3-cost-ymon-depth}
  \end{subfigure}
  \\
  \begin{subfigure}[b]{\textwidth}
    \centering
    \includegraphics[scale=0.5]{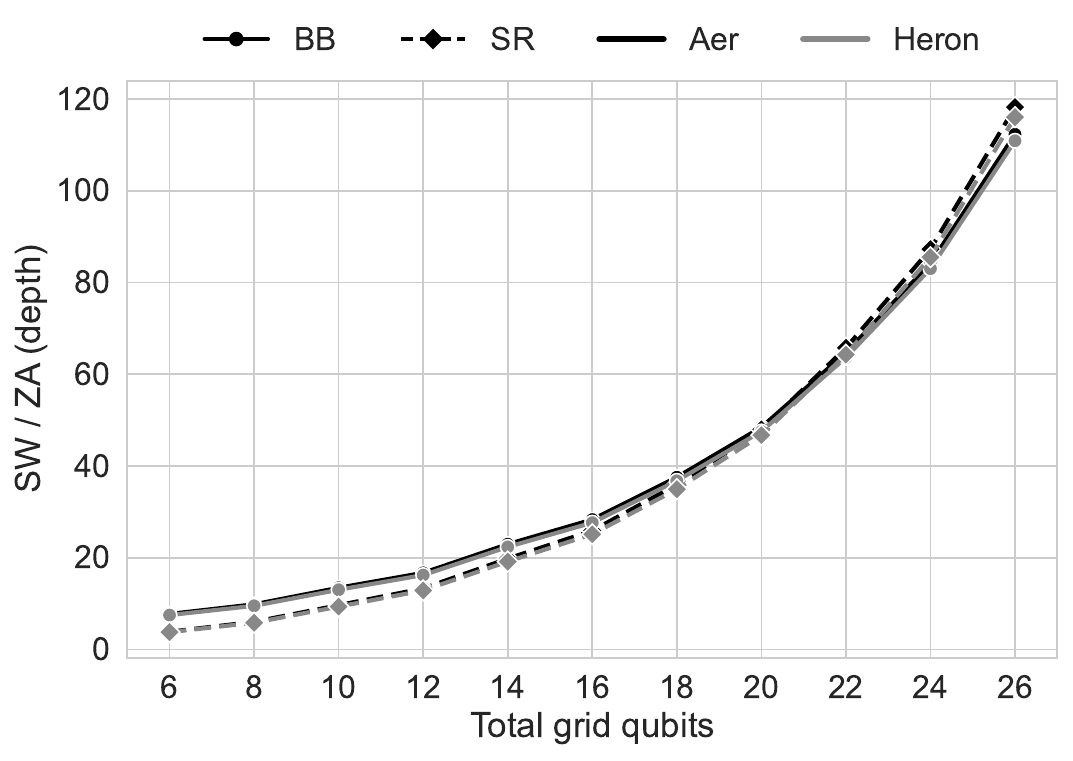}% placeholder
    \caption{Relative depth of SW compared to ZA.}
    \label{fig:bp-3-cost-ymon-ratio}
  \end{subfigure}
  \caption{Comparison of the circuit depth of the SW and ZA methods for $\Omega = \{ (x, y)~|~x > y^2 \}$
  with bounce-back (BB) and specular reflection (SR) boundary conditions,
  on grid sizes between $8^2$ and $8192^2$, transpiled to the gate sets of
  the Qiskit \textsc{AerSimulator} and IBM Heron processor.
  All-to-all connectivity is assumed.}
  \label{fig:bp-3-cost-ymon}
\end{figure}

\Cref{fig:bp-3-cost-ymon} displays the results.
\Cref{fig:bp-3-cost-ymon-depth} compares the depth of the ZA and SW circuits,
where the same trend emerges as in the analysis of the rectangular shapes:
the ZA variant outperforms its SW counterpart throughout all tested instances.
We note that the difference is far more significant, however, as
the ZA imposition is now up to two orders of magnitude shallower than the SW one.
\Cref{fig:bp-3-cost-ymon-ratio} displays the analysis of the depth ratio of the two methods.
The ratio significantly favors the ZA implementation, which requires less than $1\%$
the depth of the SW BCs for the largest grid, no matter the gate set or type BC.
The findings are consistent with the theoretical $\mathcal{O}(\sqrt{N_g})$ speedup
for the chosen arithmetic expression and predicate.
Clearly, more refined grids are relatively more advantageous for the ZA algorithm.

\section{Conclusion and future work}
\label{subsec:bp-4-conclusion}

We introduced a novel Zone-Agnostic (ZA) boundary condition imposition method
for quantum lattice Boltzmann methods.
The ZA algorithm enables
the application of boundary conditions atomically
over an entire region of space defined by an oracle.
This can lead to asymptotically better scaling compared
to previous work, which required that the region
be split into segments which are treated sequentially.
We defined the ZA algorithm for bounce-back and specular reflection
boundary conditions, and analyzed its scaling in terms of the number
of oracle applications.
We identified a small set of oracles that are polynomially
expensive to compute based on quantum arithmetic and simple logical expressions.
Finally, we demonstrated that the ZA algorithm yields physically equivalent
results through circuits that are significantly shallower than
previous work.
Nonetheless, our work suffers from several key limitations, which lead to
opportunities for further research.
We highlight four such shortcomings and suggest
avenues for future work.

\paragraph{Rigid oracles} We identified that
an efficient set of oracles consists
of simple logical queries over certain arithmetic expressions.
However, this set is far too inexpressive
to be useful for any foreseeable future industrial applications.
Deriving efficient oracles for industry-relevant cases,
either by composing known arithmetic operations
or by other means (\ie, machine learning)
would be necessary for this work to prove practically useful.
Moreover, studying how ZA and SW BCs can be complementary,
\ie, by supplementing a core oracle with SW patches
would also be beneficial.

\paragraph{More complex boundary conditions}
Our work is centered on halfway bounce-back and specular reflection
boundary conditions.
However, practitioners often employ more complex
schemes in their classical simulations.
Thus, extensions to interpolated boundary conditions,
inlet/outlet, and moving walls would all be valuable.

\paragraph{Complex specular reflection cases}
In our work, we found
an algorithm that requires only two applications
of the oracle to identify which of the four
possible causal explanations applies for reflected particles.
Yet, whether such an efficient algorithm exists in $3d$
remains an open question, as is the general
correctness of the ZA-SR algorithm.
Though all cases are sound for the \dq{2}{9}
discretization and contiguous objects,
we make no claims with regards to whether this is the
case for more complex velocity sets.
A thorough analysis of the properties of such systems
might reveal valuable insight into whether
the algorithm generalizes to practically useful scenarios.

\paragraph{Hardware analysis}
Our experiments focus on only two gate sets and do not
take into account aspects such as connectivity or error correction,
which, no doubt, will play a crucial role in determining whether
QLBMs are to reach the practical utility threshold.
Thus, extending our analyses to account for such variables,
as well as complementary optimization techniques,
would provide further insight.

\bibliographystyle{siamplain}
\bibliography{latex-bibliography/references}

\newpage
\appendix
\section{Proof of correctness of specular reflection for $2d$ unit-speed stencils}

In this section, we prove the correctness of the zone-agnostic
specular reflection (ZA-SR) boundary
condition method described in \Cref{subsec:bp-2-sr}, for unit-speed stencil discretizations.
We prove the behavior of the algorithm by enumeration over all
possible physically realizable cases.
In particular, we consider three kinds of cases:
when particles do not hit an object,
when they hit the object while traveling across
a cardinal link,
and when they hit the object through a diagonal link.
Before addressing these cases, however, we
begin by formalizing the kinds of stencils
for which our analysis holds.
We begin with the definition of a stencil.

\begin{definition}[Stencils]
\label{def:bp-sup-stencil}
Let $\mathcal{V}$ be a set of $q$ indices $\{0, \ldots, q-1\}$.
Let $\boldsymbol{\Delta}: \mathcal{V} \to \mathbb{R}^d$,
with $\boldsymbol{\Delta}(v)$ the physical displacement (or increment) associated
with index $v \in \mathcal{V}$.
Together $\left(\mathcal{V}, \boldsymbol{\Delta}\right)$ describe a velocity stencil.
\end{definition}

\Cref{def:bp-sup-stencil} provides the working definition of a stencil
as a tuple of indices and a function mapping those indices to a $d$-dimensional
real vector that describes the displacement of each velocity channel.
For the \dq{2}{9} example we have been using throughout this work,
\begin{subequations}
\label{eq:bp-sup-d2q9-deltas}
\begin{align}
  \boldsymbol{\Delta}_\text{\dq{2}{9}}(0) &= (0,0),  \label{eq:bp-sup-d2q9-delta-0} \\
  \boldsymbol{\Delta}_\text{\dq{2}{9}}(1) &= (1, 0),  \label{eq:bp-sup-d2q9-delta-1} \\
  \boldsymbol{\Delta}_\text{\dq{2}{9}}(2) &= (0, 1),  \label{eq:bp-sup-d2q9-delta-2} \\
  \boldsymbol{\Delta}_\text{\dq{2}{9}}(3) &= (-1, 0), \label{eq:bp-sup-d2q9-delta-3} \\
  \boldsymbol{\Delta}_\text{\dq{2}{9}}(4) &= (0, -1), \label{eq:bp-sup-d2q9-delta-4} \\
  \boldsymbol{\Delta}_\text{\dq{2}{9}}(5) &= (1, 1),  \label{eq:bp-sup-d2q9-delta-5} \\
  \boldsymbol{\Delta}_\text{\dq{2}{9}}(6) &= (-1, 1), \label{eq:bp-sup-d2q9-delta-6} \\
  \boldsymbol{\Delta}_\text{\dq{2}{9}}(7) &= (-1, -1),\label{eq:bp-sup-d2q9-delta-7} \\
  \boldsymbol{\Delta}_\text{\dq{2}{9}}(8) &= (1, -1). \label{eq:bp-sup-d2q9-delta-8}
\end{align}
\end{subequations}

In this section, we only consider stencils that share a particular property
of the \dq{2}{9} stencil: its unit-speed components.

\begin{definition}[Unit-speed stencils]
\label{def:bp-sup-unit-speed}
Let $S=\left(\mathcal{V}, \boldsymbol{\Delta}\right)$ be a stencil.
$S$ is a unit-speed stencil if and only if
$\forall v \in \mathcal{V}, \boldsymbol{\Delta}(v)_j \in \{-1, 0, 1\}$.
\end{definition}

\begin{figure}[h]
    \centering
    \begin{subfigure}[b]{\textwidth}
        \centering
        \input{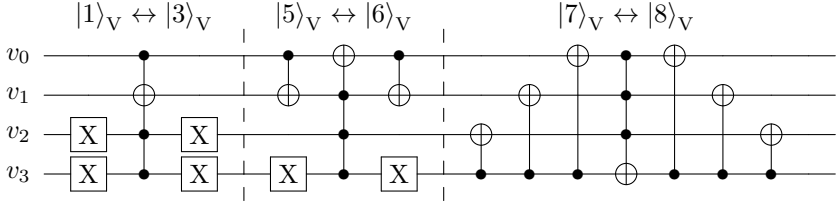}
        \caption{Quantum circuit implementation of $\mathrm{U_{P_x}}$.}
        \label{fig:bp-sup-sr-refl-x}
    \end{subfigure}

    \vspace{1em}

    \begin{subfigure}[b]{\textwidth}
        \centering
        \input{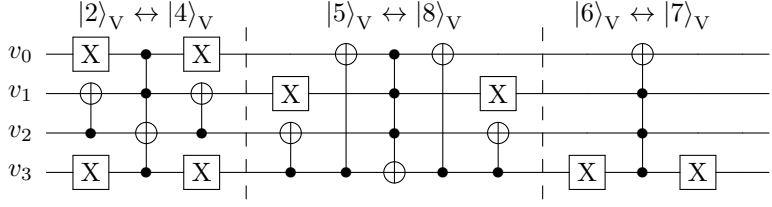}
        \caption{Quantum circuit implementation of $\mathrm{U_{P_y}}$.}
        \label{fig:bp-sup-sr-refl-y}
    \end{subfigure}

    \vspace{1em}

    \begin{subfigure}[b]{\textwidth}
        \centering
        \input{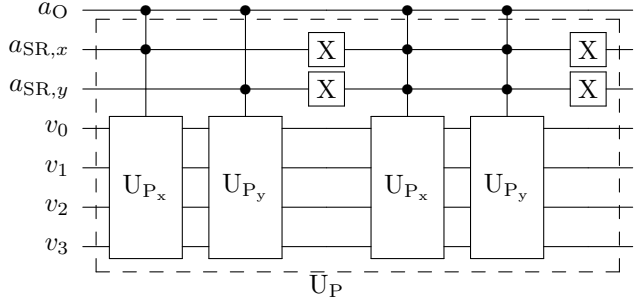}
        \caption{Quantum circuit implementation of $\mathrm{U_{P}}$, controlled on $a_\mathrm{O}$.}
        \label{fig:bp-sup-sr-refl-xy}
    \end{subfigure}
    \caption{Quantum circuits implementing SR velocity rotation.
    \label{fig:bp-sup-sr-refl}}
\end{figure}

Stencils that abide by the unit speed property include \dq{1}{3},
\dq{2}{9}, \dq{3}{15}, \dq{3}{19}, and \dq{3}{27},
which are expressive enough for many practical applications.
We note that we limit our proof to \dq{2}{9} because
it encompasses all $1d$ and $2d$ unit-speed stencils,
and our analysis thus holds for all simpler unit-speed discretizations
by simple reduction.
We use $\Omega$ to denote the solid domain, with $\mathrm{U}_\Omega$ the oracle,
and $\mathrm{U_S}$ and $\mathrm{U_P}$ the
streaming and permutation operators, respectively.
Unlike the bounce-back reflection operator, we distinguish between
three distinct specular cases.
These are:
reflection about the $x$-axis, $y$-axis,
and reflection about both axes simultaneously.
We define a permutation operator that transposes basis states
to perform these physical operations in
\Cref{fig:bp-sup-sr-refl-x}, \Cref{fig:bp-sup-sr-refl-y}, and \Cref{fig:bp-sup-sr-refl-xy},
and refer to them as $\mathrm{U_{P_x}}, \mathrm{U_{P_y}}$,
and $\mathrm{U_{P}}$, respectively.
We depict $\mathrm{U_P}$ with an additional control on the object
ancilla $a_\mathrm{O}$, which is how \Cref{alg:bp-2-sr} applies it.

We design our operator such that $\mathrm{U_P}$ applies
only $\mathrm{U_{P_x}}$ if the state of the $\mathrm{SR}$
register is $\ket{10}$, only $\mathrm{U_{P_y}}$
if it is $\ket{01}$, and $\mathrm{U_{P_x}}\mathrm{U_{P_y}}$ otherwise.
The action of $\mathrm{U_{P_x}}$ and $\mathrm{U_{P_y}}$, as realized by the
circuits in \Cref{fig:bp-sup-sr-refl-x,fig:bp-sup-sr-refl-y}, on the velocity
basis states is given by

\begin{equation}
\mathrm{U_{P_x}}\ket{k}_\mathrm{V} =
\begin{cases}
\ket{3}_\mathrm{V} & \text{if } k = 1, \\
\ket{1}_\mathrm{V} & \text{if } k = 3, \\
\ket{6}_\mathrm{V} & \text{if } k = 5, \\
\ket{5}_\mathrm{V} & \text{if } k = 6, \\
\ket{8}_\mathrm{V} & \text{if } k = 7, \\
\ket{7}_\mathrm{V} & \text{if } k = 8, \\
\ket{k}_\mathrm{V} & \text{otherwise},
\end{cases}
\qquad
\mathrm{U_{P_y}}\ket{k}_\mathrm{V} =
\begin{cases}
\ket{4}_\mathrm{V} & \text{if } k = 2, \\
\ket{2}_\mathrm{V} & \text{if } k = 4, \\
\ket{8}_\mathrm{V} & \text{if } k = 5, \\
\ket{7}_\mathrm{V} & \text{if } k = 6, \\
\ket{6}_\mathrm{V} & \text{if } k = 7, \\
\ket{5}_\mathrm{V} & \text{if } k = 8, \\
\ket{k}_\mathrm{V} & \text{otherwise}.
\end{cases}
\label{eq:bp-sup-upx-upy}
\end{equation}

Composing the two operators we get the physical reflection about both axes,
which is also equivalent to the bounce-back reflection operator,

\begin{equation}
\mathrm{U_{P_x}}\mathrm{U_{P_y}}\ket{k}_\mathrm{V} = \mathrm{U_{P_y}}\mathrm{U_{P_x}}\ket{k}_\mathrm{V} =
\begin{cases}
\ket{3}_\mathrm{V} & \text{if } k = 1, \\ 
\ket{4}_\mathrm{V} & \text{if } k = 2, \\
\ket{1}_\mathrm{V} & \text{if } k = 3, \\
\ket{2}_\mathrm{V} & \text{if } k = 4, \\
\ket{7}_\mathrm{V} & \text{if } k = 5, \\
\ket{8}_\mathrm{V} & \text{if } k = 6, \\
\ket{5}_\mathrm{V} & \text{if } k = 7, \\
\ket{6}_\mathrm{V} & \text{if } k = 8, \\
\ket{k}_\mathrm{V} & \text{otherwise}.
\end{cases}
\label{eq:bp-sup-up}
\end{equation}

We use the notation $\mathrm{U_{CF}}(\sigma)$ with $\sigma \in \{ \pm 1 \}$
to denote the realization of \Cref{alg:bp-2-flag-crossing} depicted in \Cref{fig:bp-2-za-sr-check}.
We denote the vectors $(1, 0)$ and $(0, 1)$ as $\boldsymbol{1}_0$ and $\boldsymbol{1}_1$,
and use the notation $\boldsymbol{\Delta}(v)_x$ and $\boldsymbol{\Delta}(v)_y$
to denote the first and second entries of the displacement vector of $v$.
Furthermore, we note LBM stencils may
be decomposed into their scalar components as follows.

\begin{lemma}[Decomposition of unit-speed stencils]
\label{lemma:bp-sup-comp-decom}
Let $S=\left(\mathcal{V}, \boldsymbol{\Delta}\right)$ be a unit-speed stencil.
For all $v \in \mathcal{V}$, the displacement
$\boldsymbol{\Delta}(v)$ decomposes as
$\sum_{j\in{0, \ldots, d-1}}\boldsymbol{\Delta}(v)_j \odot \boldsymbol{1}_j$, 
with $\odot$ the vector component-wise product.
The corresponding streaming operator $\mathrm{U_S}$
decomposes into $\prod_j \mathrm{U_{S}}_j$,
with $\mathrm{U_{S}}_j = \mathrm{U_{S}}\odot \boldsymbol{1}_j$ the operator performing
the incrementation/decrementation only in dimension $j$.
\end{lemma}

\Cref{lemma:bp-sup-comp-decom} states that the streaming
operator associated with any unit-speed stencil
can be decomposed into its $d$ scalar components,
which can all be streamed independently.
This is equivalent to traversing the graph that the lattice implicitly
describes by its gridpoints and velocity channels only along the
links in von Neumann neighborhoods.
In other words, any gridpoint in the neighborhood given by
the stencil is reachable through a traversal of cardinal links only.
We note that this is always the case, no matter if the cardinal links are part
of the velocity discretization.
The reason we limit our analysis to unit-speed stencils
is that we can break down traversals into at most one step in
each dimension.
This is important, because it enables us to avoid
cases where particles may \emph{overshoot} the solid boundary.
Consider, for instance, three gridpoints

\begin{equation}
  \boldsymbol{x}_0 \in \mathbb{N}^2,\quad \boldsymbol{x}_1=\boldsymbol{x}_0 + (1, 0),\quad \boldsymbol{x}_2 = \boldsymbol{x}_1 + (1, 0).
\end{equation}

Assume a particle travels with velocity $v$ with $\boldsymbol{\Delta}(v) = (2,0)$,
hence breaking the unit-speed property.
Further assume the population starts in $\boldsymbol{x}_0$, and that
$\boldsymbol{x}_0, \boldsymbol{x}_2 \notin \Omega$, but $\boldsymbol{x}_1 \in \Omega$.
In streaming two gridpoints away from its origin, the particle travels
through $\Omega$, and our ZA algorithm has no mechanism to capture this,
since performing an inverse stream, too, misses $\Omega$.
One possible generalization is to break down the streaming step
of high-speed particles into unit-speed increments, which yields
a CFL condition that strays away from typical LBM implementations,
and would require a fundamental rework of our algorithms.
We thus leave generalization to more complex stencils and $3$-dimensional
systems to future work, and in this section only consider \dq{2}{9}
and all its constituent velocities.
% For such cases, the number of feasible explanations may be pruned to increase efficiency,
% however, the general algorithm remains applicable.

In what follows, we trace the execution of \Cref{alg:bp-2-sr}
for two broad categories: particles that cross the boundary
and those that do not.
We consider the action of the algorithm on a single basis state

\begin{equation}
  \ket{\psi_0} = \ket{\boldsymbol{x}_0}\ket{v}\ket{0}_\mathrm{O}\ket{00}_\mathrm{SR},
\end{equation}

\noindent with $\boldsymbol{x}_0 \notin \Omega$.
Then, the outcome of the streaming step is

\begin{equation}
  \ket{\psi_1} = \mathrm{U_S}\ket{\psi_0} = \ket{\boldsymbol{x}_1}\ket{v}\ket{0}_\mathrm{O}\ket{00}_\mathrm{SR},
\end{equation}

\noindent with $\boldsymbol{x}_1 = \boldsymbol{x}_0 + \boldsymbol{\Delta}(v)$.
For the non-crossing case $\boldsymbol{x}_1 \notin \Omega$,
while for the crossing case, $\boldsymbol{x}_1 \in \Omega$.
We prove the correctness for these two cases independently.

\subsection{Non-crossing case}

\Cref{tab:bp-sup-correctness-sr-non-crossing} traces
\Cref{alg:bp-2-sr} throughout its nine steps for particles that
do not cross boundaries.
The correctness follows from the fact that since neither
$\boldsymbol{x}_0$, nor $\boldsymbol{x}_1$
belong to $\Omega$, none of the operations that are controlled
on the ancilla qubits alter the state.
This means that the algorithm effectively only applies
Steps \ref{alg:bp-2-sr:step1}, \ref{alg:bp-2-sr:step6}, and \ref{alg:bp-2-sr:step9}
which reduce to

\begin{equation}
  \mathrm{U_S}\mathrm{U_S}^\dagger\mathrm{U_S}\ket{\psi_0} = \mathrm{U_S}\ket{\psi_0} = \ket{\psi_1}.
\end{equation}

In other words, the algorithm is equivalent to streaming,
and we note that no further assumptions on the properties
of $\Omega$ or the stencil are necessary.
This is exactly the behavior that is expected when no boundary conditions are applied.
We note that this also includes that stand-still velocity
of any stencil, as streaming with increment $(0,0)$
implies that $\mathrm{U_S}\ket{\boldsymbol{x}_0}=\ket{\boldsymbol{x}_0}$,
and since $\boldsymbol{x}_0 \notin \Omega$ by definition,
our analysis holds.
Next, we address the crossing by considering only cardinally facing velocities.

\begin{table}[H]
\centering
\begin{tabular}{@{}c|c|c|c@{}}
\hline
Step & Operator & State & Remark \\
\hline
Init. & -- & $\ket{\boldsymbol{x}_0}\ket{v}\ket{0}_\mathrm{O}\ket{00}_\mathrm{SR}$ &  -- \\
Step \ref{alg:bp-2-sr:step1} & $\mathrm{U_S}$ & $\ket{\boldsymbol{x}_1}\ket{v}\ket{0}_\mathrm{O}\ket{00}_\mathrm{SR}$ & $\boldsymbol{x}_1 = \boldsymbol{x}_0 + \boldsymbol{\Delta}(v)$ \\
Step \ref{alg:bp-2-sr:step2} & $\mathrm{U_\Omega}$ & $\ket{\boldsymbol{x}_1}\ket{v}\ket{0}_\mathrm{O}\ket{00}_\mathrm{SR}$ & $\boldsymbol{x}_1 \notin \Omega$ \\
Step \ref{alg:bp-2-sr:step3} & $\mathrm{C^1U_{CF}(-1)}$ & $\ket{\boldsymbol{x}_1}\ket{v}\ket{0}_\mathrm{O}\ket{00}_\mathrm{SR}$ & $a_o = 0$ \\
Step \ref{alg:bp-2-sr:step4} & $\mathrm{C}^1\mathrm{U_P}$ & $\ket{\boldsymbol{x}_1}\ket{v}\ket{0}_\mathrm{O}\ket{00}_\mathrm{SR}$ & $a_o = 0$ \\
Step \ref{alg:bp-2-sr:step5} & $\mathrm{C}^1\mathrm{U_{S_x}}\mathrm{C}^1\mathrm{U_{S_y}}$ & $\ket{\boldsymbol{x}_1}\ket{v}\ket{0}_\mathrm{O}\ket{00}_\mathrm{SR}$ & $a_x = a_y = 0$ \\
Step \ref{alg:bp-2-sr:step6} & $\mathrm{U_S}^\dagger$ & $\ket{\boldsymbol{x}_0}\ket{v}\ket{0}_\mathrm{O}\ket{00}_\mathrm{SR}$ & $\boldsymbol{x}_0 = \boldsymbol{x}_1 - \boldsymbol{\Delta}(v)$ \\
Step \ref{alg:bp-2-sr:step7} & $\mathrm{C^1U_{CF}(+1)}$ & $\ket{\boldsymbol{x}_0}\ket{v}\ket{0}_\mathrm{O}\ket{00}_\mathrm{SR}$ & $a_o = 0$ \\
Step \ref{alg:bp-2-sr:step8} & $\mathrm{U_\Omega}$ & $\ket{\boldsymbol{x}_0}\ket{v}\ket{0}_\mathrm{O}\ket{00}_\mathrm{SR}$ & $\boldsymbol{x}_0 \notin \Omega$ \\
Step \ref{alg:bp-2-sr:step9} & $\mathrm{U_S}$ & $\ket{\boldsymbol{x}_1}\ket{v}\ket{0}_\mathrm{O}\ket{00}_\mathrm{SR}$ & $\boldsymbol{x}_1 = \boldsymbol{x}_0 + \boldsymbol{\Delta}(v)$ \\

\hline
\end{tabular}
\caption{Step-wise correctness check of \Cref{alg:bp-2-sr} for the non-crossing case. \label{tab:bp-sup-correctness-sr-non-crossing}}
\end{table}

\subsection{Crossing case: cardinal velocities}

Our analysis of the cardinal velocity case rests on the fact
that cardinal velocities, indexed 1 through 4, only have
one non-zero component.
In turn, this means that the application of \Cref{alg:bp-2-flag-crossing}
once crossing has occurred
only takes the population back to its pre-streaming location.
Indeed, this is exactly the same property that
allowed us to prove the correctness of the bounce-back.
Physically, bounce-back and specular reflection are identical
for cardinal velocity;
the purpose of this proof is to show that this is also the case for
our quantum implementation.

\Cref{tab:bp-sup-correctness-sr-crossing-cardinal} shows the
step-by-step analysis for the velocity indexed by $1$, with displacement $(1, 0)$.
The argument is conceptually identical to the BB case,
with the additional caveat that the applications of \Cref{alg:bp-2-flag-crossing}
appropriately reset the two trailing qubits.
Indeed, since streaming for cardinal cases is one-dimensional,
the implementation outlined in \Cref{fig:bp-2-za-sr-check} is exactly equivalent
to performing the inverse streaming we use for bounce-back.
The exact traces of both the forward and the backward passes for 
\Cref{fig:bp-2-za-sr-check} are listed in
\Cref{tab:bp-sup-sr-check-proof-forward-cardinal-1}
and
\Cref{tab:bp-sup-sr-check-proof-backward-cardinal-1},
respectively.
The remainder of the properties are straightforward and follow
directly from our assumptions and discretizations.

\begin{table}[H]
\centering
\begin{tabular}{@{}c|c|c|c@{}}
\hline
Stage & Operator & State & Remark \\
\hline
Init. &  -- & $\ket{\boldsymbol{x}_1}\ket{1}\ket{1}_\mathrm{O}\ket{00}_\mathrm{SR}$ & --\\
$x$-check & $\mathrm{C}^{n_q + 1}\mathrm{U_{S_x}}(\boldsymbol{\Delta}(1), -1)$ & $\ket{\boldsymbol{x}_0}\ket{1}\ket{1}_\mathrm{O}\ket{00}_\mathrm{SR}$ & $\boldsymbol{x}_0 = \boldsymbol{x}_1 + \overbrace{(-1)\cdot\left(\boldsymbol{\Delta}(1)\odot \boldsymbol{1}_0\right)}^{=(-1,0)}$ \\
$x$-check & $\mathrm{C}^{1}\mathrm{U}_{\bar{\Omega}}$ & $\ket{\boldsymbol{x}_0}\ket{1}\ket{1}_\mathrm{O}\ket{10}_\mathrm{SR}$ & $\boldsymbol{x}_0 \notin \Omega$ \\
$x$-check & $\mathrm{C}^{n_q + 1}\mathrm{U_{S_x}}(\boldsymbol{\Delta}(1), +1)$ & $\ket{\boldsymbol{x}_1}\ket{1}\ket{1}_\mathrm{O}\ket{10}_\mathrm{SR}$ & $\boldsymbol{x}_1 = \boldsymbol{x}_0 + \overbrace{1\cdot\left(\boldsymbol{\Delta}(1)\odot \boldsymbol{1}_0\right)}^{=(1,0)}$ \\
$y$-check & $\mathrm{C}^{n_q + 1}\mathrm{U_{S_y}}(\boldsymbol{\Delta}(1), -1)$ & $\ket{\boldsymbol{x}_1}\ket{1}\ket{1}_\mathrm{O}\ket{10}_\mathrm{SR}$ & $\boldsymbol{\Delta}(1)_{y}=0 \implies$ no $y$-stream \\
$y$-check & $\mathrm{C}^{1}\mathrm{U}_{\bar{\Omega}}$ & $\ket{\boldsymbol{x}_1}\ket{1}\ket{1}_\mathrm{O}\ket{10}_\mathrm{SR}$ & $\boldsymbol{x}_1 \in \Omega$ \\
$y$-check & $\mathrm{C}^{n_q + 1}\mathrm{U_{S_y}}(\boldsymbol{\Delta}(1), +1)$ & $\ket{\boldsymbol{x}_1}\ket{1}\ket{1}_\mathrm{O}\ket{10}_\mathrm{SR}$ & $\boldsymbol{\Delta}(1)_{y}=0 \implies$ no $y$-stream \\
\hline
\end{tabular}
\caption{Step-wise correctness check of \Cref{alg:bp-2-flag-crossing} implemented by the circuit
  shown in \Cref{fig:bp-2-za-sr-check} for the cardinal velocity $v=1$ with $\boldsymbol{\Delta}(1)=(1, 0)$ and $\sigma=-1$.
  The initial state is that after Steps \ref{alg:bp-2-sr:step1} and \ref{alg:bp-2-sr:step2} of
  \Cref{alg:bp-2-sr}. \label{tab:bp-sup-sr-check-proof-forward-cardinal-1}}
\end{table}

\begin{table}[H]
\centering
\begin{tabular}{@{}c|c|c|c@{}}
\hline
Stage & Operator & State & Remark \\
\hline
Init. &  -- & $\ket{\boldsymbol{x}_1}\ket{3}\ket{1}_\mathrm{O}\ket{10}_\mathrm{SR}$ & --\\
$x$-check & $\mathrm{C}^{n_q + 1}\mathrm{U_{S_x}}(\boldsymbol{\Delta}(1), +1)$ & $\ket{\boldsymbol{x}_0}\ket{3}\ket{1}_\mathrm{O}\ket{10}_\mathrm{SR}$ & $\boldsymbol{x}_0 = \boldsymbol{x}_1 + \overbrace{1\cdot\left(\boldsymbol{\Delta}(3)\odot \boldsymbol{1}_0\right)}^{=(-1,0)}$ \\
$x$-check & $\mathrm{C}^{1}\mathrm{U}_{\bar{\Omega}}$ & $\ket{\boldsymbol{x}_0}\ket{3}\ket{1}_\mathrm{O}\ket{00}_\mathrm{SR}$ & $\boldsymbol{x}_0 \notin \Omega$ \\
$x$-check & $\mathrm{C}^{n_q + 1}\mathrm{U_{S_x}}(\boldsymbol{\Delta}(1), -1)$ & $\ket{\boldsymbol{x}_1}\ket{3}\ket{1}_\mathrm{O}\ket{00}_\mathrm{SR}$ & $\boldsymbol{x}_1 = \boldsymbol{x}_0 + \overbrace{(-1)\cdot\left(\boldsymbol{\Delta}(3)\odot \boldsymbol{1}_0\right)}^{=(1,0)}$ \\
$y$-check & $\mathrm{C}^{n_q + 1}\mathrm{U_{S_y}}(\boldsymbol{\Delta}(1), +1)$ & $\ket{\boldsymbol{x}_1}\ket{3}\ket{1}_\mathrm{O}\ket{00}_\mathrm{SR}$ & $\boldsymbol{\Delta}(1)_{y}=0 \implies$ no $y$-stream \\
$y$-check & $\mathrm{C}^{1}\mathrm{U}_{\bar{\Omega}}$ & $\ket{\boldsymbol{x}_1}\ket{3}\ket{1}_\mathrm{O}\ket{00}_\mathrm{SR}$ & $\boldsymbol{x}_1 \in \Omega$ \\
$y$-check & $\mathrm{C}^{n_q + 1}\mathrm{U_{S_y}}(\boldsymbol{\Delta}(1), -1)$ & $\ket{\boldsymbol{x}_1}\ket{3}\ket{1}_\mathrm{O}\ket{00}_\mathrm{SR}$ & $\boldsymbol{\Delta}(1)_{y}=0 \implies$ no $y$-stream \\
\hline
\end{tabular}
\caption{Step-wise correctness check of \Cref{alg:bp-2-flag-crossing} implemented by the circuit
  shown in \Cref{fig:bp-2-za-sr-check} for the cardinal velocity $v=1$ with $\boldsymbol{\Delta}(1)=(1, 0)$ and $\sigma=+1$.
  This step takes place after the velocity has been reflected into $v=3$.
  The initial state is that after Steps \ref{alg:bp-2-sr:step1} through \ref{alg:bp-2-sr:step6} of
  \Cref{alg:bp-2-sr}. \label{tab:bp-sup-sr-check-proof-backward-cardinal-1}}
\end{table}

\begin{table}[H]
\centering
\begin{tabular}{@{}c|c|c|c@{}}
\hline
Step & Operator & State & Remark \\
\hline
Init. & -- & $\ket{\boldsymbol{x}_0}\ket{1}\ket{0}_\mathrm{O}\ket{00}_\mathrm{SR}$ &  -- \\
Step \ref{alg:bp-2-sr:step1} & $\mathrm{U_S}$ & $\ket{\boldsymbol{x}_1}\ket{1}\ket{0}_\mathrm{O}\ket{00}_\mathrm{SR}$ & $\boldsymbol{x}_1 = \boldsymbol{x}_0 + \boldsymbol{\Delta}(1)$ \\
Step \ref{alg:bp-2-sr:step2} & $\mathrm{U_\Omega}$ & $\ket{\boldsymbol{x}_1}\ket{1}\ket{1}_\mathrm{O}\ket{00}_\mathrm{SR}$ & $\boldsymbol{x}_1 \in \Omega$ \\
Step \ref{alg:bp-2-sr:step3} & $\mathrm{C^1U_{CF}(-1)}$ & $\ket{\boldsymbol{x}_1}\ket{1}\ket{1}_\mathrm{O}\ket{10}_\mathrm{SR}$ & See \Cref{tab:bp-sup-sr-check-proof-forward-cardinal-1} \\
Step \ref{alg:bp-2-sr:step4} & $\mathrm{C}^1\mathrm{U_P}$ & $\ket{\boldsymbol{x}_1}\ket{3}\ket{1}_\mathrm{O}\ket{10}_\mathrm{SR}$ & $\ket{1}_\mathrm{V} \leftrightarrow \ket{3}_\mathrm{V}$ \\
Step \ref{alg:bp-2-sr:step5} & $\mathrm{C}^1\mathrm{U_{S_x}}\mathrm{C}^1\mathrm{U_{S_y}}$ & $\ket{\boldsymbol{x}_0}\ket{3}\ket{1}_\mathrm{O}\ket{10}_\mathrm{SR}$ & $\boldsymbol{x}_0 = \boldsymbol{x}_1 + \boldsymbol{\Delta}(3)$ \\
Step \ref{alg:bp-2-sr:step6} & $\mathrm{U_S}^\dagger$ & $\ket{\boldsymbol{x}_1}\ket{3}\ket{1}_\mathrm{O}\ket{10}_\mathrm{SR}$ & $\boldsymbol{x}_0 = \boldsymbol{x}_0 + (1, 0) = \boldsymbol{x}_1$ \\
Step \ref{alg:bp-2-sr:step7} & $\mathrm{C^1U_{CF}(+1)}$ & $\ket{\boldsymbol{x}_1}\ket{3}\ket{1}_\mathrm{O}\ket{00}_\mathrm{SR}$ & See \Cref{tab:bp-sup-sr-check-proof-backward-cardinal-1} \\
Step \ref{alg:bp-2-sr:step8} & $\mathrm{U_\Omega}$ & $\ket{\boldsymbol{x}_1}\ket{3}\ket{0}_\mathrm{O}\ket{00}_\mathrm{SR}$ & $\boldsymbol{x}_1 \in \Omega$ \\
Step \ref{alg:bp-2-sr:step9} & $\mathrm{U_S}$ & $\ket{\boldsymbol{x}_0}\ket{3}\ket{0}_\mathrm{O}\ket{00}_\mathrm{SR}$ & $\boldsymbol{x}_0 = \boldsymbol{x}_1 + \boldsymbol{\Delta}(3)$ \\

\hline
\end{tabular}
\caption{Step-wise correctness check of \Cref{alg:bp-2-flag-crossing} for the cardinal
case for velocity index $1$.\label{tab:bp-sup-correctness-sr-crossing-cardinal}}
\end{table}

We only trace the algorithm for this cardinal direction,
as all others are equivalent, except for two details:
the state of the ancilla register and the permutation.
The state of the ancilla register following the application
of \Cref{alg:bp-2-flag-crossing} is $\ket{10}$ for $v\in \{1,3\}$
and $\ket{01}$ for $v\in \{2, 4\}$.
We omit the proofs for the other cardinal velocities due to their redundancy;
the structure of the trace is identical
and the arguments are only adjusted by symmetry.
For the horizontal case, the operator acts as

\begin{equation}
  \ket{\boldsymbol{x}_0}\ket{v \in \{1, 3\}}\ket{0}\ket{00} \mapsto \ket{\boldsymbol{x}_0}\overbrace{\left(\mathrm{U_{P_x}}\ket{v}\right)}^{=\ket{\bar{v}}\text{, by \ref{eq:bp-sup-upx-upy}}}\ket{0}\ket{00},
\end{equation}

\noindent while for the vertical velocity directions

\begin{equation}
  \ket{\boldsymbol{x}_0}\ket{v \in \{2, 4\}}\ket{0}\ket{00} \mapsto \ket{\boldsymbol{x}_0}\overbrace{\left(\mathrm{U_{P_y}}\ket{v}\right)}^{=\ket{\bar{v}}\text{, by \ref{eq:bp-sup-upx-upy}}}\ket{0}\ket{00}.
\end{equation}

\subsection{Crossing case: diagonal velocities}

To analyze the correctness in the diagonal case,
we trace the execution of both \Cref{alg:bp-2-flag-crossing}
and \Cref{alg:bp-2-sr} for all four possible boundary crossing explanations.
As before, we only consider, without loss of generality, the velocity
indexed $5$, as the three other cases follow from symmetry.
To understand the correctness of the method, we consider three
points in addition to $\boldsymbol{x}_0$,

\begin{subequations}
\label{eq:bp-sup-neighborhood}
\begin{align}
  \boldsymbol{x}_1 &= \boldsymbol{x}_0 + \boldsymbol{\Delta}(5) = \boldsymbol{x}_0 + (1, 1), \\
  \boldsymbol{x}_2 &= \boldsymbol{x}_0 + \boldsymbol{\Delta}(5)_x = \boldsymbol{x}_0 + (1, 0), \\
  \boldsymbol{x}_3 &= \boldsymbol{x}_0 + \boldsymbol{\Delta}(5)_y = \boldsymbol{x}_0 + (0, 1).
\end{align}
\end{subequations}

These four gridpoints are the only locations
where the population travels during the routine.
Simultaneously, whether $\boldsymbol{x}_2$ and $\boldsymbol{x}_3$
belong to $\Omega$ also determines which velocity vector
components should be reflected, and, equivalently, which
explanation is responsible for the crossing.
We recall the explanations from \Cref{tab:bp-2-example-sr-check}.

\begin{figure}[h]
    \centering
    \begin{subfigure}[b]{0.45\textwidth}
        \centering
        \centering
\begin{tikzpicture}[scale=0.4]
    % 3x3 lattice with gridpoints at (-4, 0, 4) on each axis (step 4).
    \draw[help lines, line width=0.4pt, color=gray!30, dashed]
        (-6, -6) grid[step={(4,4)}] (6, 6);

    % Background D2Q9 stencils (white) at every lattice site.
    \foreach \x in {-4, 0, 4} {
        \foreach \y in {-4, 0, 4} {
            \dtwoqeight{(\x,\y)}{0.1cm}{4}{0}{1}{white}
        }
    }

    % Obstacle geometry: rectangle [1,1]x[2,2] in lattice coords
    % (covers gridpoints (1,1), (2,1), (1,2), (2,2) in lattice -> tikz (0,0),(4,0),(0,4),(4,4)).
    \fill[gray!18] (-2,-2) rectangle (6,6);
    \foreach \x/\y in {0/0, 4/0, 0/4, 4/4} {
        \dtwoqeight{(\x,\y)}{0.1cm}{4}{0}{1}{gray}
    }

    % Boundary outline.
    \draw[black, ultra thick, rounded corners=2pt] (-2,-2) rectangle (6,6);

    % Real particle at lattice (0,0) = tikz (-4,-4), velocity index 5 (NE) emphasized.
    \dtwoqeightcolor{(-4,-4)}{0.1cm}{4}{0}{1}{red!60}{black}{black}{black}{black}{red}{black}{black}{black}

    % Ghost particle showing post-stream location at lattice (1,1) = tikz (0,0).
    \dtwoqeightcolor{(0,0)}{0.1cm}{4}{0}{1}{red!40}{black}{black}{black}{black}{red!40}{black}{black}{black}

    % Gridpoint labels centered on their lattice sites:
    % 0 at lattice (0,0), 1 at (1,1), 2 at (1,0), 3 at (0,1).
    \node[font=\tiny, anchor=center] at (-4,-4) {$\boldsymbol{x}_0$};
    \node[font=\tiny, anchor=center] at ( 0, 0) {$\boldsymbol{x}_1$};
    \node[font=\tiny, anchor=center] at ( 0,-4) {$\boldsymbol{x}_2$};
    \node[font=\tiny, anchor=center] at (-4, 0) {$\boldsymbol{x}_3$};

    % Axis labels: lattice index 0--2 along each axis.
    \foreach \i/\lab in {0/0, 1/1, 2/2} {
        \node[anchor=north, font=\scriptsize] at ({-4+\i*4},-7) {$\lab$};
        \node[anchor=east,  font=\scriptsize] at (-7,{-4+\i*4}) {$\lab$};
    }
\end{tikzpicture}
        \caption{$\boldsymbol{x}_2, \boldsymbol{x}_3 \notin \Omega$.}
        \label{fig:bp-sup-sr-proof-1}
    \end{subfigure}
    \hfill
    \begin{subfigure}[b]{0.45\textwidth}
        \centering
        \centering
\begin{tikzpicture}[scale=0.4]
    % 3x3 lattice with gridpoints at (-4, 0, 4) on each axis (step 4).
    \draw[help lines, line width=0.4pt, color=gray!30, dashed]
        (-6, -6) grid[step={(4,4)}] (6, 6);

    % Background D2Q9 stencils (white) at every lattice site.
    \foreach \x in {-4, 0, 4} {
        \foreach \y in {-4, 0, 4} {
            \dtwoqeight{(\x,\y)}{0.1cm}{4}{0}{1}{white}
        }
    }

    % Obstacle geometry: rectangle [1,1]x[2,2] + lattice point (0,1) = tikz (-4,0).
    \fill[gray!18] (-2,-2) rectangle (6,6);
    \fill[gray!18] (-6,-2) rectangle (-2,2);
    \foreach \x/\y in {0/0, 4/0, 0/4, 4/4, -4/0} {
        \dtwoqeight{(\x,\y)}{0.1cm}{4}{0}{1}{gray}
    }

    % Boundary outline (L-shape).
    \draw[black, ultra thick, rounded corners=2pt]
        (-2,-2) -- (6,-2) -- (6,6) -- (-2,6) -- (-2,2)
        -- (-6,2) -- (-6,-2) -- cycle;

    % Real particle at lattice (0,0) = tikz (-4,-4), velocity index 5 (NE) emphasized.
    \dtwoqeightcolor{(-4,-4)}{0.1cm}{4}{0}{1}{red!60}{black}{black}{black}{black}{red}{black}{black}{black}

    % Ghost particle showing post-stream location at lattice (1,1) = tikz (0,0).
    \dtwoqeightcolor{(0,0)}{0.1cm}{4}{0}{1}{red!40}{black}{black}{black}{black}{red!40}{black}{black}{black}

    % Gridpoint labels centered on their lattice sites:
    % x_0 at lattice (0,0), x_1 at (1,1), x_2 at (1,0), x_3 at (0,1).
    \node[font=\tiny, anchor=center] at (-4,-4) {$\boldsymbol{x}_0$};
    \node[font=\tiny, anchor=center] at ( 0, 0) {$\boldsymbol{x}_1$};
    \node[font=\tiny, anchor=center] at ( 0,-4) {$\boldsymbol{x}_2$};
    \node[font=\tiny, anchor=center] at (-4, 0) {$\boldsymbol{x}_3$};

    % Axis labels: lattice index 0--2 along each axis.
    \foreach \i/\lab in {0/0, 1/1, 2/2} {
        \node[anchor=north, font=\scriptsize] at ({-4+\i*4},-7) {$\lab$};
        \node[anchor=east,  font=\scriptsize] at (-7,{-4+\i*4}) {$\lab$};
    }
\end{tikzpicture}
        \caption{$\boldsymbol{x}_2 \notin \Omega$, $\boldsymbol{x}_3 \in \Omega$.}
        \label{fig:bp-sup-sr-proof-2}
    \end{subfigure}

    \vspace{1em}

    \begin{subfigure}[b]{0.45\textwidth}
        \centering
        \centering
\begin{tikzpicture}[scale=0.4]
    % 3x3 lattice with gridpoints at (-4, 0, 4) on each axis (step 4).
    \draw[help lines, line width=0.4pt, color=gray!30, dashed]
        (-6, -6) grid[step={(4,4)}] (6, 6);

    % Background D2Q9 stencils (white) at every lattice site.
    \foreach \x in {-4, 0, 4} {
        \foreach \y in {-4, 0, 4} {
            \dtwoqeight{(\x,\y)}{0.1cm}{4}{0}{1}{white}
        }
    }

    % Obstacle geometry: rectangle [1,1]x[2,2] + lattice point (1,0) = tikz (0,-4).
    \fill[gray!18] (-2,-2) rectangle (6,6);
    \fill[gray!18] (-2,-6) rectangle (2,-2);
    \foreach \x/\y in {0/0, 4/0, 0/4, 4/4, 0/-4} {
        \dtwoqeight{(\x,\y)}{0.1cm}{4}{0}{1}{gray}
    }

    % Boundary outline (L-shape, mirrored).
    \draw[black, ultra thick, rounded corners=2pt]
        (-2,-2) -- (-2,-6) -- (2,-6) -- (2,-2)
        -- (6,-2) -- (6,6) -- (-2,6) -- cycle;

    % Real particle at lattice (0,0) = tikz (-4,-4), velocity index 5 (NE) emphasized.
    \dtwoqeightcolor{(-4,-4)}{0.1cm}{4}{0}{1}{red!60}{black}{black}{black}{black}{red}{black}{black}{black}

    % Ghost particle showing post-stream location at lattice (1,1) = tikz (0,0).
    \dtwoqeightcolor{(0,0)}{0.1cm}{4}{0}{1}{red!40}{black}{black}{black}{black}{red!40}{black}{black}{black}

    % Gridpoint labels centered on their lattice sites:
    % x_0 at lattice (0,0), x_1 at (1,1), x_2 at (1,0), x_3 at (0,1).
    \node[font=\tiny, anchor=center] at (-4,-4) {$\boldsymbol{x}_0$};
    \node[font=\tiny, anchor=center] at ( 0, 0) {$\boldsymbol{x}_1$};
    \node[font=\tiny, anchor=center] at ( 0,-4) {$\boldsymbol{x}_2$};
    \node[font=\tiny, anchor=center] at (-4, 0) {$\boldsymbol{x}_3$};

    % Axis labels: lattice index 0--2 along each axis.
    \foreach \i/\lab in {0/0, 1/1, 2/2} {
        \node[anchor=north, font=\scriptsize] at ({-4+\i*4},-7) {$\lab$};
        \node[anchor=east,  font=\scriptsize] at (-7,{-4+\i*4}) {$\lab$};
    }
\end{tikzpicture}
        \caption{$\boldsymbol{x}_2 \in \Omega$, $\boldsymbol{x}_3 \notin \Omega$.}
        \label{fig:bp-sup-sr-proof-3}
    \end{subfigure}
    \hfill
    \begin{subfigure}[b]{0.45\textwidth}
        \centering
        \centering
\begin{tikzpicture}[scale=0.4]
    % 3x3 lattice with gridpoints at (-4, 0, 4) on each axis (step 4).
    \draw[help lines, line width=0.4pt, color=gray!30, dashed]
        (-6, -6) grid[step={(4,4)}] (6, 6);

    % Background D2Q9 stencils (white) at every lattice site.
    \foreach \x in {-4, 0, 4} {
        \foreach \y in {-4, 0, 4} {
            \dtwoqeight{(\x,\y)}{0.1cm}{4}{0}{1}{white}
        }
    }

    % Obstacle geometry: rectangle [1,1]x[2,2] + lattice points (1,0) and (0,1).
    \fill[gray!18] (-2,-2) rectangle (6,6);
    \fill[gray!18] (-6,-2) rectangle (-2,2);
    \fill[gray!18] (-2,-6) rectangle (2,-2);
    \foreach \x/\y in {0/0, 4/0, 0/4, 4/4, -4/0, 0/-4} {
        \dtwoqeight{(\x,\y)}{0.1cm}{4}{0}{1}{gray}
    }

    % Boundary outline (concave staircase).
    \draw[black, ultra thick, rounded corners=2pt]
        (-2,-2) -- (-2,-6) -- (2,-6) -- (2,-2)
        -- (6,-2) -- (6,6) -- (-2,6) -- (-2,2)
        -- (-6,2) -- (-6,-2) -- cycle;

    % Real particle at lattice (0,0) = tikz (-4,-4), velocity index 5 (NE) emphasized.
    \dtwoqeightcolor{(-4,-4)}{0.1cm}{4}{0}{1}{red!60}{black}{black}{black}{black}{red}{black}{black}{black}

    % Ghost particle showing post-stream location at lattice (1,1) = tikz (0,0).
    \dtwoqeightcolor{(0,0)}{0.1cm}{4}{0}{1}{red!40}{black}{black}{black}{black}{red!40}{black}{black}{black}

    % Gridpoint labels centered on their lattice sites:
    % x_0 at lattice (0,0), x_1 at (1,1), x_2 at (1,0), x_3 at (0,1).
    \node[font=\tiny, anchor=center] at (-4,-4) {$\boldsymbol{x}_0$};
    \node[font=\tiny, anchor=center] at ( 0, 0) {$\boldsymbol{x}_1$};
    \node[font=\tiny, anchor=center] at ( 0,-4) {$\boldsymbol{x}_2$};
    \node[font=\tiny, anchor=center] at (-4, 0) {$\boldsymbol{x}_3$};

    % Axis labels: lattice index 0--2 along each axis.
    \foreach \i/\lab in {0/0, 1/1, 2/2} {
        \node[anchor=north, font=\scriptsize] at ({-4+\i*4},-7) {$\lab$};
        \node[anchor=east,  font=\scriptsize] at (-7,{-4+\i*4}) {$\lab$};
    }
\end{tikzpicture}
        \caption{$\boldsymbol{x}_2, \boldsymbol{x}_3 \in \Omega$.}
        \label{fig:bp-sup-sr-proof-4}
    \end{subfigure}
    \caption{The four possible specular-reflection configurations for the
    diagonal velocity $v=5$ with $\boldsymbol{\Delta}(5)=(1, 1)$.
    The population at gridpoint $\boldsymbol{x}_0$ (red) would stream to gridpoint $\boldsymbol{x}_1$
    (faded red) in the absence of a boundary.
    \label{fig:bp-sup-sr-proof-cases}}
\end{figure}
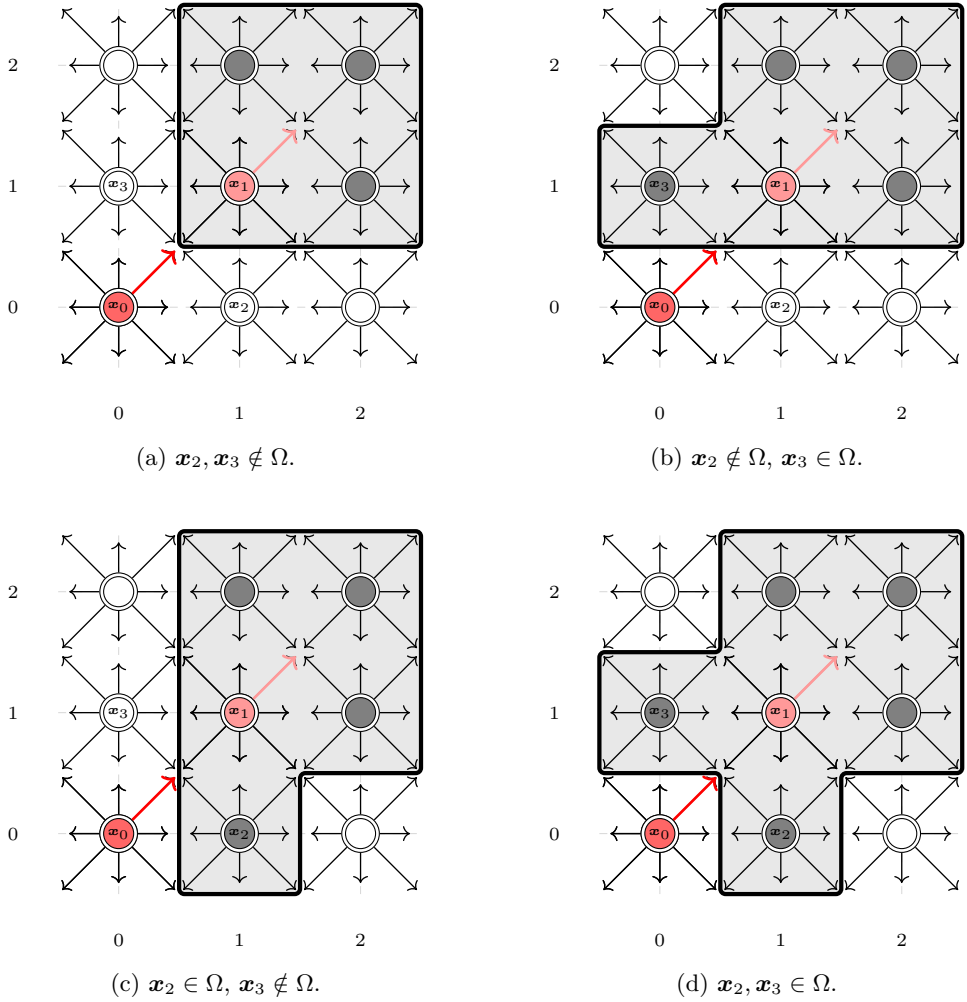

\Cref{fig:bp-sup-sr-proof-cases} enumerates the four possible reflection
configurations for the diagonal velocity $v=5$ with displacement
$\boldsymbol{\Delta}(5)=(1, 1)$.
The red population starts at gridpoint $\boldsymbol{x}_0$ and would otherwise stream into
gridpoint $\boldsymbol{x}_1$ (shown faded).
The four cases differ in whether the lateral neighbors $\boldsymbol{x}_2=(1,0)$ and
$\boldsymbol{x}_3=(0, 1)$ are part of the solid domain $\Omega$, which determines
how the specular reflection redirects the population in each of the four possible cases.
We note that our depiction on a $3\times 3$ grid is purely exemplary,
as the number and state of the gridpoints not in the neighborhood of interest
is of no consequence to our method.
In what follows, we trace both the component identification
and specular reflection algorithm for all four cases.

\subsubsection{Diagonal velocities: convex corner}

The convex corner case is visually depicted in \Cref{fig:bp-sup-sr-proof-1},
and characterized by $\boldsymbol{x}_2, \boldsymbol{x}_3 \notin \Omega$.
We recall that the quantum circuit given in 
\Cref{fig:bp-2-za-sr-check} queries the $x$- and $y$-neighbors
of the gridpoint where the population finds itself \emph{after} streaming.
Since the first application of \Cref{alg:bp-2-flag-crossing} is supplied with $\sigma=-1$,
the queried gridpoints are in in the direction opposite of $\boldsymbol{\Delta}(5)$.
Since the population has streamed from $\boldsymbol{x}_0$ to $\boldsymbol{x}_1$
in the first step of the ZA-SR algorithm, the queried gridpoints are precisely
$\boldsymbol{x}_2$ and $\boldsymbol{x}_3$.
This is, in fact, the case, for \emph{all} four diagonal cases.

We first consider the forward first pass ($\sigma = -1$) of \Cref{alg:bp-2-flag-crossing}
used in Step \Cref{alg:bp-2-sr:step3} of \Cref{alg:bp-2-sr}.
\Cref{tab:bp-sup-sr-check-proof-forward-1} traces the steps of the algorithm
and shows that the basis state $\ket{11}$ is imprinted onto the ancilla qubits.
This maps to the $\{\{x,y\}\}$ explanation, whose interpretation is
``the $x$ and $y$ components only cause the crossing
together, and neither do independently''.

\begin{table}[H]
\centering
\begin{tabular}{@{}c|c|c|c@{}}
\hline
Stage & Operator & State & Remark \\
\hline
Init. &  -- & $\ket{\boldsymbol{x}_1}\ket{5}\ket{1}_\mathrm{O}\ket{00}_\mathrm{SR}$ & --\\
$x$-check & $\mathrm{C}^{n_q + 1}\mathrm{U_{S_x}}(\boldsymbol{\Delta}(5), -1)$ & $\ket{\boldsymbol{x}_3}\ket{5}\ket{1}_\mathrm{O}\ket{00}_\mathrm{SR}$ & $\boldsymbol{x}_3 = \boldsymbol{x}_1 + \overbrace{(-1)\cdot\left(\boldsymbol{\Delta}(5)\odot \boldsymbol{1}_0\right)}^{=(-1,0)}$ \\
$x$-check & $\mathrm{C}^{1}\mathrm{U}_{\bar{\Omega}}$ & $\ket{\boldsymbol{x}_3}\ket{5}\ket{1}_\mathrm{O}\ket{10}_\mathrm{SR}$ & $\boldsymbol{x}_3 \notin \Omega$ \\
$x$-check & $\mathrm{C}^{n_q + 1}\mathrm{U_{S_x}}(\boldsymbol{\Delta}(5), +1)$ & $\ket{\boldsymbol{x}_1}\ket{5}\ket{1}_\mathrm{O}\ket{10}_\mathrm{SR}$ & $\boldsymbol{x}_1 = \boldsymbol{x}_3 + \overbrace{1\cdot\left(\boldsymbol{\Delta}(5)\odot \boldsymbol{1}_0\right)}^{=(1,0)}$ \\
$y$-check & $\mathrm{C}^{n_q + 1}\mathrm{U_{S_y}}(\boldsymbol{\Delta}(5), -1)$ & $\ket{\boldsymbol{x}_2}\ket{5}\ket{1}_\mathrm{O}\ket{10}_\mathrm{SR}$ & $\boldsymbol{x}_2 = \boldsymbol{x}_1 + \overbrace{(-1)\cdot\left(\boldsymbol{\Delta}(5)\odot \boldsymbol{1}_1\right)}^{=(0,-1)}$ \\
$y$-check & $\mathrm{C}^{1}\mathrm{U}_{\bar{\Omega}}$ & $\ket{\boldsymbol{x}_2}\ket{5}\ket{1}_\mathrm{O}\ket{11}_\mathrm{SR}$ & $\boldsymbol{x}_2 \notin \Omega$ \\
$y$-check & $\mathrm{C}^{n_q + 1}\mathrm{U_{S_y}}(\boldsymbol{\Delta}(5), +1)$ & $\ket{\boldsymbol{x}_1}\ket{5}\ket{1}_\mathrm{O}\ket{11}_\mathrm{SR}$ & $\boldsymbol{x}_1 = \boldsymbol{x}_2 + \overbrace{1\cdot\left(\boldsymbol{\Delta}(5)\odot \boldsymbol{1}_1\right)}^{=(0,1)}$ \\
\hline
\end{tabular}
\caption{Step-wise correctness check of \Cref{alg:bp-2-flag-crossing} implemented by the circuit
  shown in \Cref{fig:bp-2-za-sr-check} for the diagonal velocity $v=5$ with $\boldsymbol{\Delta}(5)=(1, 1)$ and $\sigma=-1$
  in the convex corner case.
  The initial state is that after Steps \ref{alg:bp-2-sr:step1} and \ref{alg:bp-2-sr:step2} of
  \Cref{alg:bp-2-sr}. \label{tab:bp-sup-sr-check-proof-forward-1}}
\end{table}

Symmetrically, we trace the backward pass ($\sigma = +1$)
in \Cref{tab:bp-sup-sr-check-proof-backward-1}.
The reasoning is identical, except it
is now applied to $v=7$ following the application of the permutation operator
shown in \Cref{fig:bp-sup-sr-refl}.
Finally, we use the states shown in the two traces to follow
the evolution of the end-to-end ZA-SR operator.
The state evolution of the entire operator is given
in \Cref{tab:bp-sup-sr-check-proof-full-1}.
Indeed, the trace shows that, conditioned
on the particle's trajectory intersecting a convex corner,
both velocity components are reflected by $\mathrm{U_P}$.
The remaining steps show that each basis state
is transposed as

\begin{equation}
  \ket{\boldsymbol{x}_0}\ket{v \in \{5,6,7,8\}}\ket{0}\ket{00} \mapsto \ket{\boldsymbol{x}_0}\overbrace{\left(\mathrm{U_{P_x}}\mathrm{U_{P_y}}\ket{v}\right)}^{=\ket{\bar{v}}\text{, by \ref{eq:bp-sup-up}}}\ket{0}\ket{00}.
\end{equation}

This concludes our proof of the convex corner case.
Next, we consider the concave counterpart.

\begin{table}[H]
\centering
\begin{tabular}{@{}c|c|c|c@{}}
\hline
Stage & Operator & State & Remark \\
\hline
Init. &  -- & $\ket{\boldsymbol{x}_1}\ket{7}\ket{1}_\mathrm{O}\ket{00}_\mathrm{SR}$ & --\\
$x$-check & $\mathrm{C}^{n_q + 1}\mathrm{U_{S_x}}(\boldsymbol{\Delta}(7), +1)$ & $\ket{\boldsymbol{x}_3}\ket{7}\ket{1}_\mathrm{O}\ket{11}_\mathrm{SR}$ & $\boldsymbol{x}_3 = \boldsymbol{x}_1 + \overbrace{1\cdot\left(\boldsymbol{\Delta}(7)\odot \boldsymbol{1}_0\right)}^\text{=(-1,0)}$ \\
$x$-check & $\mathrm{C}^{1}\mathrm{U}_{\bar{\Omega}}$ & $\ket{\boldsymbol{x}_3}\ket{7}\ket{1}_\mathrm{O}\ket{01}_\mathrm{SR}$ & $\boldsymbol{x}_3 \notin \Omega$ \\
$x$-check & $\mathrm{C}^{n_q + 1}\mathrm{U_{S_x}}(\boldsymbol{\Delta}(7), -1)$ & $\ket{\boldsymbol{x}_1}\ket{7}\ket{1}_\mathrm{O}\ket{01}_\mathrm{SR}$ & $\boldsymbol{x}_1 = \boldsymbol{x}_3 + \overbrace{(-1)\cdot\left(\boldsymbol{\Delta}(7)\odot \boldsymbol{1}_0\right)}^\text{=(1,0)}$ \\
$y$-check & $\mathrm{C}^{n_q + 1}\mathrm{U_{S_y}}(\boldsymbol{\Delta}(7), +1)$ & $\ket{\boldsymbol{x}_2}\ket{7}\ket{1}_\mathrm{O}\ket{01}_\mathrm{SR}$ & $\boldsymbol{x}_2 = \boldsymbol{x}_1 + \overbrace{1\cdot\left(\boldsymbol{\Delta}(7)\odot \boldsymbol{1}_1\right)}^\text{=(0,-1)}$ \\
$y$-check & $\mathrm{C}^{1}\mathrm{U}_{\bar{\Omega}}$ & $\ket{\boldsymbol{x}_2}\ket{7}\ket{1}_\mathrm{O}\ket{00}_\mathrm{SR}$ & $\boldsymbol{x}_2 \notin \Omega$ \\
$y$-check & $\mathrm{C}^{n_q + 1}\mathrm{U_{S_y}}(\boldsymbol{\Delta}(7), -1)$ & $\ket{\boldsymbol{x}_1}\ket{7}\ket{1}_\mathrm{O}\ket{00}_\mathrm{SR}$ & $\boldsymbol{x}_1 = \boldsymbol{x}_2 + \overbrace{(-1)\cdot\left(\boldsymbol{\Delta}(7)\odot \boldsymbol{1}_1\right)}^\text{=(0,1)}$ \\
\hline
\end{tabular}
\caption{Step-wise correctness check of \Cref{alg:bp-2-flag-crossing} implemented by the circuit
  shown in \Cref{fig:bp-2-za-sr-check} for the diagonal velocity $v=5$ with $\boldsymbol{\Delta}(5)=(1, 1)$ and $\sigma=+1$
  in the convex corner case.
  This step takes place after the velocity has been reflected into $v=7$.
  The initial state is that after Steps \ref{alg:bp-2-sr:step1} through \ref{alg:bp-2-sr:step6} of
  \Cref{alg:bp-2-sr}. \label{tab:bp-sup-sr-check-proof-backward-1}}
\end{table}

\begin{table}[H]
\centering
\begin{tabular}{@{}c|c|c|c@{}}
\hline
Step & Operator & State & Remark \\
\hline
Init. & -- & $\ket{\boldsymbol{x}_0}\ket{5}\ket{0}_\mathrm{O}\ket{00}_\mathrm{SR}$ &  -- \\
Step \ref{alg:bp-2-sr:step1} & $\mathrm{U_S}$ & $\ket{\boldsymbol{x}_1}\ket{5}\ket{0}_\mathrm{O}\ket{00}_\mathrm{SR}$ & $\boldsymbol{x}_1 = \boldsymbol{x}_0 + \boldsymbol{\Delta}(5)$ \\
Step \ref{alg:bp-2-sr:step2} & $\mathrm{U_\Omega}$ & $\ket{\boldsymbol{x}_1}\ket{5}\ket{1}_\mathrm{O}\ket{00}_\mathrm{SR}$ & $\boldsymbol{x}_1 \in \Omega$ \\
Step \ref{alg:bp-2-sr:step3} & $\mathrm{C^1U_{CF}(-1)}$ & $\ket{\boldsymbol{x}_1}\ket{5}\ket{1}_\mathrm{O}\ket{11}_\mathrm{SR}$ & See \Cref{tab:bp-sup-sr-check-proof-forward-1}\\
Step \ref{alg:bp-2-sr:step4} & $\mathrm{C}^1\mathrm{U_P}$ & $\ket{\boldsymbol{x}_1}\ket{7}\ket{1}_\mathrm{O}\ket{11}_\mathrm{SR}$ & $\ket{5}_\mathrm{V} \leftrightarrow \ket{7}_\mathrm{V}$ \\
Step \ref{alg:bp-2-sr:step5} & $\mathrm{C}^1\mathrm{U_{S_x}}\mathrm{C}^1\mathrm{U_{S_y}}$ & $\ket{\boldsymbol{x}_0}\ket{7}\ket{1}_\mathrm{O}\ket{11}_\mathrm{SR}$ & $a_x = a_y = 0$ and $\boldsymbol{x}_0 = \boldsymbol{x}_1 + \boldsymbol{\Delta}(7)$\\
Step \ref{alg:bp-2-sr:step6} & $\mathrm{U_S}^\dagger$ & $\ket{\boldsymbol{x}_1}\ket{7}\ket{1}_\mathrm{O}\ket{11}_\mathrm{SR}$ & $\boldsymbol{x}_1 = \boldsymbol{x}_0 - \boldsymbol{\Delta}(7)$ \\
Step \ref{alg:bp-2-sr:step7} & $\mathrm{C^1U_{CF}(+1)}$  & $\ket{\boldsymbol{x}_1}\ket{7}\ket{1}_\mathrm{O}\ket{00}_\mathrm{SR}$ & See \Cref{tab:bp-sup-sr-check-proof-backward-1}\\
Step \ref{alg:bp-2-sr:step8} & $\mathrm{U_\Omega}$ & $\ket{\boldsymbol{x}_1}\ket{7}\ket{0}_\mathrm{O}\ket{00}_\mathrm{SR}$ & $\boldsymbol{x}_1 \in \Omega$ \\
Step \ref{alg:bp-2-sr:step9} & $\mathrm{U_S}$ & $\ket{\boldsymbol{x}_0}\ket{7}\ket{0}_\mathrm{O}\ket{00}_\mathrm{SR}$ & $\boldsymbol{x}_0 = \boldsymbol{x}_1 + \boldsymbol{\Delta}(7)$ \\

\hline
\end{tabular}
\caption{Step-wise correctness check of \Cref{alg:bp-2-sr} for the convex corner
crossing case for velocity index $5$. \label{tab:bp-sup-sr-check-proof-full-1}}
\end{table}

\subsubsection{Diagonal velocities: concave corner}

\Cref{fig:bp-sup-sr-proof-4} shows the concave corner case.
This case is described by $\boldsymbol{x}_2, \boldsymbol{x}_3 \in \Omega$.
We again consider the forward and backward passes of \Cref{alg:bp-2-flag-crossing},
and the end-to-end application of \Cref{alg:bp-2-sr}
in \Cref{tab:bp-sup-sr-check-proof-forward-2},
\Cref{tab:bp-sup-sr-check-proof-backward-2},
and \Cref{tab:bp-sup-sr-check-proof-full-2}, respectively.
We note that the algorithm is nearly identical to the convex
corner we analyzed previously, except for the state of the two
ancilla qubits, which now remain in state $\ket{00}$.
This allows for the reflection by means of $\mathrm{U_P}$
to be physically identical, but distinguishable and reversible all the same.
The net effect of the operator is again

\begin{equation}
  \ket{\boldsymbol{x}_0}\ket{v \in \{5,6,7,8\}}\ket{0}\ket{00} \mapsto \ket{\boldsymbol{x}_0}\overbrace{\left(\mathrm{U_{P_x}}\mathrm{U_{P_y}}\ket{v}\right)}^{=\ket{\bar{v}}\text{, by \ref{eq:bp-sup-upx-upy}}}\ket{0}\ket{00}.
\end{equation}

\begin{table}[H]
\centering
\begin{tabular}{@{}c|c|c|c@{}}
\hline
Stage & Operator & State & Remark \\
\hline
Init. &  -- & $\ket{\boldsymbol{x}_1}\ket{5}\ket{1}_\mathrm{O}\ket{00}_\mathrm{SR}$ & --\\
$x$-check & $\mathrm{C}^{n_q + 1}\mathrm{U_{S_x}}(\boldsymbol{\Delta}(5), -1)$ & $\ket{\boldsymbol{x}_3}\ket{5}\ket{1}_\mathrm{O}\ket{00}_\mathrm{SR}$ & $\boldsymbol{x}_3 = \boldsymbol{x}_1 + \overbrace{(-1)\cdot\left(\boldsymbol{\Delta}(5)\odot \boldsymbol{1}_0\right)}^{=(-1,0)}$ \\
$x$-check & $\mathrm{C}^{1}\mathrm{U}_{\bar{\Omega}}$ & $\ket{\boldsymbol{x}_3}\ket{5}\ket{1}_\mathrm{O}\ket{00}_\mathrm{SR}$ & $\boldsymbol{x}_3 \in \Omega$ \\
$x$-check & $\mathrm{C}^{n_q + 1}\mathrm{U_{S_x}}(\boldsymbol{\Delta}(5), +1)$ & $\ket{\boldsymbol{x}_1}\ket{5}\ket{1}_\mathrm{O}\ket{00}_\mathrm{SR}$ & $\boldsymbol{x}_1 = \boldsymbol{x}_3 + \overbrace{1\cdot\left(\boldsymbol{\Delta}(5)\odot \boldsymbol{1}_0\right)}^{=(1,0)}$ \\
$y$-check & $\mathrm{C}^{n_q + 1}\mathrm{U_{S_y}}(\boldsymbol{\Delta}(5), -1)$ & $\ket{\boldsymbol{x}_2}\ket{5}\ket{1}_\mathrm{O}\ket{00}_\mathrm{SR}$ & $\boldsymbol{x}_2 = \boldsymbol{x}_1 + \overbrace{(-1)\cdot\left(\boldsymbol{\Delta}(5)\odot \boldsymbol{1}_1\right)}^{=(0,-1)}$ \\
$y$-check & $\mathrm{C}^{1}\mathrm{U}_{\bar{\Omega}}$ & $\ket{\boldsymbol{x}_2}\ket{5}\ket{1}_\mathrm{O}\ket{00}_\mathrm{SR}$ & $\boldsymbol{x}_2 \in \Omega$ \\
$y$-check & $\mathrm{C}^{n_q + 1}\mathrm{U_{S_y}}(\boldsymbol{\Delta}(5), +1)$ & $\ket{\boldsymbol{x}_1}\ket{5}\ket{1}_\mathrm{O}\ket{00}_\mathrm{SR}$ & $\boldsymbol{x}_1 = \boldsymbol{x}_2 + \overbrace{1\cdot\left(\boldsymbol{\Delta}(5)\odot \boldsymbol{1}_1\right)}^{=(0,1)}$ \\
\hline
\end{tabular}
\caption{Step-wise correctness check of \Cref{alg:bp-2-flag-crossing} implemented by the circuit
  shown in \Cref{fig:bp-2-za-sr-check} for the diagonal velocity $v=5$ with $\boldsymbol{\Delta}(5)=(1, 1)$ and $\sigma=-1$
  in the concave corner case.
  The initial state is that after Steps \ref{alg:bp-2-sr:step1} and \ref{alg:bp-2-sr:step2} of
  \Cref{alg:bp-2-sr}. \label{tab:bp-sup-sr-check-proof-forward-2}}
\end{table}

\begin{table}[H]
\centering
\begin{tabular}{@{}c|c|c|c@{}}
\hline
Stage & Operator & State & Remark \\
\hline
Init. &  -- & $\ket{\boldsymbol{x}_1}\ket{7}\ket{1}_\mathrm{O}\ket{00}_\mathrm{SR}$ & --\\
$x$-check & $\mathrm{C}^{n_q + 1}\mathrm{U_{S_x}}(\boldsymbol{\Delta}(7), +1)$ & $\ket{\boldsymbol{x}_3}\ket{7}\ket{1}_\mathrm{O}\ket{00}_\mathrm{SR}$ & $\boldsymbol{x}_3 = \boldsymbol{x}_1 + \overbrace{1\cdot\left(\boldsymbol{\Delta}(7)\odot \boldsymbol{1}_0\right)}^\text{=(-1,0)}$ \\
$x$-check & $\mathrm{C}^{1}\mathrm{U}_{\bar{\Omega}}$ & $\ket{\boldsymbol{x}_3}\ket{7}\ket{1}_\mathrm{O}\ket{00}_\mathrm{SR}$ & $\boldsymbol{x}_3 \in \Omega$ \\
$x$-check & $\mathrm{C}^{n_q + 1}\mathrm{U_{S_x}}(\boldsymbol{\Delta}(7), -1)$ & $\ket{\boldsymbol{x}_1}\ket{7}\ket{1}_\mathrm{O}\ket{00}_\mathrm{SR}$ & $\boldsymbol{x}_1 = \boldsymbol{x}_3 + \overbrace{(-1)\cdot\left(\boldsymbol{\Delta}(7)\odot \boldsymbol{1}_0\right)}^\text{=(1,0)}$ \\
$y$-check & $\mathrm{C}^{n_q + 1}\mathrm{U_{S_y}}(\boldsymbol{\Delta}(7), +1)$ & $\ket{\boldsymbol{x}_2}\ket{7}\ket{1}_\mathrm{O}\ket{00}_\mathrm{SR}$ & $\boldsymbol{x}_2 = \boldsymbol{x}_1 + \overbrace{1\cdot\left(\boldsymbol{\Delta}(7)\odot \boldsymbol{1}_1\right)}^\text{=(0,-1)}$ \\
$y$-check & $\mathrm{C}^{1}\mathrm{U}_{\bar{\Omega}}$ & $\ket{\boldsymbol{x}_2}\ket{7}\ket{1}_\mathrm{O}\ket{00}_\mathrm{SR}$ & $\boldsymbol{x}_2 \in \Omega$ \\
$y$-check & $\mathrm{C}^{n_q + 1}\mathrm{U_{S_y}}(\boldsymbol{\Delta}(7), -1)$ & $\ket{\boldsymbol{x}_1}\ket{7}\ket{1}_\mathrm{O}\ket{00}_\mathrm{SR}$ & $\boldsymbol{x}_1 = \boldsymbol{x}_2 + \overbrace{(-1)\cdot\left(\boldsymbol{\Delta}(7)\odot \boldsymbol{1}_1\right)}^\text{=(0,1)}$ \\
\hline
\end{tabular}
\caption{Step-wise correctness check of \Cref{alg:bp-2-flag-crossing} implemented by the circuit
  shown in \Cref{fig:bp-2-za-sr-check} for the diagonal velocity $v=5$ with $\boldsymbol{\Delta}(5)=(1, 1)$ and $\sigma=+1$
  in the concave corner case.
  This step takes place after the velocity has been reflected into $v=7$.
  The initial state is that after Steps \ref{alg:bp-2-sr:step1} through \ref{alg:bp-2-sr:step6} of
  \Cref{alg:bp-2-sr}. \label{tab:bp-sup-sr-check-proof-backward-2}}
\end{table}

\begin{table}[H]
\centering
\begin{tabular}{@{}c|c|c|c@{}}
\hline
Step & Operator & State & Remark \\
\hline
Init. & -- & $\ket{\boldsymbol{x}_0}\ket{5}\ket{0}_\mathrm{O}\ket{00}_\mathrm{SR}$ &  -- \\
Step \ref{alg:bp-2-sr:step1} & $\mathrm{U_S}$ & $\ket{\boldsymbol{x}_1}\ket{5}\ket{0}_\mathrm{O}\ket{00}_\mathrm{SR}$ & $\boldsymbol{x}_1 = \boldsymbol{x}_0 + \boldsymbol{\Delta}(5)$ \\
Step \ref{alg:bp-2-sr:step2} & $\mathrm{U_\Omega}$ & $\ket{\boldsymbol{x}_1}\ket{5}\ket{1}_\mathrm{O}\ket{00}_\mathrm{SR}$ & $\boldsymbol{x}_1 \in \Omega$ \\
Step \ref{alg:bp-2-sr:step3} & $\mathrm{C^1U_{CF}(-1)}$ & $\ket{\boldsymbol{x}_1}\ket{5}\ket{1}_\mathrm{O}\ket{00}_\mathrm{SR}$ & See \Cref{tab:bp-sup-sr-check-proof-forward-2}\\
Step \ref{alg:bp-2-sr:step4} & $\mathrm{C}^1\mathrm{U_P}$ & $\ket{\boldsymbol{x}_1}\ket{7}\ket{1}_\mathrm{O}\ket{00}_\mathrm{SR}$ & $\ket{5}_\mathrm{V} \leftrightarrow \ket{7}_\mathrm{V}$ \\
Step \ref{alg:bp-2-sr:step5} & $\mathrm{C}^1\mathrm{U_{S_x}}\mathrm{C}^1\mathrm{U_{S_y}}$ & $\ket{\boldsymbol{x}_0}\ket{7}\ket{1}_\mathrm{O}\ket{00}_\mathrm{SR}$ & $a_x = a_y = 0$ and $\boldsymbol{x}_0 = \boldsymbol{x}_1 + \boldsymbol{\Delta}(7)$\\
Step \ref{alg:bp-2-sr:step6} & $\mathrm{U_S}^\dagger$ & $\ket{\boldsymbol{x}_1}\ket{7}\ket{1}_\mathrm{O}\ket{00}_\mathrm{SR}$ & $\boldsymbol{x}_1 = \boldsymbol{x}_0 - \boldsymbol{\Delta}(7)$ \\
Step \ref{alg:bp-2-sr:step7} & $\mathrm{C^1U_{CF}(+1)}$  & $\ket{\boldsymbol{x}_1}\ket{7}\ket{1}_\mathrm{O}\ket{00}_\mathrm{SR}$ & See \Cref{tab:bp-sup-sr-check-proof-backward-2}\\
Step \ref{alg:bp-2-sr:step8} & $\mathrm{U_\Omega}$ & $\ket{\boldsymbol{x}_1}\ket{7}\ket{0}_\mathrm{O}\ket{00}_\mathrm{SR}$ & $\boldsymbol{x}_1 \in \Omega$ \\
Step \ref{alg:bp-2-sr:step9} & $\mathrm{U_S}$ & $\ket{\boldsymbol{x}_0}\ket{7}\ket{0}_\mathrm{O}\ket{00}_\mathrm{SR}$ & $\boldsymbol{x}_0 = \boldsymbol{x}_1 + \boldsymbol{\Delta}(7)$ \\

\hline
\end{tabular}
\caption{Step-wise correctness check of \Cref{alg:bp-2-sr} for the concave corner
crossing case for velocity index $5$. \label{tab:bp-sup-sr-check-proof-full-2}}
\end{table}

\subsubsection{Diagonal velocities: $x$-wall}

We refer to walls that reflect \emph{only} the $x$-velocity
component of populations as $x$-walls.
The position of such walls is fixed in the $x$-dimension
and may span a range in the $y$-dimension.
\Cref{fig:bp-sup-sr-proof-3} visually depicts this geometric pattern,
which is characterized by $\boldsymbol{x}_2 \in \Omega$ and $\boldsymbol{x}_3 \notin \Omega$.
We consider once more the evolution prescribed by \Cref{alg:bp-2-flag-crossing},
both forward and backwards,
and the full \Cref{alg:bp-2-sr}.
The traces are available in \Cref{tab:bp-sup-sr-check-proof-forward-3},
\Cref{tab:bp-sup-sr-check-proof-backward-3},
and \Cref{tab:bp-sup-sr-check-proof-full-3}, respectively.
Wall cases differ from their corner counterparts
by the fact that only one of the components get reflected, while the other
is streamed regularly.
Despite this difference, the backward trajectory
inference correctly queries gridpoints that abide by the
correct positional properties, as \Cref{tab:bp-sup-sr-check-proof-backward-3} describes.
The outcome of the algorithm is then

\begin{equation}
  \ket{\boldsymbol{x}_0}\ket{v \in \{5,6,7,8\}}\ket{0}\ket{00} \mapsto \ket{\boldsymbol{x}_3}\left(\mathrm{U_{P_x}}\ket{v}\right)\ket{0}\ket{00}.
\end{equation}

\begin{table}[H]
\centering
\begin{tabular}{@{}c|c|c|c@{}}
\hline
Stage & Operator & State & Remark \\
\hline
Init. &  -- & $\ket{\boldsymbol{x}_1}\ket{5}\ket{1}_\mathrm{O}\ket{00}_\mathrm{SR}$ & --\\
$x$-check & $\mathrm{C}^{n_q + 1}\mathrm{U_{S_x}}(\boldsymbol{\Delta}(5), -1)$ & $\ket{\boldsymbol{x}_3}\ket{5}\ket{1}_\mathrm{O}\ket{00}_\mathrm{SR}$ & $\boldsymbol{x}_3 = \boldsymbol{x}_1 + \overbrace{(-1)\cdot\left(\boldsymbol{\Delta}(5)\odot \boldsymbol{1}_0\right)}^{=(-1,0)}$ \\
$x$-check & $\mathrm{C}^{1}\mathrm{U}_{\bar{\Omega}}$ & $\ket{\boldsymbol{x}_3}\ket{5}\ket{1}_\mathrm{O}\ket{10}_\mathrm{SR}$ & $\boldsymbol{x}_3 \notin \Omega$ \\
$x$-check & $\mathrm{C}^{n_q + 1}\mathrm{U_{S_x}}(\boldsymbol{\Delta}(5), +1)$ & $\ket{\boldsymbol{x}_1}\ket{5}\ket{1}_\mathrm{O}\ket{10}_\mathrm{SR}$ & $\boldsymbol{x}_1 = \boldsymbol{x}_3 + \overbrace{1\cdot\left(\boldsymbol{\Delta}(5)\odot \boldsymbol{1}_0\right)}^{=(1,0)}$ \\
$y$-check & $\mathrm{C}^{n_q + 1}\mathrm{U_{S_y}}(\boldsymbol{\Delta}(5), -1)$ & $\ket{\boldsymbol{x}_2}\ket{5}\ket{1}_\mathrm{O}\ket{10}_\mathrm{SR}$ & $\boldsymbol{x}_2 = \boldsymbol{x}_1 + \overbrace{(-1)\cdot\left(\boldsymbol{\Delta}(5)\odot \boldsymbol{1}_1\right)}^{=(0,-1)}$ \\
$y$-check & $\mathrm{C}^{1}\mathrm{U}_{\bar{\Omega}}$ & $\ket{\boldsymbol{x}_2}\ket{5}\ket{1}_\mathrm{O}\ket{10}_\mathrm{SR}$ & $\boldsymbol{x}_2 \in \Omega$ \\
$y$-check & $\mathrm{C}^{n_q + 1}\mathrm{U_{S_y}}(\boldsymbol{\Delta}(5), +1)$ & $\ket{\boldsymbol{x}_1}\ket{5}\ket{1}_\mathrm{O}\ket{10}_\mathrm{SR}$ & $\boldsymbol{x}_1 = \boldsymbol{x}_2 + \overbrace{1\cdot\left(\boldsymbol{\Delta}(5)\odot \boldsymbol{1}_1\right)}^{=(0,1)}$ \\
\hline
\end{tabular}
\caption{Step-wise correctness check of \Cref{alg:bp-2-flag-crossing} implemented by the circuit
  shown in \Cref{fig:bp-2-za-sr-check} for the diagonal velocity $v=5$ with $\boldsymbol{\Delta}(5)=(1, 1)$ and $\sigma=-1$
  in the $x$-wall case.
  The initial state is that after Steps \ref{alg:bp-2-sr:step1} and \ref{alg:bp-2-sr:step2} of
  \Cref{alg:bp-2-sr}. \label{tab:bp-sup-sr-check-proof-forward-3}}
\end{table}

\begin{table}[H]
\centering
\begin{tabular}{@{}c|c|c|c@{}}
\hline
Stage & Operator & State & Remark \\
\hline
Init. &  -- & $\ket{\boldsymbol{x}_2}\ket{6}\ket{1}_\mathrm{O}\ket{10}_\mathrm{SR}$ & --\\
$x$-check & $\mathrm{C}^{n_q + 1}\mathrm{U_{S_x}}(\boldsymbol{\Delta}(6), +1)$ & $\ket{\boldsymbol{x}_0}\ket{6}\ket{1}_\mathrm{O}\ket{10}_\mathrm{SR}$ & $\boldsymbol{x}_0 = \boldsymbol{x}_2 + \overbrace{1\cdot\left(\boldsymbol{\Delta}(6)\odot \boldsymbol{1}_0\right)}^\text{=(-1,0)}$ \\
$x$-check & $\mathrm{C}^{1}\mathrm{U}_{\bar{\Omega}}$ & $\ket{\boldsymbol{x}_0}\ket{6}\ket{1}_\mathrm{O}\ket{00}_\mathrm{SR}$ & $\boldsymbol{x}_0 \notin \Omega$ \\
$x$-check & $\mathrm{C}^{n_q + 1}\mathrm{U_{S_x}}(\boldsymbol{\Delta}(6), -1)$ & $\ket{\boldsymbol{x}_2}\ket{6}\ket{1}_\mathrm{O}\ket{00}_\mathrm{SR}$ & $\boldsymbol{x}_2 = \boldsymbol{x}_0 + \overbrace{(-1)\cdot\left(\boldsymbol{\Delta}(6)\odot \boldsymbol{1}_0\right)}^\text{=(1,0)}$ \\
$y$-check & $\mathrm{C}^{n_q + 1}\mathrm{U_{S_y}}(\boldsymbol{\Delta}(6), +1)$ & $\ket{\boldsymbol{x}_1}\ket{6}\ket{1}_\mathrm{O}\ket{00}_\mathrm{SR}$ & $\boldsymbol{x}_1 = \boldsymbol{x}_2 + \overbrace{1\cdot\left(\boldsymbol{\Delta}(6)\odot \boldsymbol{1}_1\right)}^\text{=(0,1)}$ \\
$y$-check & $\mathrm{C}^{1}\mathrm{U}_{\bar{\Omega}}$ & $\ket{\boldsymbol{x}_1}\ket{6}\ket{1}_\mathrm{O}\ket{00}_\mathrm{SR}$ & $\boldsymbol{x}_1 \in \Omega$ \\
$y$-check & $\mathrm{C}^{n_q + 1}\mathrm{U_{S_y}}(\boldsymbol{\Delta}(6), -1)$ & $\ket{\boldsymbol{x}_2}\ket{6}\ket{1}_\mathrm{O}\ket{00}_\mathrm{SR}$ & $\boldsymbol{x}_2 = \boldsymbol{x}_1 + \overbrace{(-1)\cdot\left(\boldsymbol{\Delta}(6)\odot \boldsymbol{1}_1\right)}^\text{=(0,-1)}$ \\
\hline
\end{tabular}
\caption{Step-wise correctness check of \Cref{alg:bp-2-flag-crossing} implemented by the circuit
  shown in \Cref{fig:bp-2-za-sr-check} for the diagonal velocity $v=5$ with $\boldsymbol{\Delta}(5)=(1, 1)$ and $\sigma=+1$
  in the $x$-wall case.
  This step takes place after the velocity has been reflected into $v=6$.
  The initial state is that after Steps \ref{alg:bp-2-sr:step1} through \ref{alg:bp-2-sr:step6} of
  \Cref{alg:bp-2-sr}. \label{tab:bp-sup-sr-check-proof-backward-3}}
\end{table}

\begin{table}[H]
\centering
\begin{tabular}{@{}c|c|c|c@{}}
\hline
Step & Operator & State & Remark \\
\hline
Init. & -- & $\ket{\boldsymbol{x}_0}\ket{5}\ket{0}_\mathrm{O}\ket{00}_\mathrm{SR}$ &  -- \\
Step \ref{alg:bp-2-sr:step1} & $\mathrm{U_S}$ & $\ket{\boldsymbol{x}_1}\ket{5}\ket{0}_\mathrm{O}\ket{00}_\mathrm{SR}$ & $\boldsymbol{x}_1 = \boldsymbol{x}_0 + \boldsymbol{\Delta}(5)$ \\
Step \ref{alg:bp-2-sr:step2} & $\mathrm{U_\Omega}$ & $\ket{\boldsymbol{x}_1}\ket{5}\ket{1}_\mathrm{O}\ket{00}_\mathrm{SR}$ & $\boldsymbol{x}_1 \in \Omega$ \\
Step \ref{alg:bp-2-sr:step3} & $\mathrm{C^1U_{CF}(-1)}$ & $\ket{\boldsymbol{x}_1}\ket{5}\ket{1}_\mathrm{O}\ket{10}_\mathrm{SR}$ & See \Cref{tab:bp-sup-sr-check-proof-forward-3}\\
Step \ref{alg:bp-2-sr:step4} & $\mathrm{C}^1\mathrm{U_P}$ & $\ket{\boldsymbol{x}_1}\ket{6}\ket{1}_\mathrm{O}\ket{10}_\mathrm{SR}$ & $\ket{5}_\mathrm{V} \leftrightarrow \ket{6}_\mathrm{V}$ \\
Step \ref{alg:bp-2-sr:step5} & $\mathrm{C}^1\mathrm{U_{S_x}}\mathrm{C}^1\mathrm{U_{S_y}}$ & $\ket{\boldsymbol{x}_3}\ket{6}\ket{1}_\mathrm{O}\ket{10}_\mathrm{SR}$ & $a_x =1,\ a_y = 0,\ \boldsymbol{x}_3 = \boldsymbol{x}_1 + \boldsymbol{\Delta}(6)\odot \boldsymbol{1}_0$\\
Step \ref{alg:bp-2-sr:step6} & $\mathrm{U_S}^\dagger$ & $\ket{\boldsymbol{x}_2}\ket{6}\ket{1}_\mathrm{O}\ket{10}_\mathrm{SR}$ & $\boldsymbol{x}_2 = \boldsymbol{x}_3 - \boldsymbol{\Delta}(6)$ \\
Step \ref{alg:bp-2-sr:step7} & $\mathrm{C^1U_{CF}(+1)}$  & $\ket{\boldsymbol{x}_2}\ket{6}\ket{1}_\mathrm{O}\ket{00}_\mathrm{SR}$ & See \Cref{tab:bp-sup-sr-check-proof-backward-3}\\
Step \ref{alg:bp-2-sr:step8} & $\mathrm{U_\Omega}$ & $\ket{\boldsymbol{x}_2}\ket{6}\ket{0}_\mathrm{O}\ket{00}_\mathrm{SR}$ & $\boldsymbol{x}_2 \in \Omega$ \\
Step \ref{alg:bp-2-sr:step9} & $\mathrm{U_S}$ & $\ket{\boldsymbol{x}_3}\ket{6}\ket{0}_\mathrm{O}\ket{00}_\mathrm{SR}$ & $\boldsymbol{x}_3 = \boldsymbol{x}_2 + \boldsymbol{\Delta}(6)$ \\

\hline
\end{tabular}
\caption{Step-wise correctness check of \Cref{alg:bp-2-sr} for the $x$-wall case
crossing case for velocity index $5$. \label{tab:bp-sup-sr-check-proof-full-3}}
\end{table}

\subsubsection{Diagonal velocities: $y$-wall}

The final case we consider is when particles cross the solid
boundary through a $y$-wall, which is a surface
aligned with the $x$-axis that only reflects
the $y$ velocity component.
\Cref{fig:bp-sup-sr-proof-2} shows this instance,
with $\boldsymbol{x}_2 \notin \Omega$ and $\boldsymbol{x}_3 \in \Omega$.
The state evolution through the steps 
of \Cref{alg:bp-2-flag-crossing} is
traced by \Cref{tab:bp-sup-sr-check-proof-forward-4} for the forward case
and \Cref{tab:bp-sup-sr-check-proof-backward-4} for the backward case.
The full evolution of \Cref{alg:bp-2-sr}
is given by \Cref{tab:bp-sup-sr-check-proof-full-4}.
Indeed, all three algorithms are symmetric to the $x$-wall,
with the second $\mathrm{SR}$ qubit being toggled instead of the first.
The operator performs the transformation

\begin{equation}
  \ket{\boldsymbol{x}_0}\ket{v \in \{5,6,7,8\}}\ket{0}\ket{00} \mapsto \ket{\boldsymbol{x}_2}\left(\mathrm{U_{P_y}}\ket{v}\right)\ket{0}\ket{00}.
\end{equation}

This case finalizes the analysis of the diagonal boundary crossings for $v=5$.
As noted, the cases for $v\in\{6,7,8\}$ follow from symmetry
by permutations of the positions of the four relevant positions and origin.
Since all possible physical realizations have been addressed, we conclude the proof. $\square$

\begin{table}[H]
\centering
\begin{tabular}{@{}c|c|c|c@{}}
\hline
Stage & Operator & State & Remark \\
\hline
Init. &  -- & $\ket{\boldsymbol{x}_1}\ket{5}\ket{1}_\mathrm{O}\ket{00}_\mathrm{SR}$ & --\\
$x$-check & $\mathrm{C}^{n_q + 1}\mathrm{U_{S_x}}(\boldsymbol{\Delta}(5), -1)$ & $\ket{\boldsymbol{x}_3}\ket{5}\ket{1}_\mathrm{O}\ket{00}_\mathrm{SR}$ & $\boldsymbol{x}_3 = \boldsymbol{x}_1 + \overbrace{(-1)\cdot\left(\boldsymbol{\Delta}(5)\odot \boldsymbol{1}_0\right)}^{=(-1,0)}$ \\
$x$-check & $\mathrm{C}^{1}\mathrm{U}_{\bar{\Omega}}$ & $\ket{\boldsymbol{x}_3}\ket{5}\ket{1}_\mathrm{O}\ket{00}_\mathrm{SR}$ & $\boldsymbol{x}_3 \in \Omega$ \\
$x$-check & $\mathrm{C}^{n_q + 1}\mathrm{U_{S_x}}(\boldsymbol{\Delta}(5), +1)$ & $\ket{\boldsymbol{x}_1}\ket{5}\ket{1}_\mathrm{O}\ket{00}_\mathrm{SR}$ & $\boldsymbol{x}_1 = \boldsymbol{x}_3 + \overbrace{1\cdot\left(\boldsymbol{\Delta}(5)\odot \boldsymbol{1}_0\right)}^{=(1,0)}$ \\
$y$-check & $\mathrm{C}^{n_q + 1}\mathrm{U_{S_y}}(\boldsymbol{\Delta}(5), -1)$ & $\ket{\boldsymbol{x}_2}\ket{5}\ket{1}_\mathrm{O}\ket{00}_\mathrm{SR}$ & $\boldsymbol{x}_2 = \boldsymbol{x}_1 + \overbrace{(-1)\cdot\left(\boldsymbol{\Delta}(5)\odot \boldsymbol{1}_1\right)}^{=(0,-1)}$ \\
$y$-check & $\mathrm{C}^{1}\mathrm{U}_{\bar{\Omega}}$ & $\ket{\boldsymbol{x}_2}\ket{5}\ket{1}_\mathrm{O}\ket{01}_\mathrm{SR}$ & $\boldsymbol{x}_2 \notin \Omega$ \\
$y$-check & $\mathrm{C}^{n_q + 1}\mathrm{U_{S_y}}(\boldsymbol{\Delta}(5), +1)$ & $\ket{\boldsymbol{x}_1}\ket{5}\ket{1}_\mathrm{O}\ket{01}_\mathrm{SR}$ & $\boldsymbol{x}_1 = \boldsymbol{x}_2 + \overbrace{1\cdot\left(\boldsymbol{\Delta}(5)\odot \boldsymbol{1}_1\right)}^{=(0,1)}$ \\
\hline
\end{tabular}
\caption{Step-wise correctness check of \Cref{alg:bp-2-flag-crossing} implemented by the circuit
  shown in \Cref{fig:bp-2-za-sr-check} for the diagonal velocity $v=5$ with $\boldsymbol{\Delta}(5)=(1, 1)$ and $\sigma=-1$
  in the $y$-wall case.
  The initial state is that after Steps \ref{alg:bp-2-sr:step1} and \ref{alg:bp-2-sr:step2} of
  \Cref{alg:bp-2-sr}. \label{tab:bp-sup-sr-check-proof-forward-4}}
\end{table}

\begin{table}[H]
\centering
\begin{tabular}{@{}c|c|c|c@{}}
\hline
Stage & Operator & State & Remark \\
\hline
Init. &  -- & $\ket{\boldsymbol{x}_3}\ket{8}\ket{1}_\mathrm{O}\ket{01}_\mathrm{SR}$ & --\\
$x$-check & $\mathrm{C}^{n_q + 1}\mathrm{U_{S_x}}(\boldsymbol{\Delta}(8), +1)$ & $\ket{\boldsymbol{x}_1}\ket{8}\ket{1}_\mathrm{O}\ket{01}_\mathrm{SR}$ & $\boldsymbol{x}_1 = \boldsymbol{x}_3 + \overbrace{1\cdot\left(\boldsymbol{\Delta}(8)\odot \boldsymbol{1}_0\right)}^\text{=(1,0)}$ \\
$x$-check & $\mathrm{C}^{1}\mathrm{U}_{\bar{\Omega}}$ & $\ket{\boldsymbol{x}_1}\ket{8}\ket{1}_\mathrm{O}\ket{01}_\mathrm{SR}$ & $\boldsymbol{x}_1 \in \Omega$ \\
$x$-check & $\mathrm{C}^{n_q + 1}\mathrm{U_{S_x}}(\boldsymbol{\Delta}(8), -1)$ & $\ket{\boldsymbol{x}_3}\ket{8}\ket{1}_\mathrm{O}\ket{01}_\mathrm{SR}$ & $\boldsymbol{x}_3 = \boldsymbol{x}_1 + \overbrace{(-1)\cdot\left(\boldsymbol{\Delta}(8)\odot \boldsymbol{1}_0\right)}^\text{=(-1,0)}$ \\
$y$-check & $\mathrm{C}^{n_q + 1}\mathrm{U_{S_y}}(\boldsymbol{\Delta}(8), +1)$ & $\ket{\boldsymbol{x}_0}\ket{8}\ket{1}_\mathrm{O}\ket{01}_\mathrm{SR}$ & $\boldsymbol{x}_0 = \boldsymbol{x}_3 + \overbrace{1\cdot\left(\boldsymbol{\Delta}(8)\odot \boldsymbol{1}_1\right)}^\text{=(0,-1)}$ \\
$y$-check & $\mathrm{C}^{1}\mathrm{U}_{\bar{\Omega}}$ & $\ket{\boldsymbol{x}_0}\ket{8}\ket{1}_\mathrm{O}\ket{00}_\mathrm{SR}$ & $\boldsymbol{x}_0 \notin \Omega$ \\
$y$-check & $\mathrm{C}^{n_q + 1}\mathrm{U_{S_y}}(\boldsymbol{\Delta}(8), -1)$ & $\ket{\boldsymbol{x}_3}\ket{8}\ket{1}_\mathrm{O}\ket{00}_\mathrm{SR}$ & $\boldsymbol{x}_3 = \boldsymbol{x}_0 + \overbrace{(-1)\cdot\left(\boldsymbol{\Delta}(8)\odot \boldsymbol{1}_1\right)}^\text{=(0,1)}$ \\
\hline
\end{tabular}
\caption{Step-wise correctness check of \Cref{alg:bp-2-flag-crossing} implemented by the circuit
  shown in \Cref{fig:bp-2-za-sr-check} for the diagonal velocity $v=5$ with $\boldsymbol{\Delta}(5)=(1, 1)$ and $\sigma=+1$
  in the $y$-wall case.
  This step takes place after the velocity has been reflected into $v=8$.
  The initial state is that after Steps \ref{alg:bp-2-sr:step1} through \ref{alg:bp-2-sr:step6} of
  \Cref{alg:bp-2-sr}. \label{tab:bp-sup-sr-check-proof-backward-4}}
\end{table}

\begin{table}[H]
\centering
\begin{tabular}{@{}c|c|c|c@{}}
\hline
Step & Operator & State & Remark \\
\hline
Init. & -- & $\ket{\boldsymbol{x}_0}\ket{5}\ket{0}_\mathrm{O}\ket{00}_\mathrm{SR}$ &  -- \\
Step \ref{alg:bp-2-sr:step1} & $\mathrm{U_S}$ & $\ket{\boldsymbol{x}_1}\ket{5}\ket{0}_\mathrm{O}\ket{00}_\mathrm{SR}$ & $\boldsymbol{x}_1 = \boldsymbol{x}_0 + \boldsymbol{\Delta}(5)$ \\
Step \ref{alg:bp-2-sr:step2} & $\mathrm{U_\Omega}$ & $\ket{\boldsymbol{x}_1}\ket{5}\ket{1}_\mathrm{O}\ket{00}_\mathrm{SR}$ & $\boldsymbol{x}_1 \in \Omega$ \\
Step \ref{alg:bp-2-sr:step3} & $\mathrm{C^1U_{CF}(-1)}$ & $\ket{\boldsymbol{x}_1}\ket{5}\ket{1}_\mathrm{O}\ket{01}_\mathrm{SR}$ & See \Cref{tab:bp-sup-sr-check-proof-forward-4}\\
Step \ref{alg:bp-2-sr:step4} & $\mathrm{C}^1\mathrm{U_P}$ & $\ket{\boldsymbol{x}_1}\ket{8}\ket{1}_\mathrm{O}\ket{01}_\mathrm{SR}$ & $\ket{5}_\mathrm{V} \leftrightarrow \ket{8}_\mathrm{V}$ \\
Step \ref{alg:bp-2-sr:step5} & $\mathrm{C}^1\mathrm{U_{S_x}}\mathrm{C}^1\mathrm{U_{S_y}}$ & $\ket{\boldsymbol{x}_2}\ket{8}\ket{1}_\mathrm{O}\ket{01}_\mathrm{SR}$ & $a_x = 0,\ a_y = 1,\ \boldsymbol{x}_2 = \boldsymbol{x}_1 + \boldsymbol{\Delta}(8)\odot \boldsymbol{1}_1$\\
Step \ref{alg:bp-2-sr:step6} & $\mathrm{U_S}^\dagger$ & $\ket{\boldsymbol{x}_3}\ket{8}\ket{1}_\mathrm{O}\ket{01}_\mathrm{SR}$ & $\boldsymbol{x}_3 = \boldsymbol{x}_2 - \boldsymbol{\Delta}(8)$ \\
Step \ref{alg:bp-2-sr:step7} & $\mathrm{C^1U_{CF}(+1)}$  & $\ket{\boldsymbol{x}_3}\ket{8}\ket{1}_\mathrm{O}\ket{00}_\mathrm{SR}$ & See \Cref{tab:bp-sup-sr-check-proof-backward-4}\\
Step \ref{alg:bp-2-sr:step8} & $\mathrm{U_\Omega}$ & $\ket{\boldsymbol{x}_3}\ket{8}\ket{0}_\mathrm{O}\ket{00}_\mathrm{SR}$ & $\boldsymbol{x}_3 \in \Omega$ \\
Step \ref{alg:bp-2-sr:step9} & $\mathrm{U_S}$ & $\ket{\boldsymbol{x}_2}\ket{8}\ket{0}_\mathrm{O}\ket{00}_\mathrm{SR}$ & $\boldsymbol{x}_2 = \boldsymbol{x}_3 + \boldsymbol{\Delta}(8)$ \\

\hline
\end{tabular}
\caption{Step-wise correctness check of \Cref{alg:bp-2-sr} for the $y$-wall case.
crossing case for velocity index $5$. \label{tab:bp-sup-sr-check-proof-full-4}}
\end{table}

% \section*{Acknowledgments}
% We would like to acknowledge the assistance of volunteers in putting
% together this example manuscript and supplement.
\end{document}